\pdfoutput=1
\documentclass[PAPER, cernpreprint, atlasdraft=false, texlive=2023, UKenglish, texmf, orcidlogo]{atlasdoc}
 
\usepackage{atlaspackage}
\usepackage{atlasbiblatex}
 
\usepackage{atlasphysics}

\usepackage{xcolor}
\usepackage{tabularx}
 
\graphicspath{{logos/}{figures/}}
 
\usepackage{ANA-EXOT-2022-39-PAPER-defs}

\AtlasTitle{Search for leptoquarks decaying into the b$\tau$ final state in $pp$ collisions at $\sqrt{s} =$~13 TeV with the ATLAS detector}

\AtlasAbstract{
A search for leptoquarks decaying into the $b\tau$ final state  is performed using Run 2 proton--proton collision data from the Large Hadron Collider, corresponding to an integrated luminosity of 139~fb$^{-1}$ at $\sqrt{s} = 13$ TeV recorded by the ATLAS detector.
The benchmark models considered in this search are vector leptoquarks with electric charge of 2/3e
and scalar leptoquarks with an electric charge of 4/3e.
No significant excess above the Standard Model prediction is observed, and 95\% confidence level upper limits are set on the cross-section times branching fraction of leptoquarks decaying into $b\tau$.
For the vector leptoquark production two models are considered: the Yang--Mills and Minimal coupling models.
In the Yang--Mills (Minimal coupling) scenario, vector leptoquarks with a mass below 1.58 (1.35) TeV are excluded for a gauge coupling of 1.0 and below 2.05 (1.99) TeV for a gauge coupling of 2.5.
In the case of scalar leptoquarks, masses below 1.28 (1.53) TeV are excluded for a Yukawa coupling of 1.0 (2.5).
Finally, an interpretation of the results with minimal model dependence is performed for each of the signal region categories, and limits on the  visible cross-section for beyond the Standard Model processes are provided.
}

\AtlasRefCode{EXOT-2022-39}
 
\PreprintIdNumber{CERN-EP-2023-033}

\AtlasJournal{JHEP}
\AtlasJournalRef{\JHEP 10 (2023) 001}
\AtlasDOI{10.1007/JHEP10(2023)001}

\hypersetup{pdftitle={ATLAS document},pdfauthor={The ATLAS Collaboration}}
 
\begin{document}
 
\maketitle
 
\tableofcontents

\section{Introduction}
 
The existing similarities between the structure of the quark and lepton sectors in the Standard Model (SM) suggest the possibility of a new underlying symmetry in particle physics.
Leptoquarks (LQs)  that couple to both quarks and leptons, with non-zero baryon and lepton numbers, and fractional electric charges are predicted by several beyond the SM theories that attempt to unify the fundamental interactions, such as technicolour~\cite{Dimopoulos1979,Dimopoulos:1979sp,Farhi:1980xs}, composite models~\cite{Schrempp:1984nj}, and grand unification~\cite{Pati:1973uk,PatiSalam1974,GeorgiGlashow1974}.
 
Recent results reported by BaBar~\cite{BaBar:1,BaBar:2}, Belle~\cite{Belle:1} and LHCb~\cite{LHCb:2} show hints of deviations from lepton-flavour universality in \textit{B}-meson decays into final states with $D^{(*)}$ mesons, which could be caused by the existence of LQs.
The 4.2 standard deviation disagreement with respect to the SM prediction observed in the anomalous muon magnetic moment measurement~\cite{Muong22021}, though significantly reduced when updated lattice quantum chromodynamics (QCD) calculations ~\cite{Borsanyi_2021} are considered, could be caused by LQ contributions to the muon magnetic moment~\cite{Greljo:2021xmg}.

\begin{figure}[!h]
\centering
\subfloat[]{\includegraphics[width=0.3\textwidth]{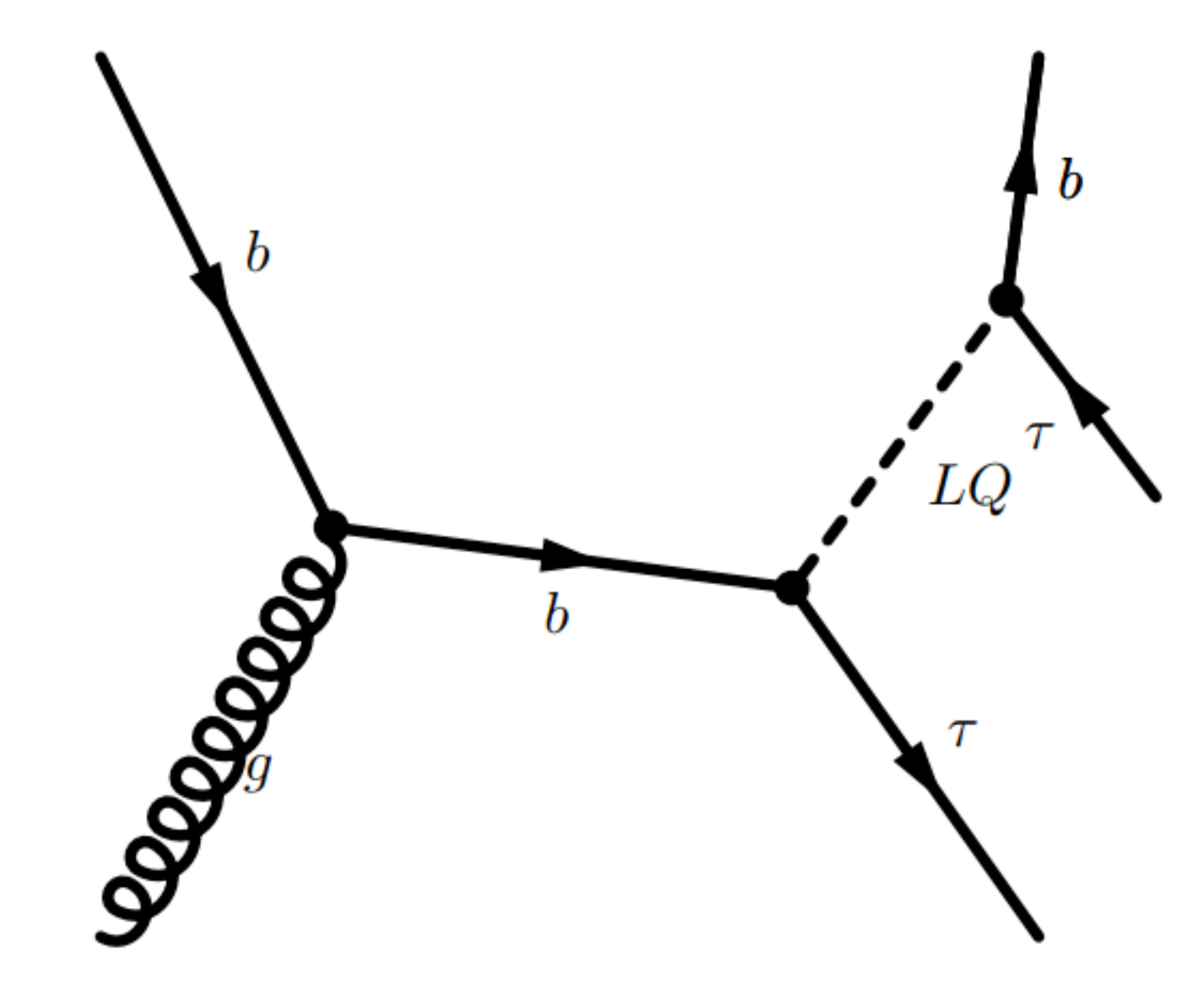} }
\subfloat[]{\includegraphics[width=0.3\textwidth]{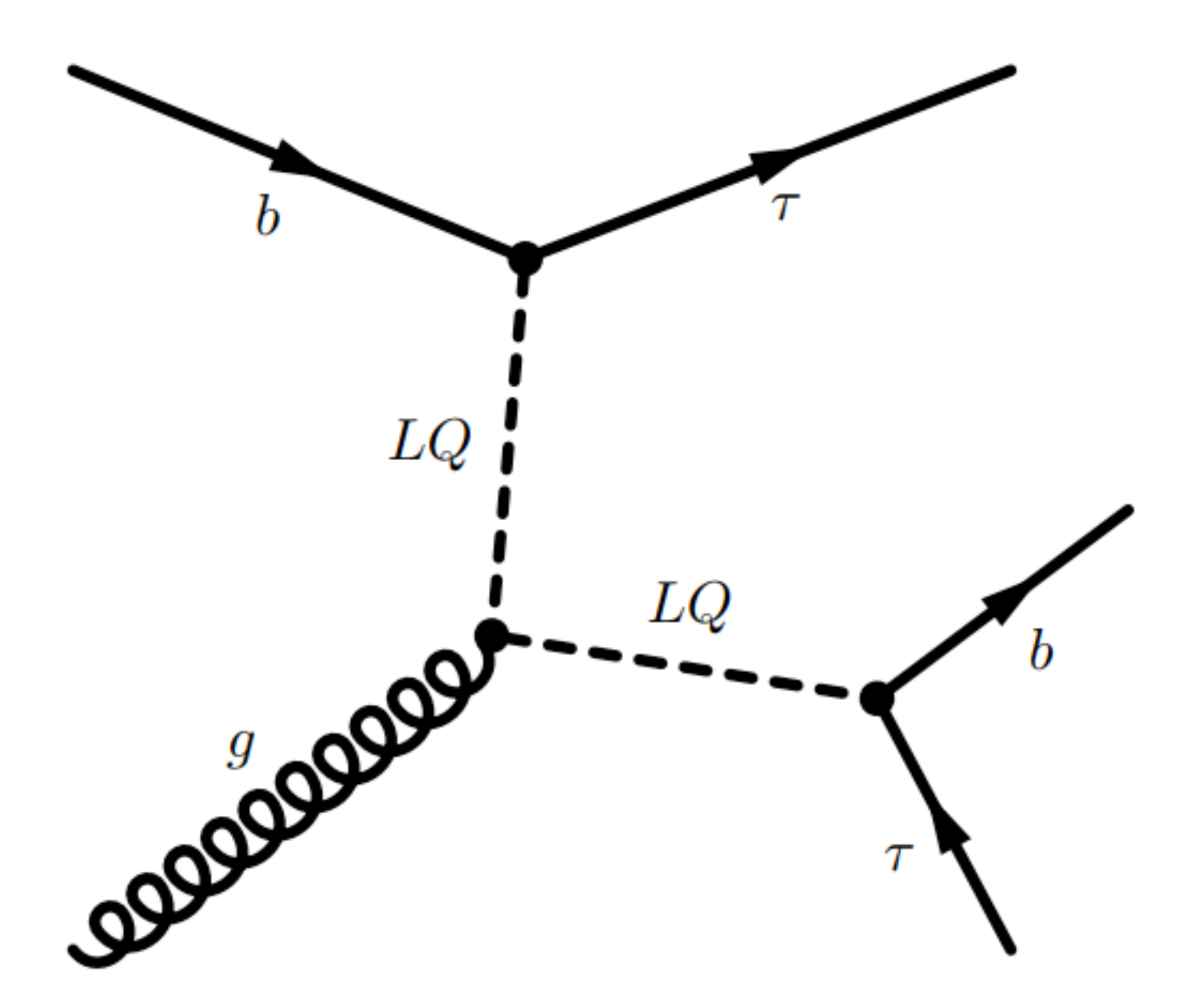} } \\
\subfloat[]{\includegraphics[width=0.3\textwidth]{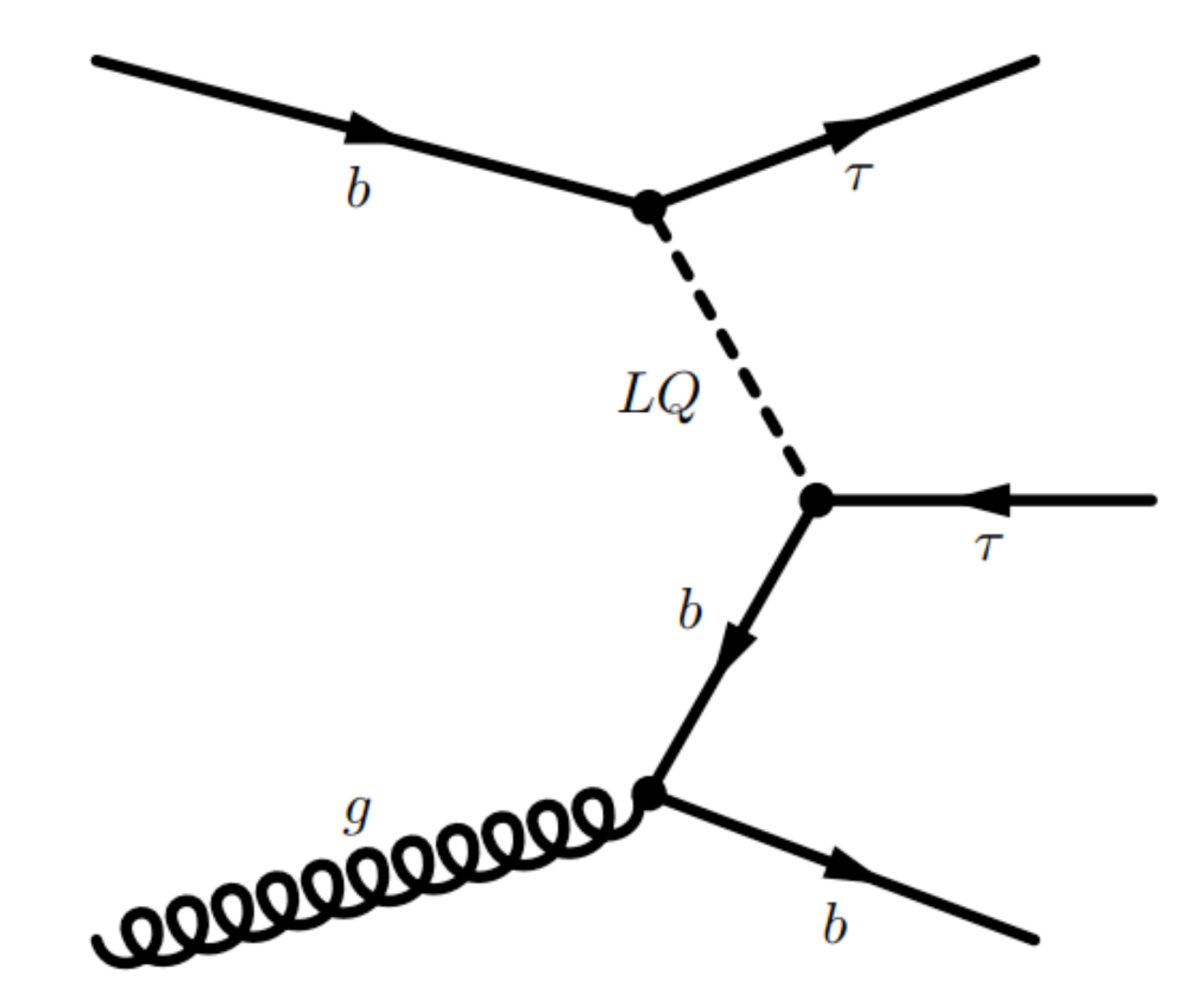} }
\subfloat[]{\includegraphics[width=0.3\textwidth]{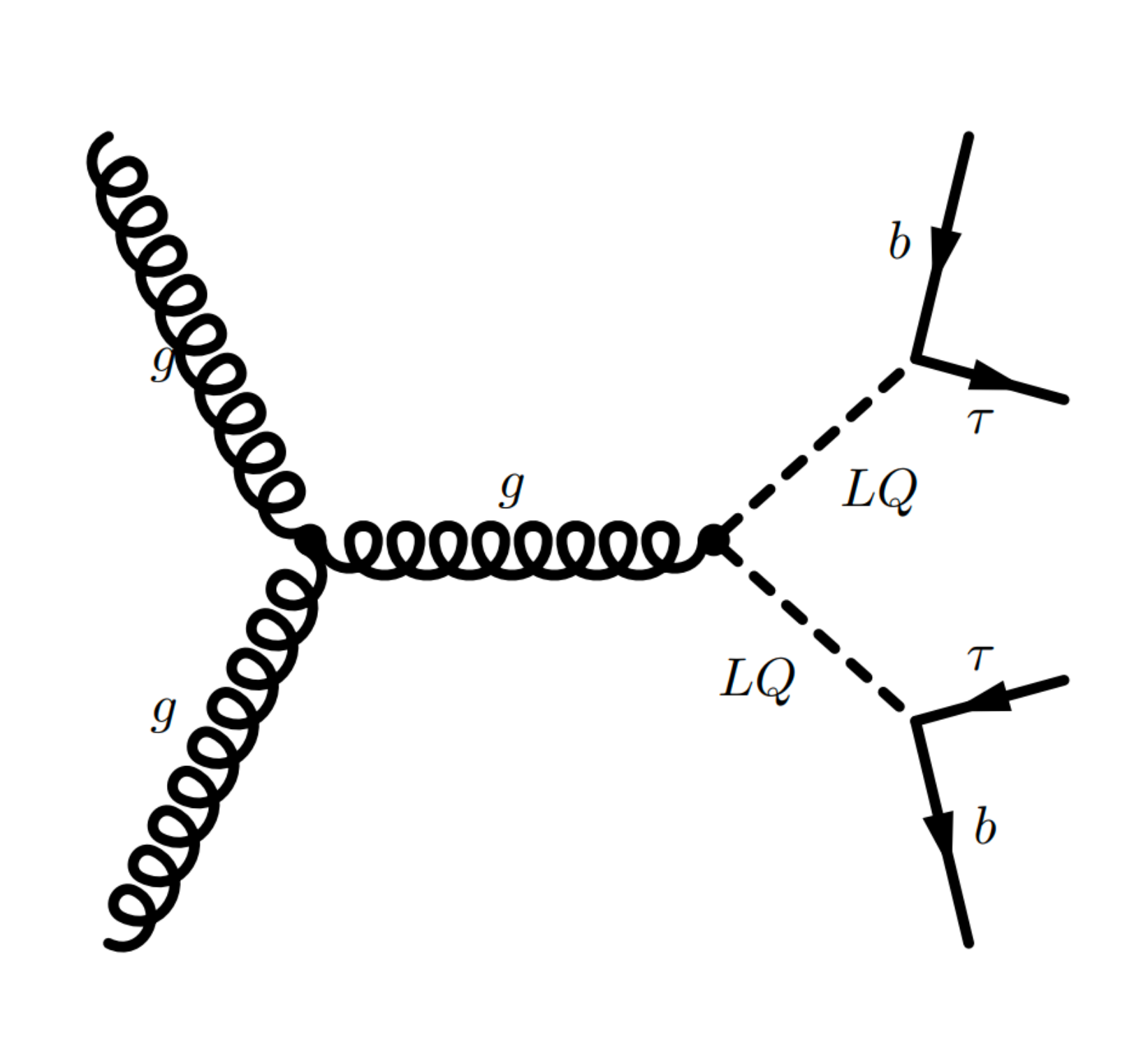} } \\
\caption{Illustrative Feynman diagrams of (a,b) single LQ production, (c) non-resonant LQ production, and (d) LQ pair production.
}
\label{fig:LQ_feyn}
\end{figure}

In light of the lepton-flavour universality anomalies observed in the  \textit{B}-meson decays into $D^{(*)}\tau\nu$ final states, the couplings of LQs to third-generation quarks and leptons are expected to be large~\cite{Buttazzo_2017}. 
At the LHC, third-generation LQs can be produced singly  via quark--gluon fusion and quark--gluon scattering or in pairs via the gluon--gluon fusion process, as shown in the Feynman diagrams in Figure~\ref{fig:LQ_feyn}. 
The search presented in this paper is optimised for the single production of third-generation LQ via the $bg \rightarrow \mathrm{LQ}\tau \rightarrow b\tau\tau$ channel, while LQ pair and non-resonant production processes are also considered since they can also contribute to the $b\tau\tau$ final state.
The single LQ production contribution becomes larger than that from LQ pair production at high LQ mass and coupling values.
The results are obtained from  proton--proton collision data at a centre-of-mass energy of $\sqrt{s}$~=~13~\TeV{} collected by the ATLAS detector~\cite{PERF-2007-01, PIX-2018-001} at the LHC~\cite{Evans:2008zzb} during Run 2 between 2015 and 2018, corresponding to a total integrated luminosity of 139~fb$^{-1}$.
 
The vector LQ model chosen for this search is the $U_{1}$ model~\cite{U1Model}, a $\text{SU(2)}_{\text{L}}$ singlet with fermion number $F = 3B + L = 0$, where $B$ and $L$ are the baryon and lepton numbers respectively, and an electric charge of $2/3e$.
The interaction part of $U_{1}$ model Lagrangian is:
 
\begin{equation*}
\mathcal{L}_{U_{1}} \supset -i g_s (1-\kappa) U_{1\mu}^{\dag} T^a U_{1\nu} G^{a\mu\nu} + \frac{g_U}{\sqrt{2}} [ U_1^\mu (\beta_{\text{L}}^{ij} \bar{q}_{\text{L}}^{i} \gamma_\mu \ell_{\text{L}}^{j} + \beta_{\text{R}}^{ij} \bar{d}_{\text{R}}^i \gamma_\mu e_{\text{R}}^j) + \text{h.c.}] ,
\end{equation*}
 
where $T^{a} = \lambda^{a}/2$ with $\lambda^{a}$($a=$1,$\cdots$,8) are the Gell-Mann matrices, $g_{s}$ is the QCD coupling, $q_{\text{L}}$ ($\ell_{\text{L}}$) denotes the left-handed quark (lepton) doublets and $d_{\text{R}}$ ($e_{\text{R}}$)  denotes the right-handed down-type quark (charged-lepton) singlets. The $i$ and $j$ indices represent the flavour generation. A summation over the colour indices is performed and omitted for clarity.
The term $-ig_{s}(1-\kappa)U_{1\mu}^{\dagger}T^{a}U_{1\nu}G^{a\mu\nu}$ describes the interaction between $U_{1}$ leptoquarks and SM gluon gauge fields $G^{a\mu\nu}$.
In this analysis, two vector LQ scenarios are considered: the Yang--Mills (\YMLQ) coupling scenario, $\kappa=0$, and the  Minimal (\minLQ) coupling scenario, $\kappa=1$.
The $\beta_{\text{L}}^{ij}$ and $\beta_{\text{R}}^{ij}$ parameters describe the coupling between $U_{1}$ leptoquarks and left-handed or right-handed charged leptons and quarks, respectively.
In the framework of the \YMLQ{} and \minLQ{} scenarios, the probability to decay into the $b$-quark and $\tau$-lepton final state is predicted to be the same as to decay into the top-quark and neutrino final state. Hence, the branching fraction $\mathcal{B}$ of the LQ decays into a $b$-quark and a $\tau$-lepton is set to 0.5.
In this search, all of $\beta_{\text{R}}^{ij}$ are set to zero, $\beta_{\text{L}}^{33}$ is set to one and other $\beta_{\text{L}}^{ij}$ are set to zero, such that each LQ decays into a $b$-quark and a $\tau$-lepton or into a top-quark and a neutrino.
Due to these choices, the gauge coupling ($\lambda$) between $U_{1}$ leptoquarks and third-generation charged leptons and quarks can be written as $\lambda = g_U \beta_{\text{L}}^{33} / \sqrt{2}$.
 
The scalar LQ model $\widetilde{S_{1}}$ is also considered, with $F = 3B + L = -2$ and electric charge of $4/3e$~\cite{BUCHMULLER1987442,DorsnerLQ2016}.
There are three parameters in this model: the branching fraction $\mathcal{B}$ into charged leptons, the LQ to $\tau b$ Yukawa coupling parameter $\lambda$, and the mass term of the LQ\@.
Following Ref.~\cite{DorsnerLQ2016}, the Lagrangian terms for $\widetilde{S_{1}}$ LQs related to this analysis are:
 
\begin{equation*}
\mathcal{L}_{\widetilde{S_{1}}} \supset +\lambda^{ij}\bar{d}_{\text{R}}^{Ci}\widetilde{S_{1}}e_{\text{R}}^{j}+\text{h.c.} ,
\end{equation*}
 
\noindent where $C$ in the superscript stands for the charge conjugation operation.
The terms $e_{\text{R}}$ and $d_{\text{R}}$ are the right-handed charged leptons and down-type quarks and
$\lambda^{ij}$ represents the Yukawa couplings between $\widetilde{S_{1}}$, charged leptons, and quarks, where the $ij$ refers to the generations of the quark and charged lepton.
In the framework of the  $\widetilde{S_{1}}$ model, the only non-zero Yukawa coupling considered in this paper is the coupling to a $b$-quark and a $\tau$-lepton.
Since only the coupling to the third-generation charged lepton and quark is considered, $\lambda^{33}=\lambda$ is assumed to be different from zero, while the rest of the $\lambda^{ij}$ are set to zero.
 
Most of the previous searches for LQs performed by the ATLAS~\cite{SUSY-2018-12,SUSY-2018-34,SUSY-2019-18,EXOT-2019-13,EXOT-2019-15,EXOT-2019-19,ATLAS:2023uox} and CMS~\cite{CMS-B2G-16-027,CMS-B2G-16-028,CMS-EXO-17-015,CMS-EXO-17-016} collaborations have been conducted on different final states compared to this search. 
For third generation LQs, the CMS Collaboration has recently published  results of searches for LQs decaying into $t\nu$ and $b\tau$~\cite{CMS-EXO-19-015-corr}. 
The ATLAS Collaboration performed a search for pair produced scalar LQs in $b\tau b\tau$ final states with 36~fb$^{-1}$ of proton--proton collision data at $\sqrt{s}=13~\TeV{}$ that excluded scalar LQs with masses below 1~\TeV{}, assuming a LQ to $b\tau$ branching fraction equal to one ~\cite{EXOT-2017-30}.

The analysis described in this paper is the first search by the ATLAS Collaboration for singly produced LQs decaying into $b\tau$.
The search is performed over a LQ mass ($\mlq$) in the range of 0.4~\TeV{} to 2.5~\TeV.
The $\lambda$ range is chosen to be between 0.5 and 2.5 to cover possible regions where LQs could explain the anomalies observed in the $B$-meson decay and is extended to large $\lambda$ where the single LQ production channel provides a significant contribution compared to the pair-production process.
In the case of the vector LQ production, the contribution from LQs decaying into $t\nu$ 
is neglected.

The analysis starts from the selection of a pair of oppositely charged $\tau$-leptons produced in association with a jet identified as containing a $b$-hadron ($b$-jet). The main backgrounds to the search are the \ttbar{} and $tW$ production processes.
Two signatures are considered, containing either a $\lephad$ or $\hadhad$ pair, where $\thad$ ($\tlep$) refers to a $\tau$-lepton decaying into hadrons and a neutrino (two neutrinos and an electron or a muon).
In each of these two analysis channels, events are classified, based on the transverse momentum ($\pt$) of the $b$-jet, in two categories of low and high $b$-jet $\pt$.
The search for LQs is only performed in the high $b$-jet $\pt$ category, where the effect from the interference of non-resonant LQ production with SM processes is expected to be small ~\cite{PhysRevD.99.055028}. The non-resonant contribution can be significantly modified by the interference contribution, which depends on the signal model parameters \cite{PhysRevD.99.055028,Bessaa:2014jya}. The effect of the interference with SM diagrams, such as those from $Z/\gamma^{*}(\to\tau\tau)+b$-jet, is neglected.
Due to the lack of detailed studies of the interference effect,
an additional search is performed considering both high and low $b$-jet $\pt$ categories and not relying on a specific LQ model choice for its signal description (henceforth called ‘model-independent’), and covering a wider range of beyond-SM signatures. For this search, the results are expressed in terms of the visible cross-section of the beyond-SM signal.
 
The next sections discuss the ATLAS detector in Section~\ref{sec:detector}, data and simulated samples in Section~\ref{sec:samples}, followed by the object reconstruction and definitions in Section~\ref{sec:recoid}.
Section~\ref{sec:eventsel} discusses the overall analysis strategy and event selection, then Section~\ref{sec:background} goes into more details of the background estimation methods.
The systematic uncertainties are discussed in Section~\ref{sec:systematics} followed by the results in Section~\ref{sec:results}, with the conclusion in Section~\ref{sec:conclusion}.


\section{The ATLAS Detector}
\label{sec:detector}
 
\newcommand{\AtlasCoordFootnote}{
ATLAS uses a right-handed coordinate system with its origin at the nominal interaction point (IP)
in the centre of the detector and the \(z\)-axis along the beam pipe.
The \(x\)-axis points from the IP to the centre of the LHC ring,
and the \(y\)-axis points upwards.
Cylindrical coordinates \((r,\phi)\) are used in the transverse plane,
\(\phi\) being the azimuthal angle around the \(z\)-axis.
The pseudorapidity is defined in terms of the polar angle \(\theta\) as \(\eta = -\ln \tan(\theta/2)\).
Angular distance is measured in units of \(\Delta R \equiv \sqrt{(\Delta\eta)^{2} + (\Delta\phi)^{2}}\).}
 
The ATLAS detector~\cite{PERF-2007-01} at the LHC covers nearly the entire solid angle around the collision point.\footnote{\AtlasCoordFootnote}
It consists of an inner tracking detector surrounded by a thin superconducting solenoid, electromagnetic and hadron calorimeters,
and a muon spectrometer incorporating three large superconducting air-core toroidal magnets.
 
The inner-detector system (ID) is immersed in a \SI{2}{\tesla} axial magnetic field
and provides charged-particle tracking in the range of \(|\eta| < 2.5\).
The high-granularity silicon pixel detector covers the vertex region and typically provides four measurements per track,
the first hit normally being in the insertable B-layer installed before Run~2~\cite{ATLAS-TDR-19,PIX-2018-001}.
It is followed by the silicon microstrip tracker, which usually provides eight measurements per track.
These silicon detectors are complemented by the transition radiation tracker (TRT),
which enables radially extended track reconstruction up to \(|\eta| = 2.0\).
The TRT also provides electron identification information
based on the fraction of hits (typically 30 in total) above a higher energy-deposit threshold corresponding to transition radiation.
 
The calorimeter system covers the pseudorapidity range of \(|\eta| <
4.9\).  Within the region \(|\eta|= 3.2\), electromagnetic calorimetry
is provided by barrel and endcap high-granularity lead/liquid-argon
(LAr) calorimeters, with an additional thin LAr presampler covering
\(|\eta| < 1.8\) to correct for energy loss in material upstream of
the calorimeters.  Hadron calorimetry is provided by the
steel/scintillator-tile calorimeter, segmented into three barrel
structures within \(|\eta| = 1.7\), and two copper/LAr hadron endcap
calorimeters.  The solid angle coverage is completed by forward
copper/LAr and tungsten/LAr calorimeter modules optimised for
electromagnetic and hadronic energy measurements respectively.
 
The muon spectrometer (MS) comprises separate trigger and
high-precision tracking chambers measuring the deflection of muons in
a magnetic field generated by the superconducting air-core toroidal
magnets.  The field integral of the toroids ranges between \num{2.0}
and \SI{6.0}{\tesla\metre} across most of the detector.  A set of
precision chambers covers the region \(|\eta| < 2.7\) with three
layers of monitored drift tubes, complemented by cathode-strip
chambers in the forward region, where the background is highest.  The
muon trigger system covers the range of \(|\eta| < 2.4\) with
resistive-plate chambers in the barrel, and thin-gap chambers in the
endcap regions.
 
Interesting events are selected by the first-level trigger system
implemented in custom hardware, followed by selections made by
algorithms implemented in software in the high-level
trigger~\cite{TRIG-2016-01}.  The first-level trigger accepts events
from the \SI{40}{\MHz} bunch crossings at a rate below \SI{100}{\kHz},
which the high-level trigger reduces in order to record events
to disk at about \SI{1}{\kHz}.
 
An extensive software suite~\cite{ATL-SOFT-PUB-2021-001} is used in data simulation, in
the reconstruction and analysis of real and simulated data, in detector
operations, and in the trigger and data acquisition systems of the experiment.


\section{Data and Monte Carlo Samples}
\label{sec:samples}
The data were collected using unprescaled single-lepton and single $\thad$ triggers. A more detailed description of the triggers used in the analysis for each data-taking period is given in \cref{sec:eventsel}.
Quality criteria are applied to events to ensure that the data were not affected by any hardware- or software-related issues~\cite{DAPR-2018-01}.
 
Monte Carlo (MC) simulated events of single LQs decaying into $b\tau$ were produced for masses ranging from 0.4~\TeV{} to 2.5~\TeV. The signal samples were produced at leading order (LO) in QCD in the five-flavour scheme  using the \MGNLO[2.8.1]~\cite{Alwall:2014hca} generator with the \NNPDF[3.0nnlo]~\cite{Ball:2012cx} parton distribution function (PDF) followed by parton shower (PS) and hadronisation with \PYTHIA[8.244]~\cite{Sjostrand:2014zea} using the A14 set of tuned parameters (tune)~\cite{ATL-PHYS-PUB-2014-021}   and the \NNPDF[2.3lo] PDF set. Single scalar and vector LQ signal samples were produced with coupling parameters $\lambda$ from 0.5 to 2.5.  The intrinsic width of the LQs increases quadratically with  $\lambda$ and linearly as a function of $\mlq$. In the considered range of parameters, the LQ width is 16\% or less of the LQ mass.
The simulated signal events do not include interference effects with the SM processes.
For the vector LQ signal two samples were produced for each $\lambda$, one with Yang--Mills coupling ($\kappa=0$) and the other with minimal coupling ($\kappa=1$). 
The implementation of the signal model is based on that described in Refs.~\cite{DorsnerLQ2016, BUCHMULLER1987442, LQToolbox}.

Simulated events with pair produced scalar LQs were generated at next-to-leading order (NLO) in QCD with \MGNLO[2.6.0], using the LQ model described in Ref.~\cite{Mandal_2016}, which adds PS to the fixed-order NLO QCD calculations~\cite{Kr_mer_2005, Kr_mer_1997} interfaced to \PYTHIA[8.230] for the PS and hadronisation.  Parton luminosities are provided by the five-flavour scheme \NNPDF[3.0nlo] PDF set with a value of the strong coupling constant $\alpha_{s}=0.118$, and the underlying event was modelled with the A14 tune. The LQ pair-production cross-sections were obtained from the calculation of direct top-squark pair production assuming that all other supersymmetric particles are heavier, since the production modes of this process are the same as the LQ pair production. The cross-sections were computed at approximate next-to-next-to-leading order (NNLO) in QCD with resummation of next-to-next-to-leading logarithmic (NNLL) soft gluon terms~\cite{Beenakker_2016,Beenakker_1998,Beenakker_2010,Beenakker_2016_v1}. The cross-sections do not include lepton $t$-channel contributions, which are neglected in Ref.~\cite{Mandal_2016} and may lead to corrections at the percent level~\cite{Borschensky_2020}.
 
Simulated events with pair-produced vector LQs were generated with \MGNLO[2.6.0] at LO in QCD, using the LQ model of Ref.~\cite{U1Model} and the  \NNPDF[3.0nlo] PDF set with  $\alpha_{s}=0.118$.  Decays of the LQs were performed with \textsc{MadSpin}, while PS and hadronisation were simulated using  \PYTHIA[8.244] with the A14 tune. Since no higher-order cross-sections are available for this model, the LO  \MGNLO[] cross-sections were used.
 
Several simulation samples are used to model the expected background processes. These include \ttbar, \singletop, $Z$ and $W$ bosons produced in association with jets (\Zjets\ and \Wjets), and diboson events.

The production of \ttbar\ simulated events was performed with the \POWHEGBOX[v2]~\cite{Frixione:2007nw,Nason:2004rx,Frixione:2007vw,Alioli:2010xd} generator at NLO with the \NNPDF[3.0nlo] PDF set ~\cite{Ball:2014uwa} and the \hdamp parameter\footnote{The \hdamp\ parameter is a resummation damping factor and one of the parameters that controls the matching of \POWHEG matrix elements to the PS and thus effectively regulates the high-\pt radiation against which the \ttbar\ system recoils.} set to 1.5\,\mtop~\cite{ATL-PHYS-PUB-2016-020}, with $\mtop = 172.5~\GeV$.
The events were interfaced to \PYTHIA[8.230]~\cite{Sjostrand:2014zea} to model the PS, hadronisation, and underlying event, with parameters set according to the A14 tune and using the \NNPDF[2.3lo] set of PDFs.
The \ttbar{} sample was normalised to the cross-section prediction at NNLO in QCD including the resummation of NNLL calculated using \TOPpp[2.0]~\cite{Beneke:2011mq,Cacciari:2011hy,Baernreuther:2012ws,Czakon:2012zr,Czakon:2012pz,Czakon:2013goa,Czakon:2011xx}.
 
Single top-quark $s$-channel ($t$-channel) production was simulated using the \POWHEGBOX[v2] generator at NLO in QCD in the five-flavour (four-flavour) scheme with the \NNPDF[3.0nlo] set of PDFs.
The events were interfaced with \PYTHIA[8.230] using the A14 tune and the \NNPDF[2.3lo] PDF set.
The samples were normalised to the theory prediction calculated at NLO in QCD with \HATHOR~2.1~\cite{Aliev:2010zk,Kant:2014ohav2}.
Similarly, the associated production of top quarks with $W$ bosons ($tW$) was modelled by the \POWHEGBOX[v2] generator at NLO in QCD using the five-flavour scheme and the \NNPDF[3.0nlo] set of PDFs.
The diagram removal scheme~\cite{Frixione:2008yi} was used to remove interference and overlap with \ttbar\ production. The related uncertainty is estimated by comparison with an alternative sample
generated using the diagram subtraction scheme~\cite{Frixione:2008yi,ATL-PHYS-PUB-2016-020}.
The events were interfaced to \PYTHIA[8.230] using the A14 tune and the \NNPDF[2.3lo] set of PDFs.
 
For the production of $Z/\gamma^{*}$ sample, the \POWHEGBOX[v1]~\cite{Nason:2004rx,Frixione:2007vw,Alioli:2010xd,Alioli:2008gx} generator was used for the simulation at NLO accuracy of the hard-scattering processes of $Z$ boson production and decay into the electron, muon, and $\tau$-lepton channels.
It was interfaced to \PYTHIA[8.186]~\cite{Sjostrand:2007gs} for the modelling of the PS, hadronisation, and underlying event, with parameters set according to the \AZNLO tune~\cite{STDM-2012-23}.
The \CT[10nlo] PDF set~\cite{Lai:2010vv} was used for the hard-scattering processes, whereas the \CTEQ[6L1] PDF set~\cite{Pumplin:2002vw} was used for the PS\@.
The effect of QED final-state radiation was simulated with \PHOTOSpp[3.52]~\cite{Golonka:2005pn,Davidson:2010ew}.
 
The production of \Wjets\ events was generated with \SHERPA[2.2.1]~\cite{Bothmann:2019yzt}.
In this set-up, NLO-accurate matrix elements for up to two partons and LO-accurate matrix elements for up to four partons were calculated with the Comix~\cite{Gleisberg:2008fv} and \OPENLOOPS~\cite{Buccioni:2019sur,Cascioli:2011va,Denner:2016kdg} libraries. The default \SHERPA PS~\cite{Schumann:2007mg} based on Catani--Seymour dipole factorisation and the cluster hadronisation model~\cite{Winter:2003tt_fixed} were used. They employed the dedicated set of tuned parameters developed by the \SHERPA authors and the \NNPDF[3.0nnlo] PDF set~\cite{Ball:2014uwa}.
The NLO matrix elements for a given jet multiplicity were matched to the PS using a colour-exact variant of the MC@NLO algorithm~\cite{Hoeche:2011fd}. Different jet multiplicities were then merged into an inclusive sample using an improved CKKW matching procedure~\cite{Catani:2001cc,Hoeche:2009rj} that is extended to NLO accuracy using the \MEPSatNLO prescription~\cite{Hoeche:2012yf}.
 
Diboson production was simulated with the \SHERPA[2.2.1] or 2.2.2 generator depending on the process.
Fully leptonic final states and semileptonic final states, where one boson decays leptonically and the other hadronically, were generated using matrix elements at NLO accuracy in QCD for up to one additional parton emission and at LO accuracy for up to three additional parton emissions.
Samples for the loop-induced processes $gg \to VV$ were generated using LO-accurate matrix elements for up to one additional parton emission for both the fully leptonic and semileptonic final states.
The matrix element calculations were matched and merged with the \SHERPA PS based on Catani--Seymour dipole factorisation  using the MEPS@NLO prescription.
The virtual QCD corrections were provided by the \OPENLOOPS library.
The \NNPDF[3.0nnlo] set of PDFs were used, along with the dedicated set of tuned PS parameters developed by the \SHERPA authors.
The samples were normalised to a NLO prediction.
\begin{table}[h]
\caption{Overview of the MC generators used for the main signal and background samples. The last column specifies the order in QCD for the cross-section calculation used for the normalisation of the simulated samples.}
\label{tab:generators}
\resizebox{\textwidth}{!}{
\begin{tabular}{l c c c c l l}
\toprule
Process                                & \multicolumn{2}{c}{Generator}                                   & \multicolumn{2}{c}{PDF set}                                      & Tune  & Normalisation        \\
& ME                                & PS                          & ME                                        & PS                   &       &              \\
\midrule
LQ$\rightarrow b\tau$                  & MadGraph5\_aMC@NLO                & \PYTHIA8.244               & \NNPDF[3.0nnlo]                           & \NNPDF[2.3lo] & A14 & LO \\
Scalar LQLQ$\rightarrow b\tau  b\tau$  & MadGraph5\_aMC@NLO                & \PYTHIA8.230               & \NNPDF[3.0nnlo]                           & \NNPDF[2.3lo] & A14 & NNLO + NNLL  \\
Vector LQLQ$\rightarrow b\tau  b\tau$ & MadGraph5\_aMC@NLO                 & \PYTHIA8.244               & \NNPDF[3.0nnlo]                           & \NNPDF[2.3lo] & A14 & LO  \\
\midrule
\ttbar                & \POWHEGBOX[v2]                           & \PYTHIA8.230                    & \NNPDF[3.0nnlo] & \NNPDF[2.3lo]               & A14                  & NNLO + NNLL          \\
Single top            & \POWHEGBOX[v2]                           & \PYTHIA8.230                    & \NNPDF[3.0nnlo] & \NNPDF[2.3lo]               & A14                  & NLO                  \\
$Z/\gamma^{*}$        & \POWHEGBOX[v1]                           & \PYTHIA8.186                   & \CT[10nlo]      & \CTEQ[6L1]                  & \AZNLO  & NLO \\
\Wjets                & \multicolumn{2}{c}{\SHERPA[2.2.1]} & \multicolumn{2}{c}{\NNPDF[3.0nnlo]} & \SHERPA                                   & NNLO                                      \\
Diboson               & \multicolumn{2}{c}{\SHERPA[2.2.1]/\SHERPA[2.2.2]} & \multicolumn{2}{c}{\NNPDF[3.0nnlo]} & \SHERPA                                   & NLO                                        \\
\bottomrule
\end{tabular}
}
\end{table}
 
A summary of all the features used for the simulation of the signal and background processes is shown in \cref{tab:generators}.
In all samples except those produced with  \SHERPA[2.2.1] or  \SHERPA[2.2.2], decays of heavy-flavour hadrons were modelled with \EVTGEN[1.2.0] or \EVTGEN[1.6.0] program~\cite{Lange:2001uf}, depending on the process.
All samples of simulated events were processed through the ATLAS detector simulation~\cite{SOFT-2010-01} based on $\GEANT$~\cite{Agostinelli:2002hh}.
The effects of multiple interactions in the same and nearby bunch crossings (pile-up) were modelled by overlaying minimum-bias events simulated using the soft QCD processes of \PYTHIA[8.186]~\cite{Sjostrand:2007gs} with the A3 tune~\cite{ATL-PHYS-PUB-2016-017} and the \NNPDF[2.3lo] PDF set.
 
\FloatBarrier


\section{Object Reconstruction and Identification}
\label{sec:recoid}
 
Tracks measured in the ID are used to reconstruct the interaction vertices~\cite{ATL-PHYS-PUB-2015-026}. The primary vertex of the hard interaction is chosen as the proton--proton vertex candidate with the highest sum of the squared transverse momenta of the associated tracks.
 
Electrons are reconstructed from topological clusters of energy deposits in the electromagnetic calorimeter that are matched to a track reconstructed in the ID~\cite{EGAM-2018-01}.
In the $\lephad$ ($\hadhad${}) final state, the selected (rejected) electrons are required to satisfy the \objectsc{medium} (\objectsc{loose}) identification criteria and have $\pT>\SI{20}{\GeV}$ ($\SI{15}{\GeV}$). Moreover, electrons are required to be within $|\eta_\text{cluster}|=2.47$ with the exclusion of the region between the barrel and endcap calorimeters ($1.37 < |\eta_\text{cluster}| < 1.52$).
An additional \objectsc{loose} isolation criterion~\cite{EGAM-2018-01} is also required, which has an efficiency of \SI{90}{\percent} for candidates with $\pT>\SI{15}{\GeV}$, increasing to more than \SI{98}{\percent} for candidates with $\pT>\SI{30}{\GeV}$.
 
Muons are reconstructed from signals in the MS matched with tracks inside the ID\@.
In the $\lephad$  final state, the selected  muons are required to satisfy the \objectsc{medium} identification criteria with an average efficiency of  \SI{97}{\percent}, and have $\pT>\SI{25}{\GeV}$ and $|\eta|<2.5$. In the $\hadhad$ chanel, muons having $\pT>\SI{7}{\GeV}$ are rejected if they satisfy the \objectsc{loose} identification criteria.
A \objectsc{tight} isolation criterion~\cite{MUON-2018-03} based on track information and having an average efficiency of \SI{89}{\percent} is also applied.
 
Jets are reconstructed with a particle-flow algorithm, which combines energy deposits in the calorimeter with ID tracks~\cite{ATLAS:2017ghe}, using the \antikt algorithm~\cite{Cacciari:2008gp,Fastjet} with a radius parameter $R=0.4$. Only jets with $|\eta| < 2.5$ and $\pT>\SI{25}{\GeV}$ are considered. The \objectsc{tight} working point of the jet vertex tagger (JVT)~\cite{PERF-2014-03} algorithm is selected to remove jets with $\pT<\SI{60}{\GeV}$ and $\abseta<2.4$ that are identified as not being associated with the primary vertex of the hard interaction. Jets 
containing $b$-hadrons are identified using the DL1r $b$-tagging algorithm~\cite{FTAG-2019-07,ATL-PHYS-PUB-2017-013}. A \SI{70}{\percent} efficiency working point is used, with the efficiencies being measured in simulated $t\bar{t}$ events.
The corresponding rejection factors (defined as the reciprocal of the efficiency values) for $b$-tagged jets initiated by $c$-quarks and light partons are 9.4 and 390 respectively.
 
The $\thad$ decays are composed of a neutrino and a set of visible decay products, most frequently one or three charged pions and up to two neutral pions. The visible decay products of the  $\thad$ decay are denoted by \tauhadvis.
The reconstruction of the \tauhadvis is seeded by jets reconstructed by  the \antikt algorithm~\cite{Cacciari:2008gp}, using calibrated topological clusters~\cite{PERF-2014-07} as inputs, with a radius parameter of $R=0.4$~\cite{ATLAS-CONF-2017-029}.
Reconstructed tracks are matched to \tauhadvis candidates and a multivariate discriminant is used to assess whether these tracks are likely to have been produced by the charged \tauhad decay products, rejecting tracks originating from other interactions, nearby jets, photon conversions or misreconstructed tracks.
The \tauhadvis objects are required to have one or three associated charged-particle tracks selected by this discriminant.
Their charge ($q$) is defined as the sum of the measured electric charges of these associated tracks and is required to be $|q|=1$.
The \tauhadvis objects are also required to satisfy $\pT>\SI{20}{\GeV}$ and $|\eta|< 2.5$, excluding the region $1.37 < |\eta| < 1.52$.
To separate the \tauhadvis candidates produced by hadronic $\tau$-lepton decays from those due to jets initiated by quarks or gluons, a recurrent neural network (RNN) identification algorithm~\cite{ATL-PHYS-PUB-2019-033} ($\thad$-ID) is constructed using information from reconstructed charged-particle tracks and calorimeter-energy clusters associated with \tauhadvis candidates. This analysis uses two $\thad$-ID working points: \objectsc{medium}, which has a 75\% (60\%) acceptance efficiency and a background rejection of 35 (240) and \objectsc{loose}, which has a 85\% (75\%) acceptance efficiency and a background rejection of 21 (90) for $\thad$ with one (three) charged-particle tracks. A  \objectsc{very loose} working point, having a 95\% acceptance efficiency, is also used for background estimation.
A separate boosted decision tree discriminant (\objectsc{eBDT}) is also used to reject backgrounds arising from electrons misidentified as \tauhadvis.
This discriminant is built using information from the calorimeter and the ID, most notably transition radiation information from the TRT system and variables sensitive to the ratio of the energy deposited in the calorimeter to the visible momentum measured from the reconstructed tracks.
 
The reconstructed objects used in this analysis are not built from disjoint sets of tracks or calorimetric clusters.
It is therefore possible that two different objects share most of their constituents.
An overlap removal procedure is applied to resolve this ambiguity.
This procedure is summarised in \cref{tab:object:orl}.

The missing transverse momentum vector, $\vecmet$, is reconstructed as the negative vector sum of the transverse momenta of leptons, \tauhadvis and jets, and a \enquote{soft-term}~\cite{PERF-2016-07}.
The soft-term is calculated as the vectorial sum of the $\vec{\pT}$ of tracks matched to the primary vertex but not associated with a reconstructed lepton, \tauhadvis or jet.
The magnitude of $\vecmet$ is referred to as the missing transverse energy, \met.
 
\begin{table}[tbhp]
\caption{
Criteria applied to overlapping reconstructed objects. The criteria are listed in the order they are applied.
}
\label{tab:object:orl}
\begin{center}
\begin{tabularx}{\textwidth}{ccX}
\toprule
 
Object to keep & Object to remove & Criteria \\
\midrule
Electron   & Electron & If they share the same track, the electron with the highest transverse momentum is kept.\\
Electron   & \tauhadvis & If $\Delta R <0.2$, the electron is kept. \\
Muon       & \tauhadvis &  If $\Delta R <0.2$, the muon is kept. \\
Muon       & Electron &  If they share a track, the electron is removed if the muon is associated with a signature in the MS, otherwise the muon is removed. \\
Electron   & Jet & Any jet within $\Delta R =0.2$  of an electron is removed. \\
Muon       & Jet & Any jet within $\Delta R =0.2$ of a muon is removed if it has fewer than three associated tracks. \\
Jet        & Electron & Any electron within $\Delta R =0.4$  of a jet is removed. \\
Jet        & Muon & Any muon within $\Delta R =0.4$ of a jet is removed. \\
\tauhadvis & Jet & Any jet within $\Delta R =0.2$ of a \tauhadvis is removed. \\
\bottomrule
\end{tabularx}
\end{center}
\end{table}
 
\FloatBarrier


\section{Event Selection}
\label{sec:eventsel}
 
Events are required to contain at least one primary vertex with at least two associated tracks.
 
For the $\lephad$ channel events were selected by single-lepton triggers. In 2015  single-electron triggers were simultaneously active with \pT thresholds of 24, 60 and 120~\GeV{}~\cite{TRIG-2018-05}. For data from 2016 onward the \pT thresholds are 26, 60 and 140~\GeV{}. Similarly to the single-electron triggers, the single-muon triggers had \pT thresholds of 20 and 50~\GeV{} for 2015, and 26 and 50~\GeV{} from 2016~\cite{TRIG-2018-01}. The trigger thresholds were raised to keep the trigger rates sufficiently low as the luminosity was increased. The lowest \pT threshold electron and muon triggers also have an isolation requirement. The lepton isolation and identification requirements loosen as the trigger $\pt$ thresholds increase. Events must contain at least one \thad candidate and exactly one electron or one muon. The electron or muon must be isolated and satisfy the medium lepton identification. Events with more than one lepton satisfying the medium identification are rejected, considering  electrons (muons)  with a $\pt$ greater than  15 (7) ~\GeV{}. This helps to reject $Z/\gamma^{*}\rightarrow ee/\mu\mu$ events and $Z/\gamma^{*}\rightarrow \tlep\tlep$. Furthermore, the electron and muon candidates are required to have $\pt > 30$~\GeV{}, and be matched to the trigger object that caused the event to be selected. The \thad candidate is required to have $\pt > 50$~\GeV{}, satisfy the medium $\thad$-ID selection and have $|\eta| < 2.3$. The pseudorapidity selection requirement rejects events with  \thad candidates in a region with a higher background contamination and large uncertainties in the determination of the rate of electrons misidentified as $\thad$.

In the $\hadhad$ channel, events were selected by a single \thad trigger~\cite{ATLAS-CONF-2017-061}. For 2015 and 2016, three single \thad triggers were available with \pT thresholds of 80, 125 and 160~\GeV{}. 
In 2017 and 2018, due to higher instantaneous luminosity, only the $\pT > 160$~\GeV{} trigger threshold was used. The  \thad identification requirements become less stringent as the trigger $\pt$ thresholds rise.
Events must contain at least two \thad candidates where the leading \thad candidate in \pT must be matched to the trigger within an angular distance of $\Delta R = 0.2$ and have $\pt$  that is at least 5~\GeV{} above the trigger threshold. The subleading-\pT \thad candidate is required to have $\pt > 65$~\GeV{}. Identification requirements are applied to both \thad{} candidates; the leading-\pT \thad  must satisfy the medium selection and the subleading-\pT  \thad the loose selection. Events that contain any electron or muon that satisfies the loose identification requirements are rejected, which ensures orthogonality to the $\lephad$ channel.

Events passing the previous requirements are then selected with criteria that are similar between the two channels. The two $\thad$ or the electron/muon (denoted by $\ell$) and $\thad$ must have opposite electric charges and at least one $b$-tagged jet is required. The invariant mass of the visible decay products of the two $\tau$-leptons, $m_{\mathrm{vis}}(\ell,\tauhad)$ or $m_{\mathrm{vis}}(\tauhad,\tauhad)$, is required to be above 100~\GeV{}, which is effective at reducing the \DY{} background. 
An additional requirement $\Delta \phi(\ell, \met) < 1.5$ is applied in the $\lephad$ channel to reduce single top and \ttbar\ events.
 
The variable $\st$ is defined as the scalar $\pt$ sum of the two $\tauhadvis$ (or $\ell$ and $\tauhadvis$) and the leading-$\pt$ $b$-jet. 
A minimum requirement of $\st{} > 300$~\GeV{} is applied, as there is almost no improvement in the expected results of the analysis, discussed in \cref{sec:results}, by adding events with lower  \st{} values.
 
The selection criteria described above define the signal region (SR) of the analysis. The signal acceptance times efficiency of the event selection varies between 3\% and 10\%, depending on the LQ mass and coupling. The efficiency is defined as the ratio of events passing the selection  in each channel  with respect to the signal events of the $b\taulep\tauhad$ and $b\hadhad$ final states, respectively.
Events in the SR of each channel are assigned to two categories of low ($<200$~\GeV{}) and high ($>200$~\GeV{}) transverse momentum of the leading-$\pt$ $b$-jet. The two categories are called high and low $b$-jet $\pt$ SRs, respectively.
The high $b$-jet $\pt$ SR is found to perform better for low-mass singly produced LQs, where the resonant contribution is dominant. Conversely, the low $b$-jet $\pt$ SR has a better acceptance for high mass signals, where the  non-resonant contribution is dominant for signals with $\mlq \ge 0.9~\TeV{}$.
This split into two categories improves  the expected results of the analysis, discussed in  \cref{sec:results}, by up to 30\%.
 
Alternative selections define the control regions (CR), used to evaluate the contribution of the main background processes in the SR, and the validation regions (VR), used to verify the good modelling of the backgrounds. The selection requirements used for the signal regions are summarised in Table~\ref{tab:sregions_lephad_hadhad}. The use of the control and validation regions in the background estimation methods is discussed in \cref{sec:background}.

\begin{table}[h]
\footnotesize
\begin{center}
\caption{Definition of signal regions (SR)  used in the $\lephad$ and $\hadhad$ channel. The symbol
$\ell$ represents the selected electron or muon candidate and $\tauhadvis$
represents the leading \tauhadvis candidate.  The symbol
$\tau_{1}$ ($\tau_{2}$) represents the leading (sub-leading) \tauhadvis candidate.}
\label{tab:sregions_lephad_hadhad}
 
\begin{tabular}{lll}
\toprule
$\lephad$ Signal Regions & \multicolumn{2}{l}{Selection} \\
\midrule
SR  & \multicolumn{2}{l}{$\ell$ (trigger, isolated), $\tauhadvis$ (medium $\thad$-ID), $q(\ell)\times q(\tauhadvis)<0$,} \\
& \multicolumn{2}{l}{$\Delta\phi(\ell,\met)<1.5$, $m_{\mathrm{vis}}(\ell,\tauhadvis) > 100$~\GeV{}, $\st > 300$~\GeV{},} \\
& \multicolumn{2}{l}{at least one $b$-jet} \\
High $b$-jet $\pt$ SR &  SR selection, leading $b$-jet $\pt$ $>$ 200~\GeV{} \\
Low  $b$-jet $\pt$ SR &  SR selection, leading $b$-jet $\pt$ $<$ 200~\GeV{} \\
\midrule
$\hadhad$ Signal Regions & \multicolumn{2}{l}{Selection} \\
\midrule
SR & \multicolumn{2}{l}{$\tau_{1}$ (trigger, medium $\thad$-ID), $\tau_{2}$ (loose $\thad$-ID), $q(\tau_{1})\times q(\tau_{2})<0$, $m_{\mathrm{vis}}(\tau_{1}, \tau_{2}) > 100$~\GeV{},} \\
& $\st >$~300~\GeV{}, at least one $b$-jet \\
High $b$-jet $\pt$ SR &  SR selection, leading $b$-jet $\pt$ $>$ 200~\GeV{} \\
Low  $b$-jet $\pt$ SR &  SR selection, leading $b$-jet $\pt$ $<$ 200~\GeV{} \\
 
\bottomrule
\end{tabular}
\end{center}
\end{table}

\FloatBarrier


\section{Background Estimation}
\label{sec:background}
 
\subsection{\lephad channel}
\label{subsec:lephad_background}
 
Control and validation regions are used in the analysis to estimate and study the modelling of the main background processes. The selection requirements used for the CR and VR in the $\lephad$ channel are summarised in Table~\ref{tab:control_regions_lephad}.
 
\begin{table}[h]
\footnotesize
\begin{center}
\caption{Definition of the background-enriched control regions (CR) and validation regions (VR) used in the $\lephad$ channel. The symbol
$\ell$ represents the selected electron or muon candidate and $\tauhadvis$
represents the leading \tauhadvis candidate. }
\label{tab:control_regions_lephad}
 
\begin{tabular}{lll}
\toprule
$\lephad$ Control/Validation  & Selection & Purpose \\
Regions              & \\
\midrule
Multijet-CR & $\ell$ (trigger, pass/fail offline isolation), $\mt(\ell,\met) < \SI{30}{\GeV}$, & Measure lepton \\
& one $b$-jet, $\thad$-ID score $<$ 0.01, $\met < 50$~\GeV{} & fake-factor \\
Top-CR            & Satisfy SR except: $\Delta\phi(\ell,\met)>2.5$, no $\st$ and lead. $b$-jet $\pt$ req. & Derive top correction \\
SS-CR           & Satisfy SR except:  $q(\ell)\times q(\tauhadvis)>0$, no $\Delta\phi(\ell,\met)$,    & Measure jet$\rightarrow \tau$ background  \\
& and $\st$ req.                                                                   & scale factor \\
High $b$-jet $\pt$ VR  & Satisfy high $b$-jet $\pt$ SR except: $1.5<\Delta\phi(\ell,\met)<2.5$,                   & Background modelling validation \\
&  300~\GeV{} $<$ $\st$ $<$ 600~\GeV{} & \\
Low  $b$-jet $\pt$ VR  & Satisfy low $b$-jet $\pt$ SR except: $1.5<\Delta\phi(\ell,\met)<2.5$,                     & Background modelling validation \\
&  300~\GeV{} $<$ $\st$ $<$ 600~\GeV{} & \\
$b$-jet $Z$-CR       & Satisfy SR except: 45 \GeV{} $< m_{\mathrm{vis}}(\ell,\tauhadvis) < $ 80 \GeV{},                                & $Z$+heavy-flavour jets \\
&  $\pt(\ell)/\pt(b\text{-jet}) > 0.8$, $|\Delta\phi(\ell,\tauhadvis)|>2.4$, no $\st$ req.                     &  normalisation factor \\
\bottomrule
\end{tabular}
\end{center}
\end{table}
 
In the $\lephad$ channel the dominant background contributions are from \ttbar{} and \singletop{} events. Processes involving top quarks can produce real $\tau$-leptons, or jets that are misidentified as $\thad$, and are estimated by using simulation with data-driven corrections. The \ttbar{} and $tW$ contributions are treated as one combined top-quark background due to their similar kinematics and final states. In the low (high) $b$-jet $\pt$ SR, \ttbar{} accounts for 90\% (86\%) of all top-quark processes and 96\% (97\%) of the single top-quarks are from $tW$. To ensure that this background is accurately modelled, a top-quark control region (Top-CR) is defined. With respect to the SR selection, the requirements on the leading $b$-jet $\pt$ and the \st{} are removed, and the condition $\Delta \phi(\ell, \met)< 1.5$ is replaced by $\Delta \phi(\ell, \met)>2.5$. This results in a region with a purity of 91\% in top-quark processes and negligible signal contamination. Out of all top-quark events in the Top-CR, 91\% are from \ttbar{} processes and $\sim$97\% of the \singletop{} events are from $tW$, which is compatible with the composition of the SRs.
 
\begin{figure}[!h]
\centering
\subfloat[]{\includegraphics[width=0.45\textwidth]{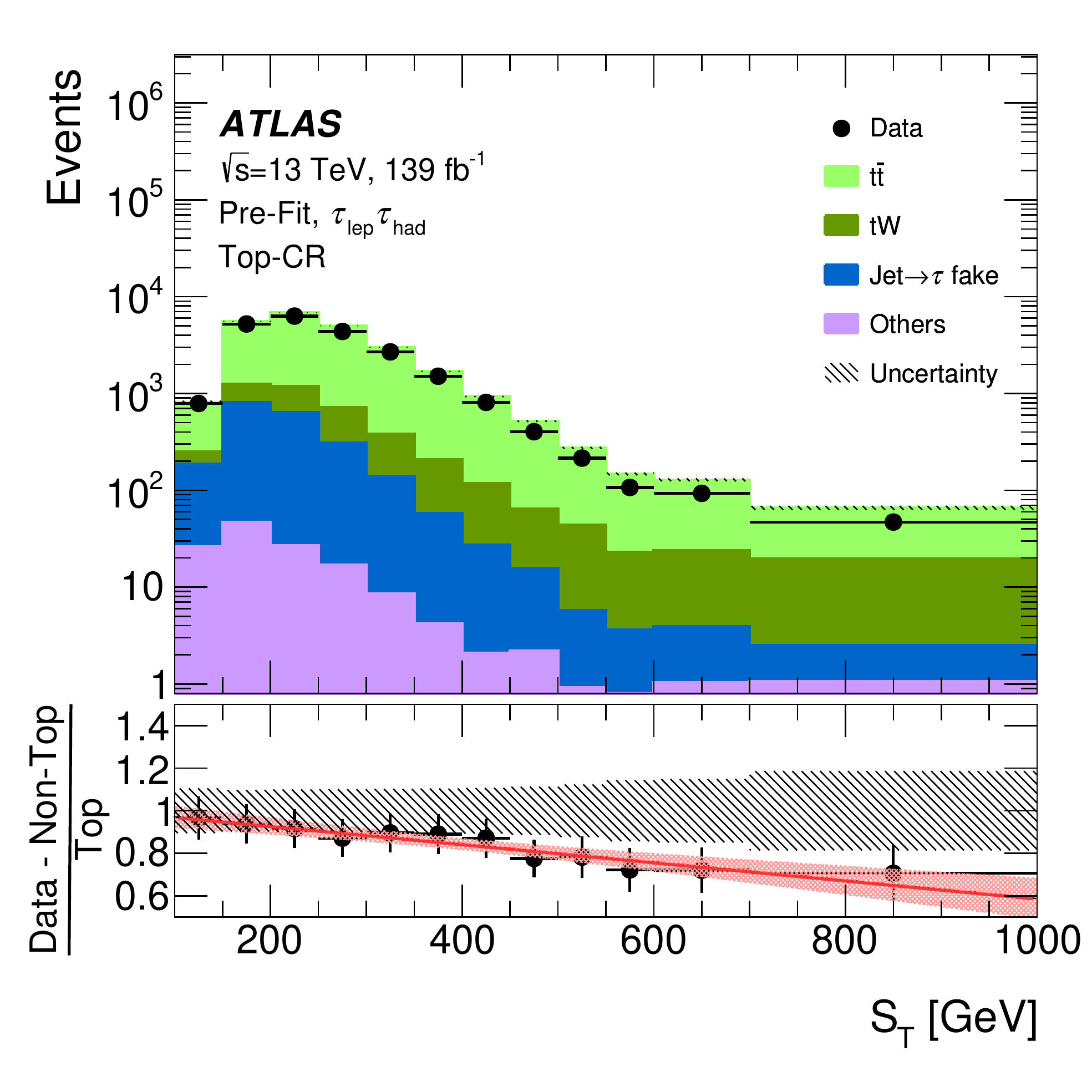}}
\subfloat[]{\includegraphics[width=0.45\textwidth]{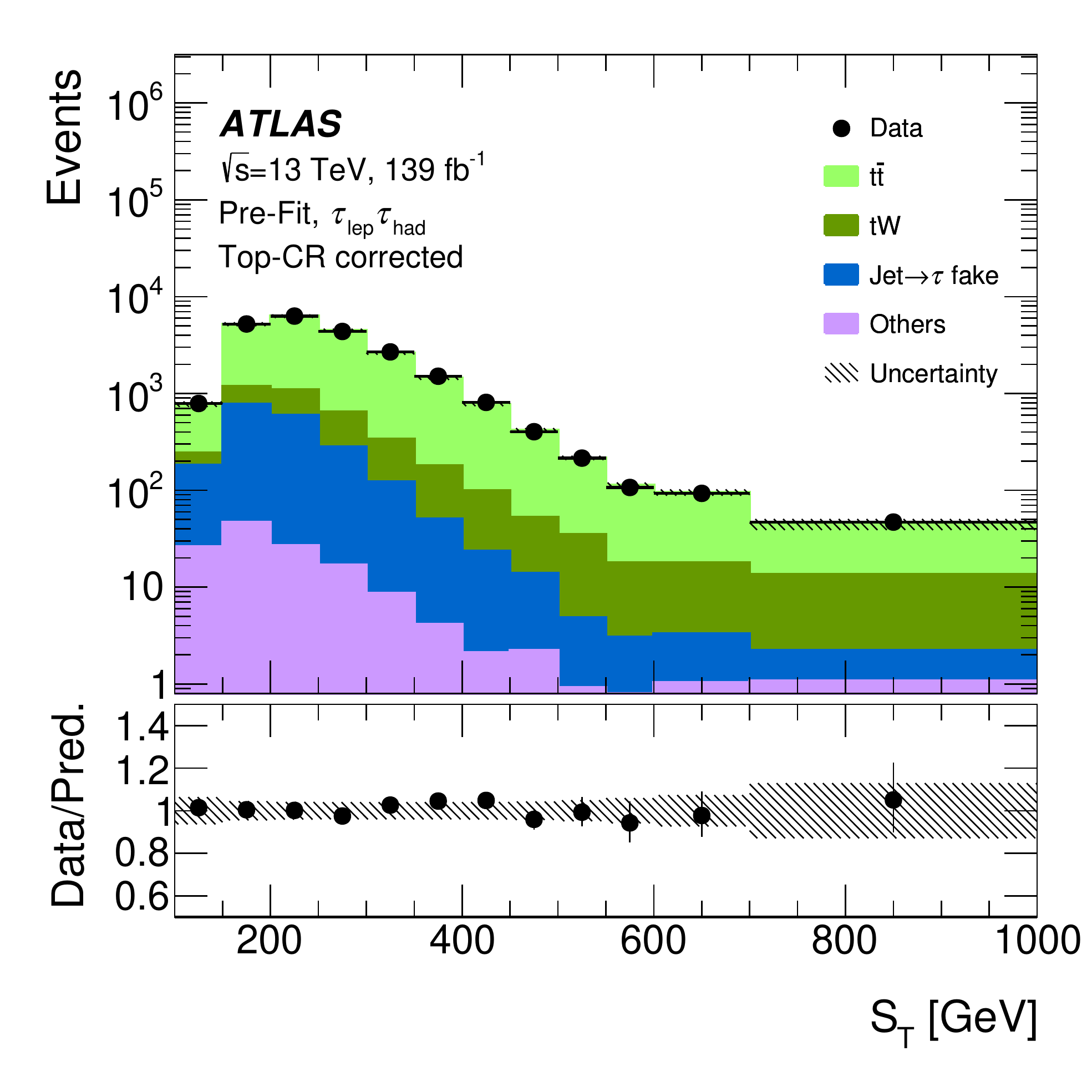}} \\
\caption{(a) Comparison between data and the background prediction for the $\st$ distribution in the Top-CR in the $\lephad$ channel. The $\ttbar$ and $tW$ contributions only include events with true leptons and $\thad$ in the final state. The label \emph{Jet}$\rightarrow\tau$ \emph{fake} corresponds to events with a lepton and a quark- or gluon-initiated jet misidentified as $\thad$; this contribution is estimated by using simulation. The bottom panel shows the ratio of the data to the prediction where the uncertainty shown by both the points and hatched band includes the statistical uncertainty in the data and background predictions, the theoretical uncertainty in the MC simulation predictions, and the MC subtraction uncertainty. Finally, the line and the cross dashed band show the resulting fit and the associated uncertainty. The label `Top' in the bottom panel denotes the sum of \ttbar{} and $tW$ processes, while `Non-Top' refers to all other processes considered. Entries with values above the $x$-axis range are included in the last bin of the distribution. (b) Shows the same distribution after the top-quark correction scale factor is applied to the $\ttbar$ and $tW$ simulated events. Only the uncertainty associated to the top-quark correction and the statistical uncertainty on data and simulation are considered.}
\label{fig:lephad_topcr}
\end{figure}

A discrepancy between the data and simulation prediction is observed in the Top-CR, with the simulation overestimating the background contribution. Recent measurements of differential cross-sections have demonstrated that the current simulations of $\ttbar$
processes overestimate the upper tail of the top-quark $\pT$ spectrum~\cite{TOPQ-2018-15,TOPQ-2018-17}.
The discrepancy varies  depending on $\st$; for this reason, a correction is derived as a function of $\st$ in this region based on the ratio between data and simulation. A top-quark correction scale factor is defined in Eq.~(\ref{eq:lephadSFt}) and is applied to all \ttbar{} and \singletop{} simulated events. The comparison between data and the background prediction in the Top-CR and the derived correction as a function of $\st$ are shown in Figure~\ref{fig:lephad_topcr}, where $\ttbar$ and $tW$ events with a generated lepton reconstructed as a lepton (a `true' lepton) and a jet misidentified as a $\thad$ are included under the \emph{Jet}$\rightarrow\tau$ \emph{fake} contribution. This demonstrates that the Top-CR is dominated by $\ttbar{}$ and $tW$ events with true leptons and $\thad$ in the final state, thus this correction does not account for mismodelling due to jets being misidentified as $\thad$. The modelling of events with a true lepton and jet misidentified as a $\thad$ in the final state is discussed later in this section.

The top-quark correction scale factor is defined as a function of $\st$:
\begin{equation}
\label{eq:lephadSFt}
SF_{\textrm{Top}} (\st) = \frac{(N_{\mathrm{data}} - N_{\mathrm{non-}\textrm{Top}})(\st)}{N_{\textrm{Top}}(\st)} ,
\end{equation}
 
\noindent where $N_{\mathrm{data}}$ and $N_{\textrm{Top}}$ represent respectively the number of data events and of \ttbar{} plus \singletop{} events predicted by simulation, $N_{\textrm{Top}}$ includes events with both true and misidentified $\thad$ in the final state and $N_{\mathrm{non-}\textrm{Top}}$ includes all the other backgrounds estimated by using simulation. The resulting correction is well fitted by a linear function, which is used to derive the correction scale factors. The correction is also derived with an alternative logarithmic function: $SF_{\textrm{Top}}(\st) = a \ln(\st) + b$, and the difference between the two corrections is taken as an uncertainty on the correction. Additional uncertainties related to the cross-section and acceptance of the top-quark processes, as well as the statistical and cross-section uncertainties related to the subtraction of the contribution from the other processes, are applied to account for the slight difference between the fractions of top-quark events that are due to \ttbar{} production in the Top-CR and SRs and for the extrapolation to the SR\@. The scale factor is applied at the per-event level to the  \ttbar{} plus \singletop{} events passing the selections of the signal, control or validation regions.
The total uncertainty in the scale factor varies between 4\% and 7\% for $\st$ in the range of 300--700~\GeV.
Different binning choices for the $\st$ distribution were also considered, but the impact on the $SF_{\textrm{Top}} (\st)$ uncertainty is found to be below 5$\%$ and therefore not considered as an additional source of uncertainty.
 
Another source of background in the \lephad channel stems from multi-jet events, where jets can mimic both the $\tlep$ and $\thad$. This type of background from multi-jet events is estimated via a data-driven fake-factor method by deriving a lepton fake-factor. The lepton fake-factor is measured in the multi-jet control region (multijet-CR) that is enriched in multi-jet events, but is similar kinematically to the SR\@. The events are still required to satisfy the single lepton trigger and to have exactly one $b$-jet, but the identification algorithm to reject jets misidentified as $\thad$ is instead used to select multi-jet events by requiring an extremely low value for $\thad$ RNN identification score (corresponding to only 1\% acceptance for true $\thad$). 
Additional selection criteria on $m_{\textrm{T}}(\ell, \met) < 30$~\GeV{} and $\met < 50$~\GeV{} are applied  to increase the purity of multi-jet events relative to other backgrounds. The fake-factor is measured with a requirement on the leading $b$-jet $\pt > 25~\GeV$ and is defined as:

\begin{equation*}
FF_{\mathrm{lep}} (\pt(\tlep), \eta(\tlep)) = \frac{(N_{\mathrm{data}} - N_{\mathrm{MC}})^{\mathrm{pass-iso}} (\pt(\tlep), \eta(\tlep)) }{(N_{\mathrm{data}} - N_{\mathrm{MC}})^{\mathrm{fail-iso}} (\pt(\tlep), \eta(\tlep))  } .
\end{equation*}
 
The variable $N_\mathrm{data}$ is the total number of data events and $N_{\mathrm{MC}}$ is the number of background events predicted by simulation that contain a true $\tlep$. Events are split between the numerator and denominator based on whether the $\tlep$ satisfied the lepton isolation requirement or not. The fake-factor is parameterised as a function of the \tlep \pt and split into central ($|\eta| < 1.52$) and forward ($|\eta| > 1.52$) regions. The statistical uncertainty on the  $FF_{\mathrm{lep}} (\pt(\tlep), \eta(\tlep))$ fake-factor and simulation-related uncertainties on $N_{\mathrm{MC}}$ are considered as systematic uncertainties and propagated to the estimate of the multi-jet background. The uncertainties are in the range of 6--230\% as a function of the $\tauhadvis$ $\eta$ and $\pt$. 
A control region is defined by having the same selection as the SR, except that the lepton isolation requirement inverted.
Applying the fake-factor at the per-event level, the multi-jet estimate in each SR is then obtained by scaling the distribution in the corresponding control region where the isolation criteria are not satisfied.
 
An additional source of background are events where a lepton is produced in association with a jet that is misidentified as a $\thad$ (\emph{Jet}$\rightarrow\tau$ \emph{fake}). These contribute approximately 20\% to the expected background in the SR and are mostly from $\ttbar$ with contributions from $\Wjets$, $\Zjets$, and diboson events. To ensure that these are well modelled, a `same-sign' control region (SS-CR) is defined by taking the same selection as the SR, but with a light lepton with the same electric charge as $\thad$. The requirements on $\Delta \phi(\ell, \met)$, $\st$ and the leading $b$-jet $\pt$  are removed to increase the number of events in the CR. The top-quark correction scale factor derived in Eq.~(\ref{eq:lephadSFt}) is applied to top-quark events in this region (approximately 81\% of the total). As the Top-CR used to derive that scale factor is dominated by $\ttbar$ and $tW$ events with true $\tau$-leptons in the final state, it does not correct for mismodelling of jets that are misidentified as a $\thad$. As a difference between the simulation prediction and the data is still observed, another scale factor is derived to account for any remaining differences from those backgrounds with a lepton and misidentified $\thad$ (approximately 60\% of the events in this region). 
The remaining events contain true $\thad$ and are subtracted, before calculating the scale factor, by applying the top-quark correction scale factor. Then, the scale factor for events with a lepton and a jet misidentified as a $\thad$ is defined as:
 
\begin{equation*}
SF_{\mathrm{fake}-\tau} (\pt(\tauhadvis), n_{\mathrm{track}}) = \frac{(N_{\mathrm{data}} - N_{\mathrm{true}-\tau}) (\pt(\tauhadvis), n_{\mathrm{track}}) }{N_{\mathrm{fake}-\tau} (\pt(\tauhadvis), n_{\mathrm{track}}) } ,
\end{equation*}
 
\noindent where $N_{\mathrm{true}-\tau}$ is the total number of events predicted by simulation where both the \thad and \tlep are true and $N_{\mathrm{fake}-\tau}$ is the number of predicted events with a jet misidentified as a \thad and a true $\tlep$. The scale factor is parameterised as a function of $\pt (\tauhadvis)$ and the number of charged-particle tracks ($n_{\mathrm{track}}$). It is applied to any MC background event with a true lepton and a jet misidentified as a $\thad$. The correction is derived in the SS-CR of the $\muhad$ channel and then applied to both $\ehad$ and $\muhad$, because the $\ehad$ SS-CR contains events with misidentified electrons, which are not well modelled by simulation. The $SF_{\mathrm{fake-}\tau}$ correction values are in the range of 1--1.2 (1--1.5) for $\thad$ with one (three) charged-particle tracks. The statistical uncertainty on the $SF_{\mathrm{fake-}\tau}$ correction and the simulation-related uncertainties affecting  $N_{\mathrm{true}-\tau}$ and $N_{\mathrm{fake}-\tau}$ are propagated to the background estimate as systematic uncertainties. As a function of the $\tauhadvis$ $\pt$, the uncertainties amount to 15--20\% (22--140\%) for $\tauhadvis$ candidates with one (three) associated charged-particle track.

\begin{figure}[!h]
\centering
\subfloat[]{\includegraphics[width=0.45\textwidth]{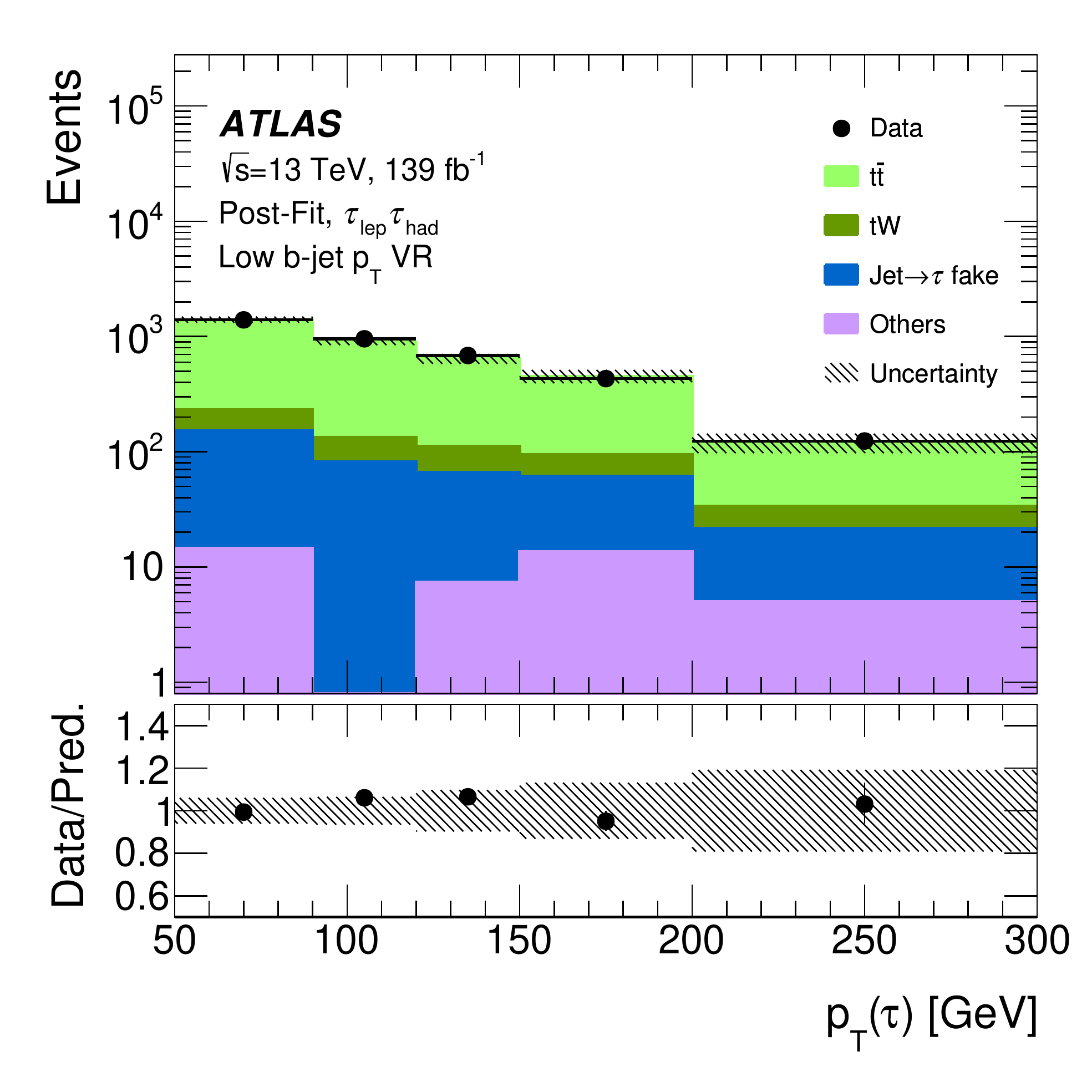}}
\subfloat[]{\includegraphics[width=0.45\textwidth]{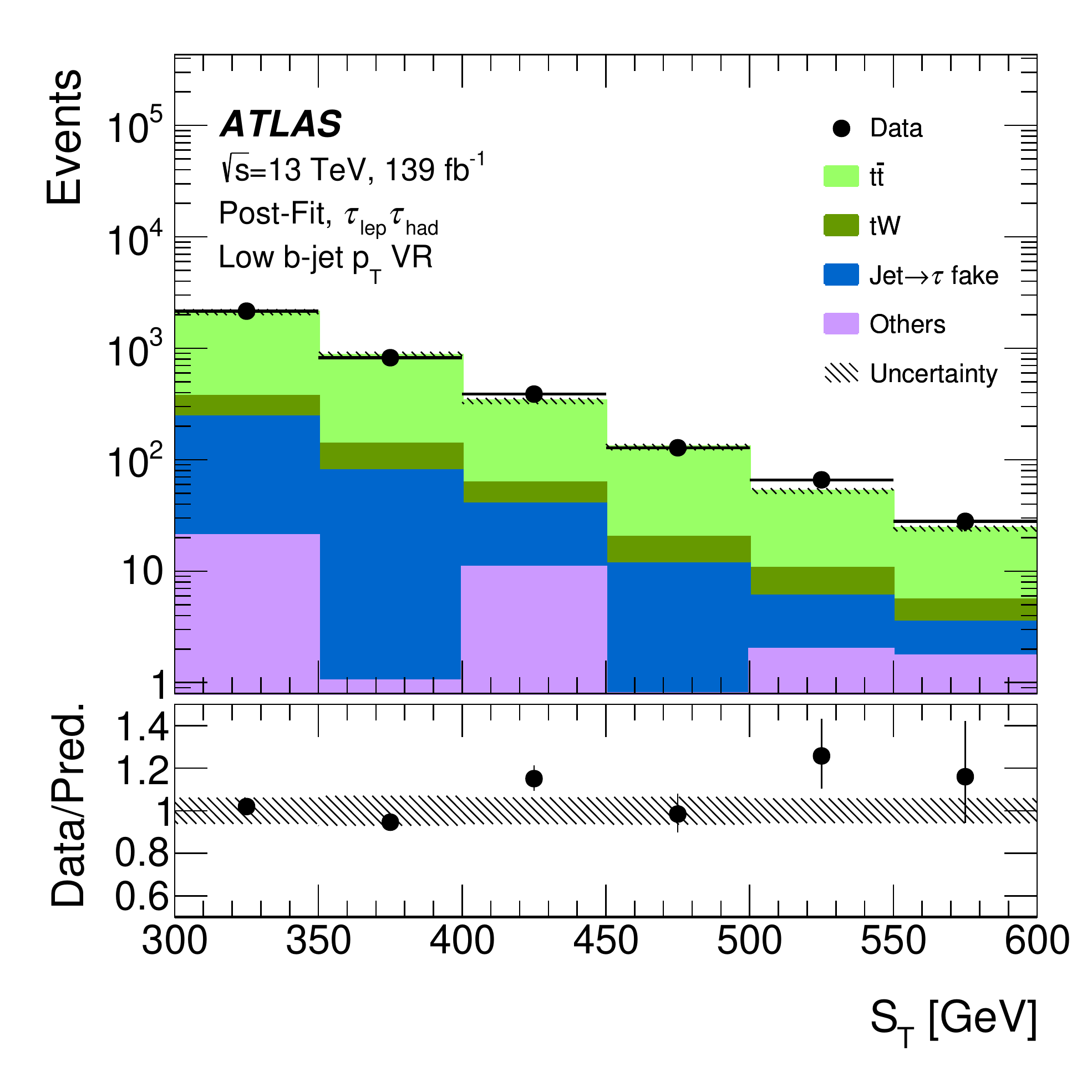}} \\
\subfloat[]{\includegraphics[width=0.45\textwidth]{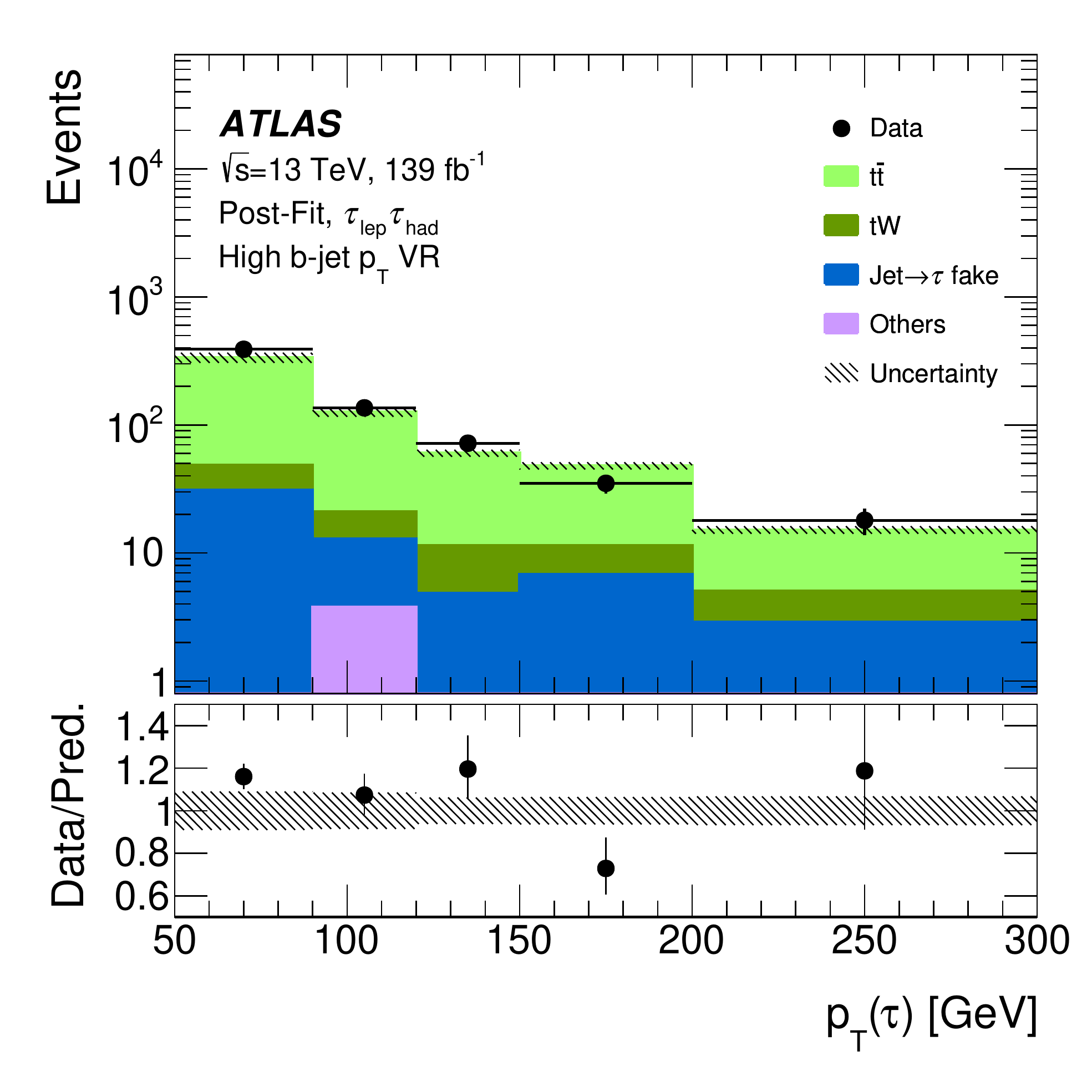}}
\subfloat[]{\includegraphics[width=0.45\textwidth]{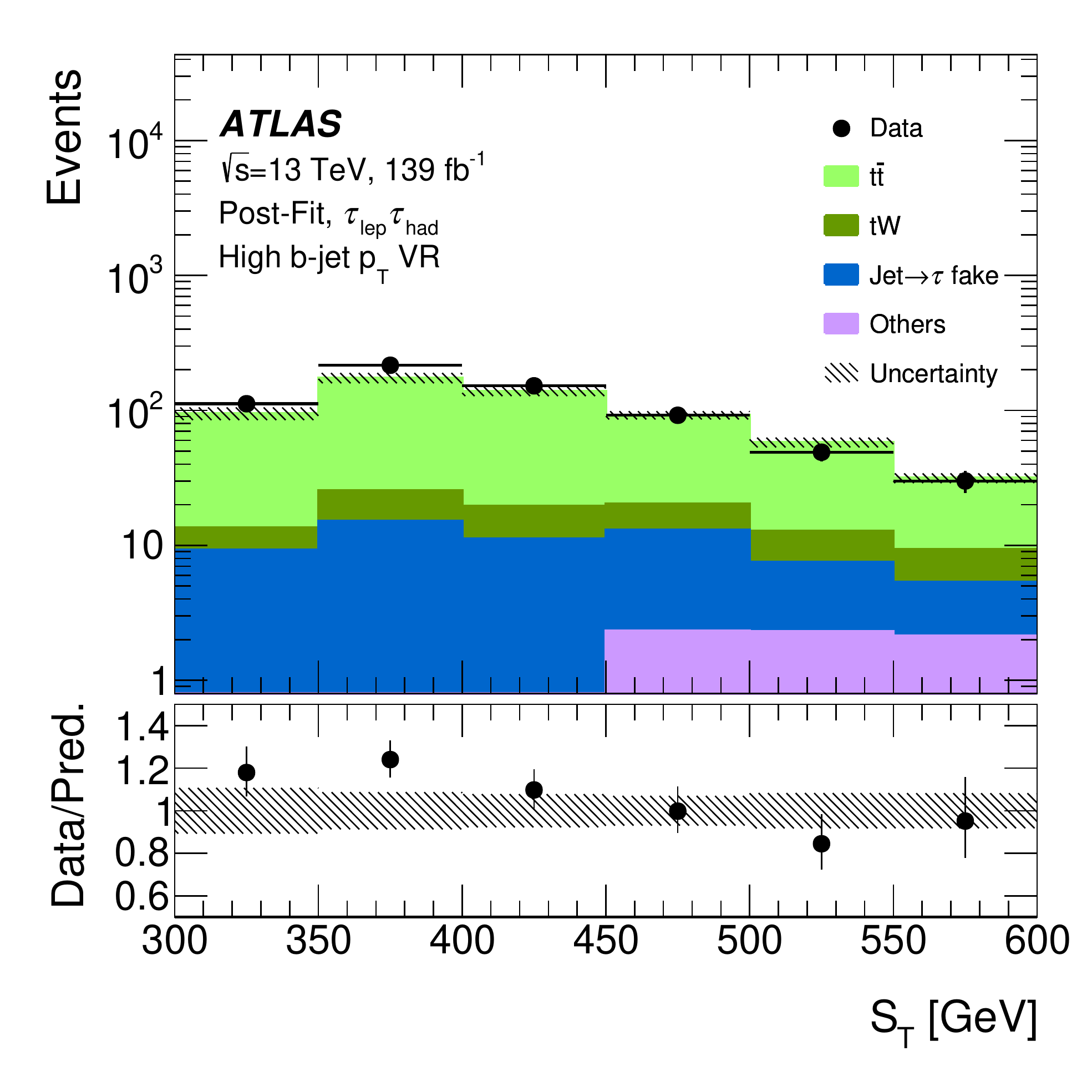}} \\
\caption{Comparison between data and the background prediction for the \lephad{} channel validation regions after applying the scale factors discussed in  Section ~\ref{subsec:lephad_background}. (a) and (b) show the $p_{\text{T}}(\tau)$ and $\st$ respectively for the low $b$-jet $\pt$ category, (c) and (d) show the same for the high $b$-jet $\pt$ category. The uncertainty band includes both statistical and systematic uncertainties evaluated in the fit described in \cref{sec:results}. Entries with values above the $x$-axis range are included in the last bin of each distribution. The lower panels show the ratio of the data to the predictions.}
\label{fig:lephad_vr}
\end{figure}

To validate the background modelling in a region depleted in signal, high and low $b$-jet $\pt$ validation regions (high and low $b$-jet $\pt$ VR) are defined by applying the SR requirements, with the exceptions of the $\Delta \phi(\ell, \met)<1.5$ and the $\st>300$ ~\GeV{} criteria, that are modified into $1.5 < \Delta \phi(\ell, \met) < 2.5$ and $300 < \st < 600$~\GeV{}. The low (high) $b$-jet $\pt$ VR consists of 82\% (80\%) \ttbar{}  events,  of 8\% (10\%) single top-quark events, and  of 9\% (9\%) of events where a jet is misidentified as a $\thad$. Good modelling of the background is found in the validation regions; the background estimate agrees with data within the total uncertainty, as shown in Figure~\ref{fig:lephad_vr}.

\subsection{\hadhad channel}

In the $\hadhad$ channel, the main background sources are $Z/\gamma^{*}\rightarrow \thad\thad$ events, as well as top-quark processes, with $W$ bosons decaying to $\thad$, or to electrons, muons or jets misidentified as $\thad$. Both $Z/\gamma^{*}\rightarrow \thad\thad$ and top-quark backgrounds are estimated using simulation with data-driven corrections, which are discussed further below.  The selection requirements used to define the CR and VR in the $\hadhad$ channel are summarised in Table~\ref{tab:control_regions_hadhad}.
\begin{table}[h]
\footnotesize
\begin{center}
\caption{Definition of the background-enriched control regions (CR) and validation regions (VR) used in the $\tauhad\tauhad$ channel. The symbol
$\tau_{1}$ ($\tau_{2}$) represents the leading (sub-leading) \tauhadvis candidate.
}
\label{tab:control_regions_hadhad}
\begin{tabular}{lll}
\toprule
$\hadhad$ Control/Validation & Selection & Purpose \\
Regions            &                   \\
\midrule
Dijet-CR & Satisfy SR except: $\tau_{1}$ and $\tau_{2}$ satisfy very loose $\thad$-ID, & Measure $\tauhadvis$ fake-factor \\
& $\tau_{1}$ fail medium $\thad$-ID & \\
CR-1  & Satisfy SR except: $\tau_{2}$ fail loose $\thad$-ID & Apply $\tauhadvis$ fake-factor \\
SS-VR & Satisfy SR except: $q(\tau_{1})\times q(\tau_{2})>0$ & Multijet modelling check \\
$Z$+light flavour jets VR & Satisfy SR except: 0 $b$-jets, $\Delta \phi(\tau_{1},\tau_{2})$ $>$ 0.25, & $Z$+light jets \\
& $m_{\mathrm{vis}}(\tau_{1},\tau_{2})$ $<$ 100~\GeV{}, $\met$ $>$ 60~\GeV{} & modelling \\
\bottomrule
\end{tabular}
\end{center}
\end{table}
 
As the mismodelling of the kinematic distributions observed in the $\lephad$ channel originates from the underlying top-quark process rather than the $\thad$ decay, it is also expected to be present in the $\hadhad$ channel. However, due to small number of events in the $\hadhad$ channel the statistical uncertainty in the top-quark processes is comparable with the expected mismodelling. This makes it difficult to select a $\hadhad$-only control region to quantify this mismodelling. Therefore, the $\st$-dependent top-quark correction scale factor from Eq.~(\ref{eq:lephadSFt}) derived in the $\lephad$ channel is also applied to the $\hadhad$ channel. The shape of the $\st$ distribution is checked and  found to be compatible between the  $\lephad$ and  $\hadhad$ channels.
 
\begin{figure}[!h]
\centering
\includegraphics[width=0.50\textwidth]{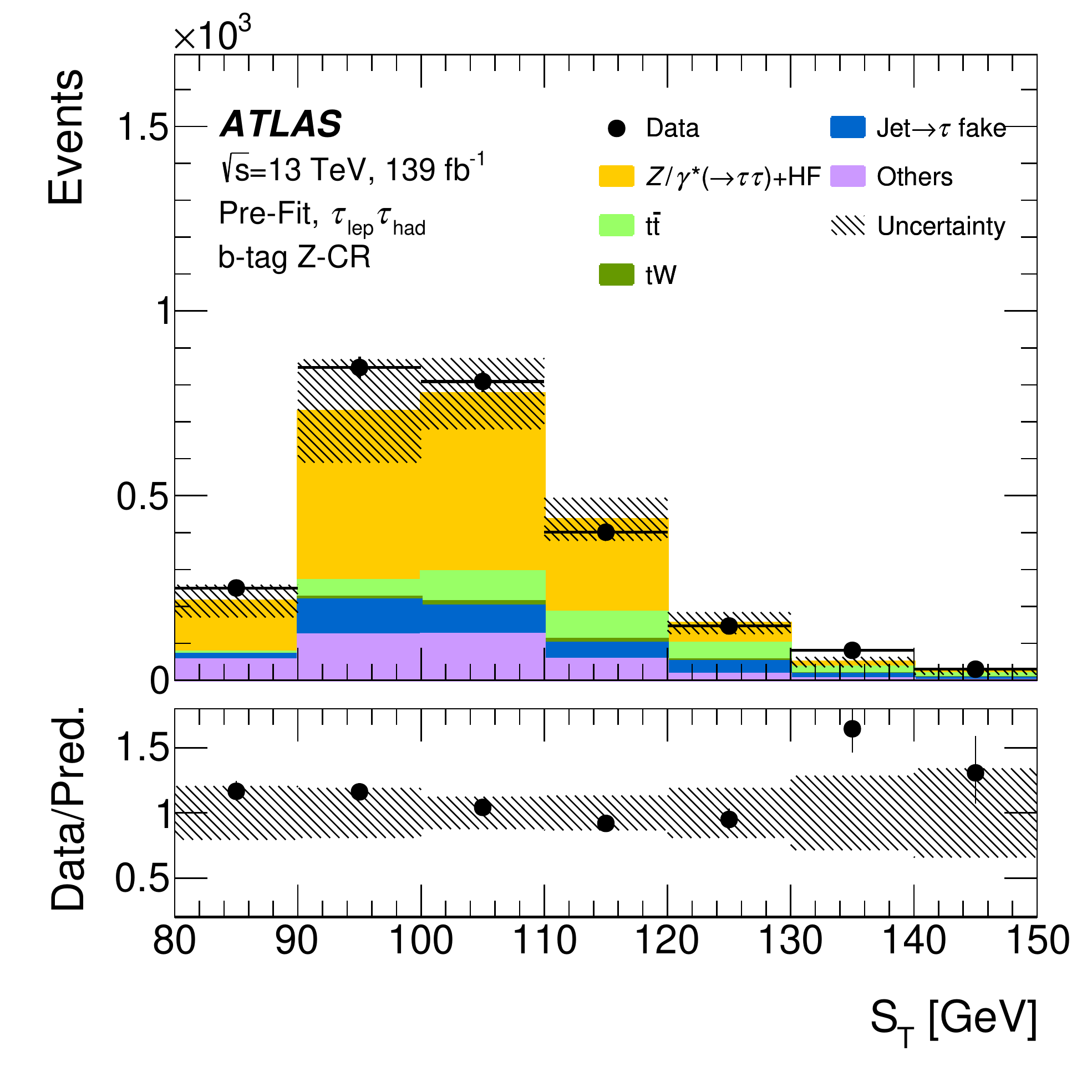}\\
\caption{Comparison between data and the background prediction for the  $\st$ distribution in the control region used for correcting the $Z(\rightarrow \tau\tau) \,+ $ heavy-flavour jets process. The uncertainty band includes both statistical and systematic components. Entries with values above the $x$-axis range are included  in the last bin of the distribution. The lower panel shows the ratio of the data to the predictions.}
\label{fig:hadhad_zhf}
\end{figure}

The $Z/\gamma^{*}\rightarrow \thad\thad$ background is also modelled using simulation. Due to a known discrepancy in the simulation compared with the data for $Z(\rightarrow \tau\tau) \,+ $ heavy-flavour jets with at least one $b$- or $c$-jet  ($Z$+HF)~\cite{ATL-PHYS-PUB-2017-006}, a correction factor for the normalisation of this background is derived in the \lephad $b$-jet $Z$-CR, defined in Table~\ref{tab:control_regions_lephad}, which has a purity of around 60\% for the $Z$+HF processes and is inclusive in $\pt$ of the leading $b$-jet. A comparison between the data and the prediction from the simulation in the $b$-jet $Z$-CR before deriving the correction factor is shown in Figure~\ref{fig:hadhad_zhf}.
 
The scale factor is derived by subtracting backgrounds estimated from simulation that are not from the $Z$+HF process ($N_{\mathrm{non-ZHF}}$):
 
\begin{equation*}
SF_{\mathrm{ZHF}} = \frac{N_{\mathrm{data}} - N_{\mathrm{non-ZHF}} }{N_{\mathrm{ZHF}} } ,
\end{equation*}
 
\noindent where $N_{\mathrm{ZHF}}$ is the number of $Z$+HF events predicted by simulation.
 
The scale factor is applied as a normalisation to the total $Z$+HF contribution, with a value of 1.13~$\pm$~0.23 obtained from the control region. The uncertainty includes the statistical uncertainty, the uncertainty in the subtraction of the simulation events and the extrapolation uncertainty from the control region to the SRs. The extrapolation uncertainty is obtained by repeating the scale factor calculation in the $\lephad$ channel using selection criteria for the control and signal regions equivalent to the ones used in the $\hadhad$ channel.
 
For $Z (\rightarrow \tau\tau) \,+ $ light-flavour jets ($Z$+LF, no $b$- or $c$-jets), the modelling is validated in a $b$-veto region ($Z$+LF VR). This region has the same event selection as the SR, except that zero $b$-jets are required. In addition, the requirements  $m_{\mathrm{vis}} < 100$~\GeV{}, $\met > 60$~\GeV{} and $\Delta \phi(\tau,\tau) > 0.25$ are applied to ensure a high $Z$+LF purity. The data is found to be in agreement with the simulation within the statistical uncertainty in the data. 
 
Finally the background originating from multi-jet events, where jets are misidentified as $\thad$, is estimated by using a data-driven fake-factor method.
A control region dominated by multi-jet events, called Dijet-CR, is defined by taking events that satisfy one of the single-jet triggers (with thresholds between 15 and 420~\GeV{}). The leading $\tauhadvis$ candidate is required to not satisfy the medium $\thad$ identification and the subleading \tauhadvis candidate is used as a probe. Both $\tauhadvis$ candidates are still required to pass the very loose $\thad$ identification requirement. As for the $\lephad$ channel, the fake-factor is measured inclusively in the leading $b$-jet $\pt$. The leading and subleading $\tauhadvis$ candidates are required to have opposite electric charges and have a $\pt > 65$~\GeV{}. At least one additional $b$-tagged jet is required, but no selection is made on the leading $b$-jet $\pt$. Then, the fake-factor is defined as:
 
\begin{equation*}
\ftauid(\pt,\numTrack) \equiv \left.\frac{(N_{\mathrm{data}} - N_{\mathrm{MC}})^{\mathrm{pass\,}\thad\mathrm{-ID}}(\pt, \numTrack) }{(N_{\mathrm{data}} - N_{\mathrm{MC}})^{\mathrm{fail\,}\thad\mathrm{-ID}} (\pt, \numTrack) } \right|_\mathrm{dijet} \, ,
\end{equation*}
 
\noindent where $N_{\mathrm{MC}}$ includes all simulated background events. The pass or fail $\thad$-ID superscript refers to whether the subleading $\tauhadvis$ candidate satisfies the loose $\thad$-ID or not, while still satisfying the very loose requirement. For each of the SRs, the multi-jet estimate is then obtained from the a control region, called CR-1, composed of events in which the subleading $\tauhadvis$ candidate fails the loose $\thad$-ID, using the fake-factor:
 
\begin{equation*}
N_\mathrm{multi-jet}(\pt,\numTrack)  =  \ftauid(\pt,\numTrack) \times ( N_{\mathrm{data}} - N_{\mathrm{MC}} )^{\mathrm{fail\,}\thad\mathrm{-ID}} (\pt, \numTrack) .
\end{equation*}
 
The fake-factor is parameterised as a function of the \pt and number of charged-particle tracks of the subleading \tauhadvis candidate.  The statistical uncertainty on the  $\ftauid(\pt,\numTrack)$ fake-factor is considered as systematic uncertainty, and it varies in the range of 4--15\% as a function of the  $\tauhadvis$ $\pt$, for $\tauhadvis$ candidates with both one and three associated charged-particle tracks. For this method to be accurate, it is important that the fail-$\thad$-ID and multi-jet control regions have a similar composition of quark- and gluon-initiated jets. This is obtained by inspecting the shape of subleading $\tauhad$ identification scores in the two regions, which depends on the quark--gluon fraction.
A lower threshold
than the very loose requirement  is applied on this score, which ensures that the shapes of the distributions are compatible.
 
After the selection, the multi-jet contribution to the SR is expected to be small. The modelling of the multi-jet background is verified in the same-sign validation regions (SS-VR). The SS-VR has the same selection as the SR, but the electric charges of the $\thad$ candidates are required to be the same. 
For key distributions in the low $b$-jet \pt category, the data is found to agree with the background prediction, with approximately half of the events being from $\ttbar$ and half from multi-jet background. The selection in the high $b$-jet $\pt$ SS-VR leads to low statistics with 2.8 expected events (mostly from \ttbar) and 4 observed data events.


\section{Systematic Uncertainties}
\label{sec:systematics}

Systematic uncertainties arise from the reconstruction of the various physics objects and from theoretical or modelling uncertainties affecting the predictions for both the backgrounds and signals. These uncertainties manifest themselves in both the overall yield and shape of the final observable, and can be divided into two main groups: the experimental uncertainties and the modelling uncertainties.

The experimental uncertainties include the uncertainties related to the trigger, reconstruction, calibration and identification of electrons~\cite{EGAM-2018-01}, muons~\cite{MUON-2018-03}, taus~\cite{ATLAS-CONF-2017-029}  and jets~\cite{JETM-2018-05-corr,PERF-2014-03,FTAG-2019-07}; for electron and muons, additional uncertainties in the lepton isolation are considered. Uncertainties related to background with misidentified $\tau$-leptons are described in \cref{sec:background}. Another source of experimental uncertainties is given by the luminosity measurement, whose primary measurement is obtained using the LUCID-2 detector~\cite{Avoni:2018iuv}. An uncertainty value of 1.7$\%$~\cite{ATLAS-CONF-2019-021} is assigned for the combined 2015--2018 integrated luminosity.
 
Among the experimental uncertainties, the ones with the highest impact on the analysis sensitivity are the \tauhadvis related uncertainties, with an impact on the results in the range of 30--40$\%$ depending on the LQ coupling and mass values considered. 
The uncertainties in the \tauhadvis identification efficiency are in the range of \SI{2}{\percent} to \SI{6}{\percent}, while the eBDT efficiency uncertainties are of the order of \SI{1}{\percent} to \SI{2}{\percent}. These uncertainties are parameterised as a function of the \tauhadvis \pt\ and the number of associated tracks for the  \tauhadvis identification efficiency, and as a function of the $\tauhad$ decay mode for the eBDT efficiency. In both cases, the uncertainties are derived in dedicated tag and probe measurements~\cite{ATLAS-CONF-2017-029}.
The \tauhadvis reconstruction efficiency uncertainty is derived from comparisons between simulations using different detector geometries or $\GEANT$ physics lists; this uncertainty is parameterised as a function of true \tauhadvis \pt\  and is between \SI{1}{\percent} and \SI{1.5}{\percent}.
For the \tauhadvis energy scale, the total uncertainty is in the range of \SI{1}{\percent} to \SI{4}{\percent} of the  \tauhadvis{} \pt, arising from a combination of measurements: a direct measurement with \Ztaumutauhad events, measurements of the calorimeter response to single particles, and comparisons between simulations using different detector geometries or $\GEANT$ physics lists.
This uncertainty is also parameterised as a function of the \tauhadvis \pt\ and the number of associated tracks.
 
The uncertainties in the background modelling include uncertainties in the top-quark, $Z$+jets and diboson backgrounds, as well as multijet events in which quark- or gluon-initiated jets are misidentified as a $\tauhad$.
Among the background modelling uncertainties, the ones related to the top-quark background
have the largest impact on the analysis sensitivity, with an impact on the results in the range of 40--50$\%$ depending on the LQ coupling and mass values.
This uncertainty is extracted by comparing nominal and alternative $\ttbar$ and single top-quark MC samples in the phase space of the SR and Top-CR\@.
For each sample, a dedicated data-driven $\st$-dependent correction is applied before the comparison.
The difference between the nominal and alternative samples in the $\st$ distribution is taken as the uncertainty in the top-quark processes.
The alternative samples have variations of the initial/final-state radiation, matrix element and PS compared to the nominal sample.
To derive the initial/final-state radiation uncertainty, the generator parameters used to produce the nominal samples are varied.
The matrix element to PS NLO matching uncertainty is derived by comparing the \MGNLO{} and \POWHEG{} predictions while keeping the same generator for the PS component.
For the PS, the uncertainty is derived by a comparison with an alternative sample generated by using \Herwig{} for the PS while keeping the same generator for the hard-scattering simulation component.
The uncertainties in the background modelling originating from the  PDF and $\alpha_{S}$ uncertainties are found to be less than 1\% and are neglected.
Finally, an uncertainty in the $tW$ interference for the single top-quark background is estimated by comparing the nominal sample, where the diagram removal scheme is applied, to an alternative sample that uses the diagram subtraction scheme~\cite{White_2009}.
 
The uncertainties in the signal modelling include those from the signal cross-section and acceptance due to renormalisation scale ($\mu_{R}$) and factorisation scale ($\mu_{F}$) variations, PDF and $\alpha_{S}$.
The $\mu_{R}$ and $\mu_{F}$ uncertainties are estimated through an envelope of the variations obtained from scaling $\mu_{R}$ and $\mu_{F}$ by a factor between 0.5 and 2, while keeping their ratio between 0.5 and 2. 
The uncertainties due to the \NNPDF[3.0nlo] PDF set and $\alpha_{S}$ are evaluated following the PDF4LHC recommendation~\cite{Butterworth:2015oua}.

\FloatBarrier


\section{Results}
\label{sec:results}
The distribution of the $\st$ variable for the events of the signal regions defined in Table~\ref{tab:sregions_lephad_hadhad} for the $\lephad$ and $\hadhad$ channels, is used as final discriminant between the leptoquark signal and the background.
The statistical analysis of the data is performed using the profile likelihood ratio method~\cite{Read:2002hq}, to test whether a model can be rejected given the observed data.
As the model under test is a signal plus background hypothesis, the chosen parameter of interest is the signal strength, $\mu$, defined as the ratio of the observed to the predicted value of the signal cross-section times branching fraction.
The likelihood function $\mathcal{L}(\mu,\theta)$ is then constructed as a product of Poisson probability terms for each bin of the distributions.
It depends on $\mu$ and on the nuisance parameters $\theta$, which encode systematic uncertainties that can affect the signal and background distributions and are constrained using Gaussian probability density functions.
The asymptotic approximation is used when constructing the test statistic $\tilde{q}_u$ \cite{Cowan:2010js} from the likelihood ratio, defined as $\tilde{q}_{\mu} = -2 \ln (\mathcal{L} (\mu, \hat{\hat{\theta}}) / \mathcal{L}(\hat{\mu}, \hat{\theta}) )$ where $\hat{\mu}$ and $\hat{\theta}$ are the parameters that define the global maximum-likelihood function and $\hat{\hat{\theta}}$ are the nuisance parameters that give the maximum likelihood for a given value of $\mu$. 
To ensure a reliable estimation of the backgrounds in the fit model, the binning in $\st$ is optimised so that sufficient background events (greater than 10) are present in each bin of the pre-fit distribution.
No significant excess above the background expectations is observed and corresponding limits on production cross-sections of the LQ signals are set.
 
\begin{figure}[!h]
\centering
\subfloat[]{\includegraphics[width=0.4\textwidth]{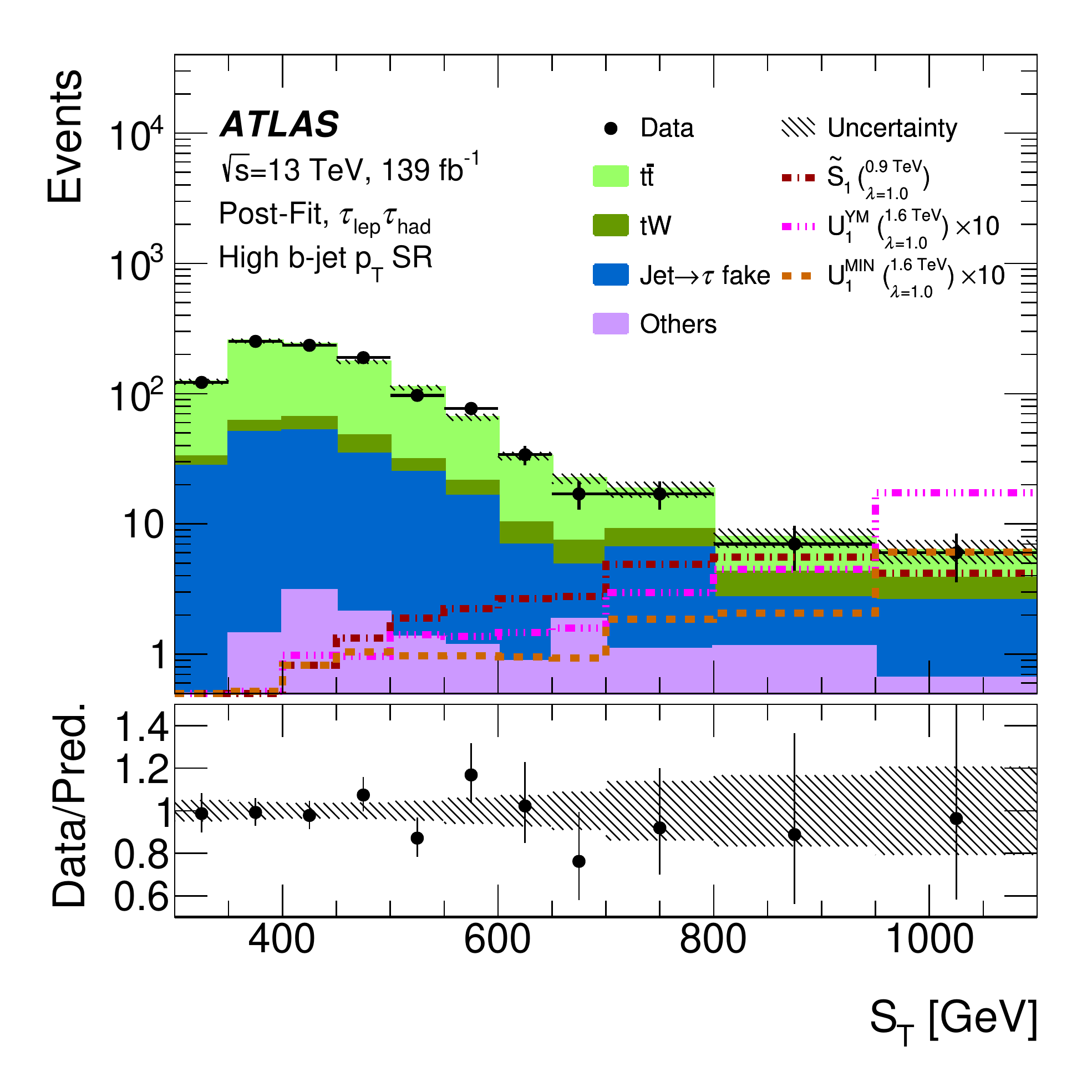} }
\subfloat[]{\includegraphics[width=0.4\textwidth]{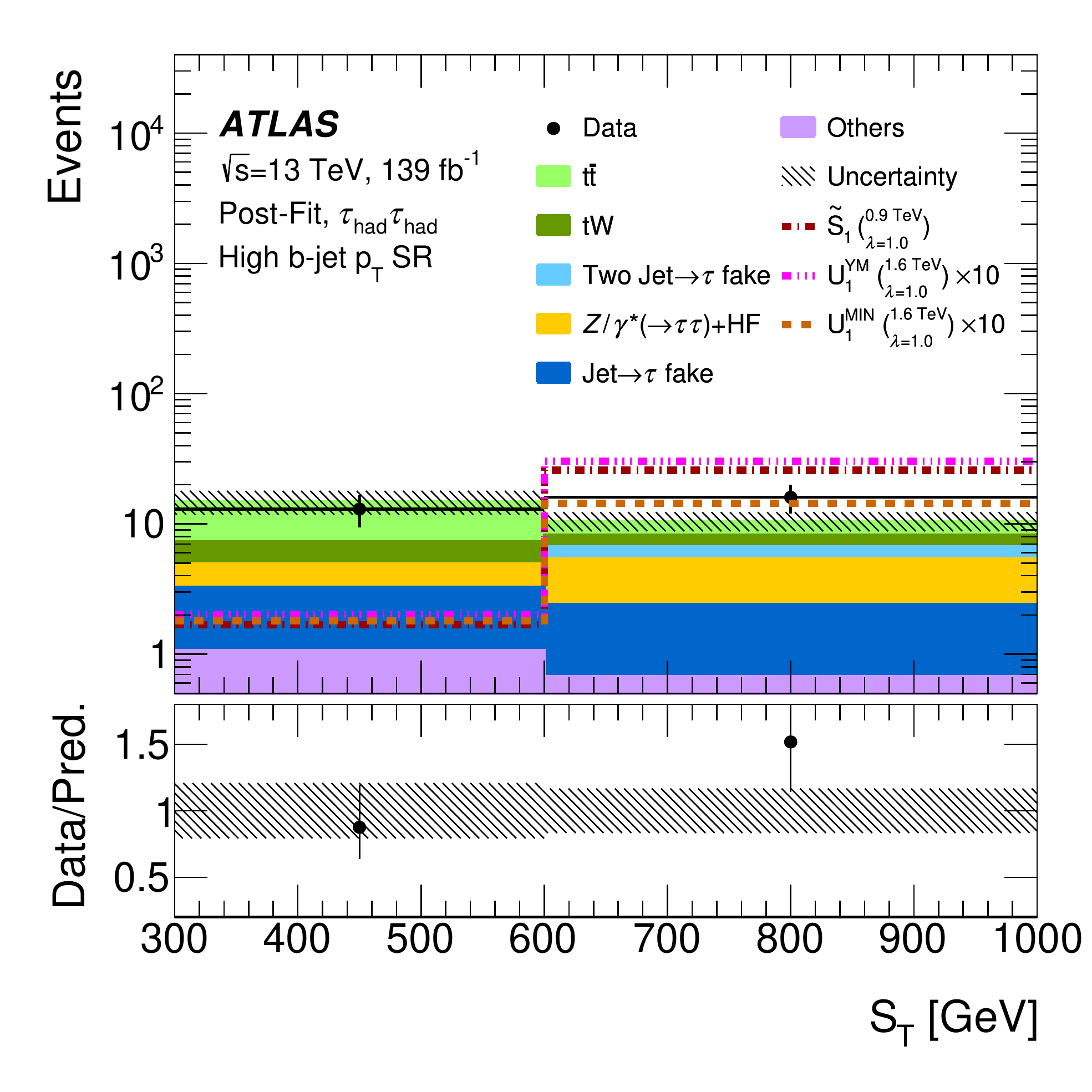} } \\
\caption{Post-fit distributions of the discriminating variable $\st$ in the high $b$-jet $\pt$ signal region for the (a) $\taulep\tauhad$ and (b) $\tauhad\tauhad$ channels. The two channels are fit simultaneously considering the background-only hypothesis.  The lower panels show the ratio of the data to the predictions, where the uncertainty band includes both statistical and systematic post-fit errors. Entries with values above the $x$-axis range are included in the last bin of each distribution. For illustrative purposes, the dotted lines show the expectation from three singly produced scalar and vector LQ signals.} 
\label{fig:bonly_postfit}
\end{figure}

For the LQ results interpretation, only the high $b$-jet $\pt$ signal regions from the $\lephad$ and $\hadhad$ channels are considered and fit simultaneously. For the high $b$-jet $\pt$ signal regions, the contribution from the non-resonant LQ production process is small. Therefore the interference between the LQ non-resonant processes and the SM processes is not expected to be substantial in these SRs, and it is neglected. 
By setting $\mu = 0$ in the profile likelihood ratio, the test statistic can be used to check for compatibility with the background-only hypothesis.
The data are first fit under the background-only hypothesis and the resulting post-fit distributions are found to be in good agreement with the data, as shown in Figure~\ref{fig:bonly_postfit}. Table~\ref{tab:postfit_yields} shows the yields for the $\lephad$ and $\hadhad$ channels, respectively.

\begin{table}[tp]
\small
\begin{center}
\caption{Post-fit background yields in the high $b$-jet $\pt$ signal region of $\taulep\tauhad$ and $\tauhad\tauhad$ channels. `Jet$\rightarrow\tau$ \emph{fake}' indicates the events with a true lepton and a quark- or gluon-initiated jet misidentified as a $\thad$. The `Two jet$\rightarrow\tau$ \emph{fake}' indicates the events where two jets are misidentified as $\thad$. `Others' in $\taulep\tauhad$ includes \ZtautauLFjets, diboson, $W$+jets and $Z$($\rightarrow ee$,$\mu\mu$)+jets while `Others' in $\tauhad\tauhad$ includes \ZtautauLFjets, diboson and $W$+jets. The results are extracted from a fit assuming the background-only hypothesis.}
\label{tab:postfit_yields}
 
\resizebox{0.45\textwidth}{!}{
\begin{tabular}{
@{}
l
S[table-format=2.1]@{$\,\pm\,$}S[table-format=2.1]
S[table-format=3.2]@{$\,\pm\,$}S[table-format=2.1]
@{}
}
\toprule
Process         &  \multicolumn{2}{c}{$\taulep\tauhad$} & \multicolumn{2}{c}{$\tauhad\tauhad$} \\
\midrule
$\ttbar$                        & 764 & 82 & 9.9 & 2.6 \\
Single top                      & 65 & 35  & 3.9 &1.0 \\
Jet$\rightarrow\tau$ \emph{fake}& 215 & 79 & 3.9 &1.0 \\
Two jet$\rightarrow\tau$ \emph{fake}& \multicolumn{2}{c}{--} & 1.34 & 0.27 \\
\ZtautauHFjets\                 & 5.5 & 0.4 & 4.6 & 1.1 \\
Others                          & 9.7 & 1.0 & 1.75 & 0.30 \\
\midrule
Total                           & 1059 & 51 & 25.4 & 4.9\\
\midrule
Data                            & \multicolumn{2}{c}{1053} & \multicolumn{2}{c}{29} \\
\bottomrule
\end{tabular}}
\end{center}
\end{table}
 
As good agreement is found between the data and the background expectation, upper limits are set on the cross-section assuming that the branching fraction $\mathcal{B}(\mathrm{LQ}\to b\tau)$ is 100\% in the case of the  $\widetilde{S_{1}}$ model and 50\% for the \YMLQ{} and \minLQ{} scenarios.
This is performed using the frequentist CL$_{\mathrm{s}}$ method~\cite{Read:2002hq}.
A production cross-section for a given signal scenario is excluded at the 95\% confidence level (CL) when CL$_{\mathrm{s}} < 0.05$.
 
\begin{figure}[!h]
\centering
\subfloat[]{\includegraphics[width=0.33\textwidth]{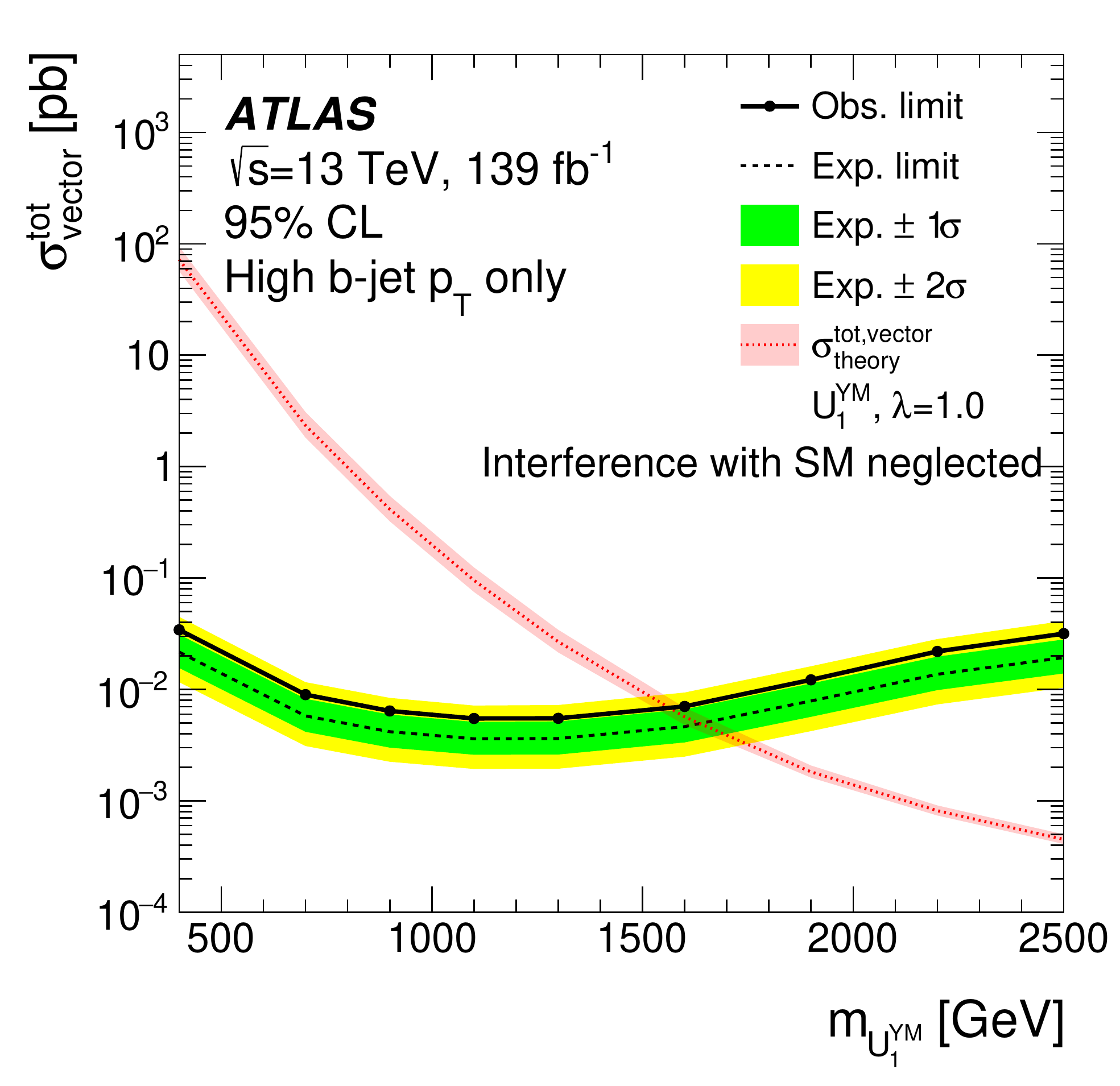}}
\subfloat[]{\includegraphics[width=0.33\textwidth]{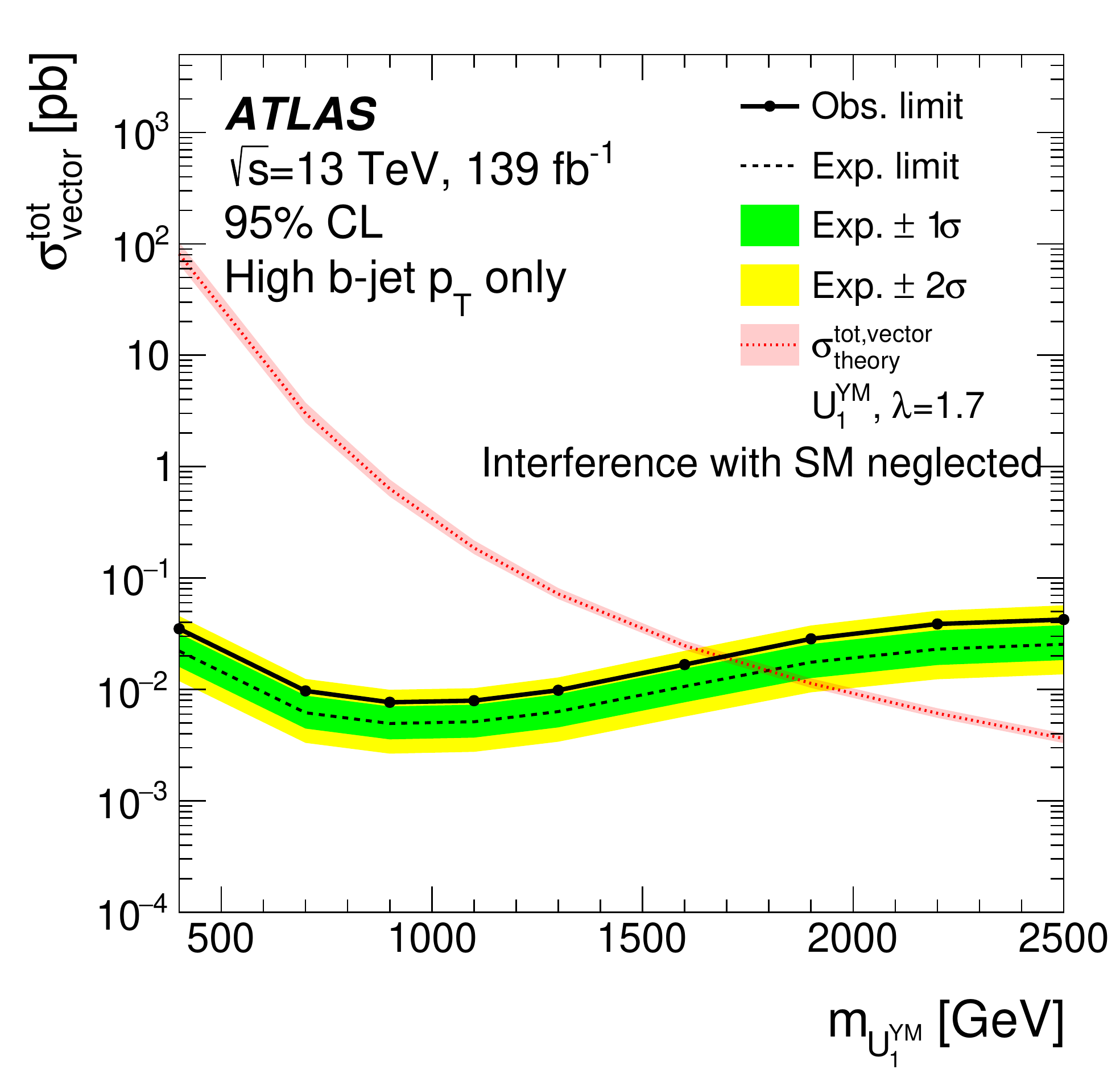}}
\subfloat[]{\includegraphics[width=0.33\textwidth]{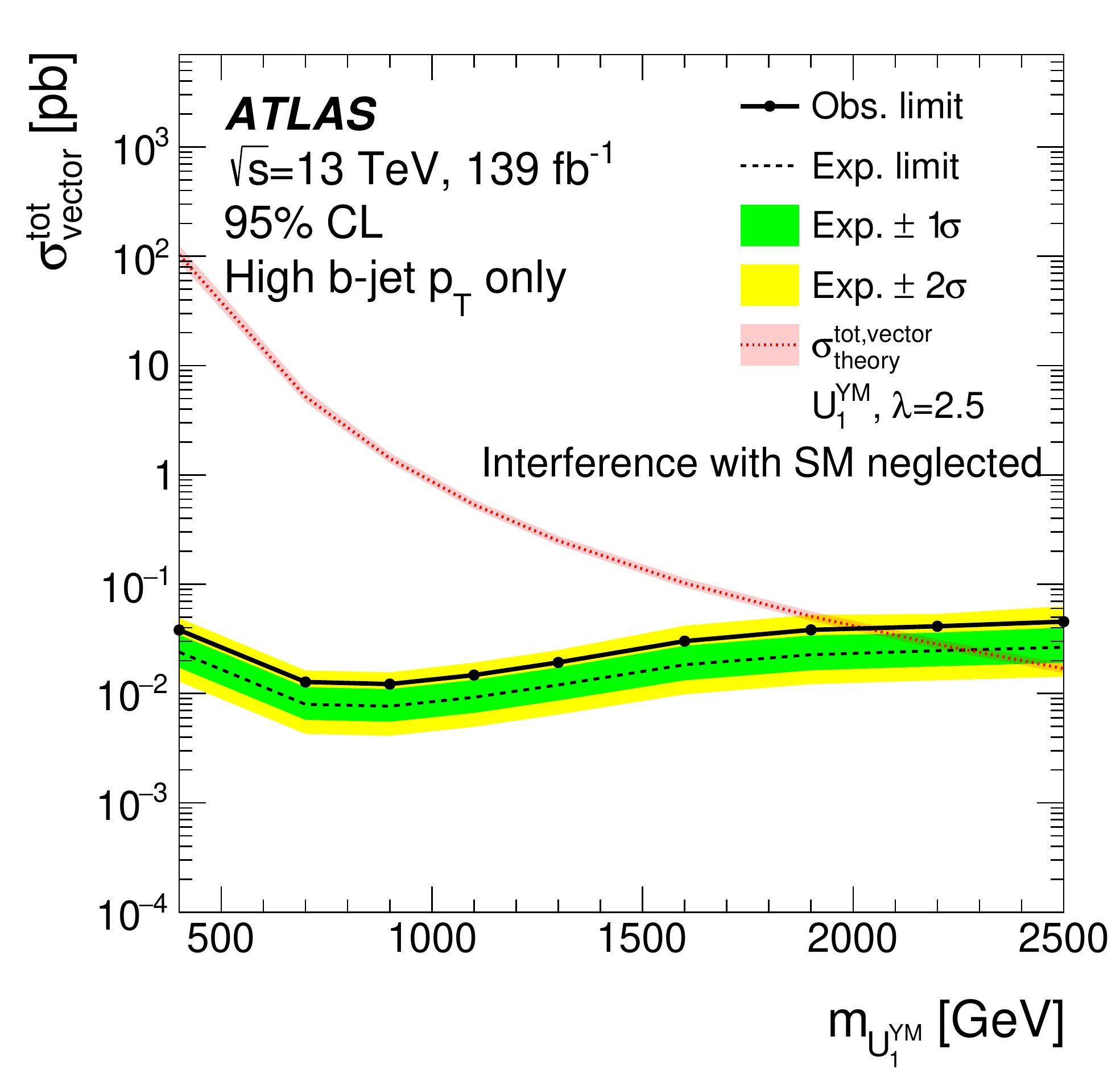}} \\
\subfloat[]{\includegraphics[width=0.33\textwidth]{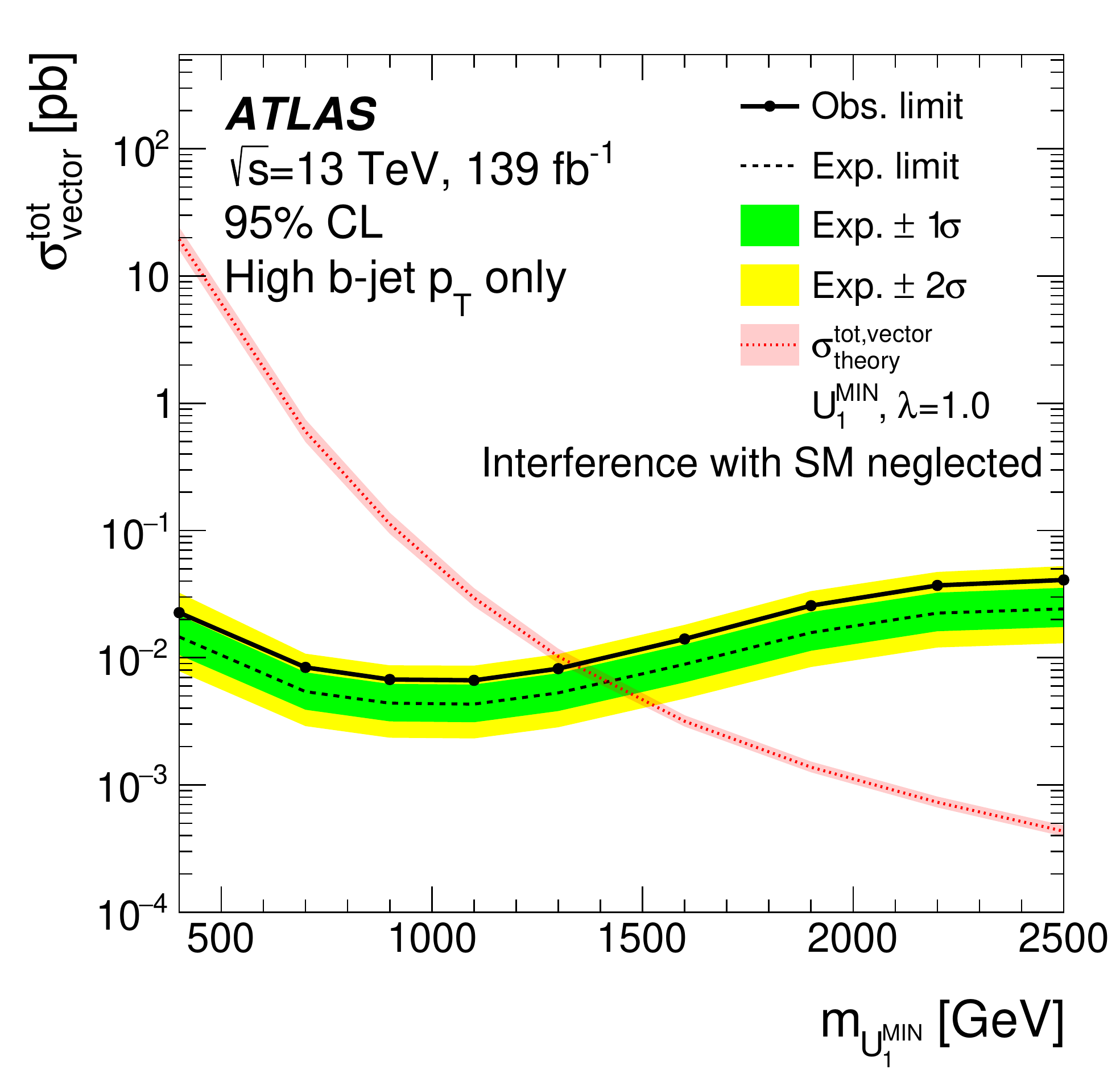}}
\subfloat[]{\includegraphics[width=0.33\textwidth]{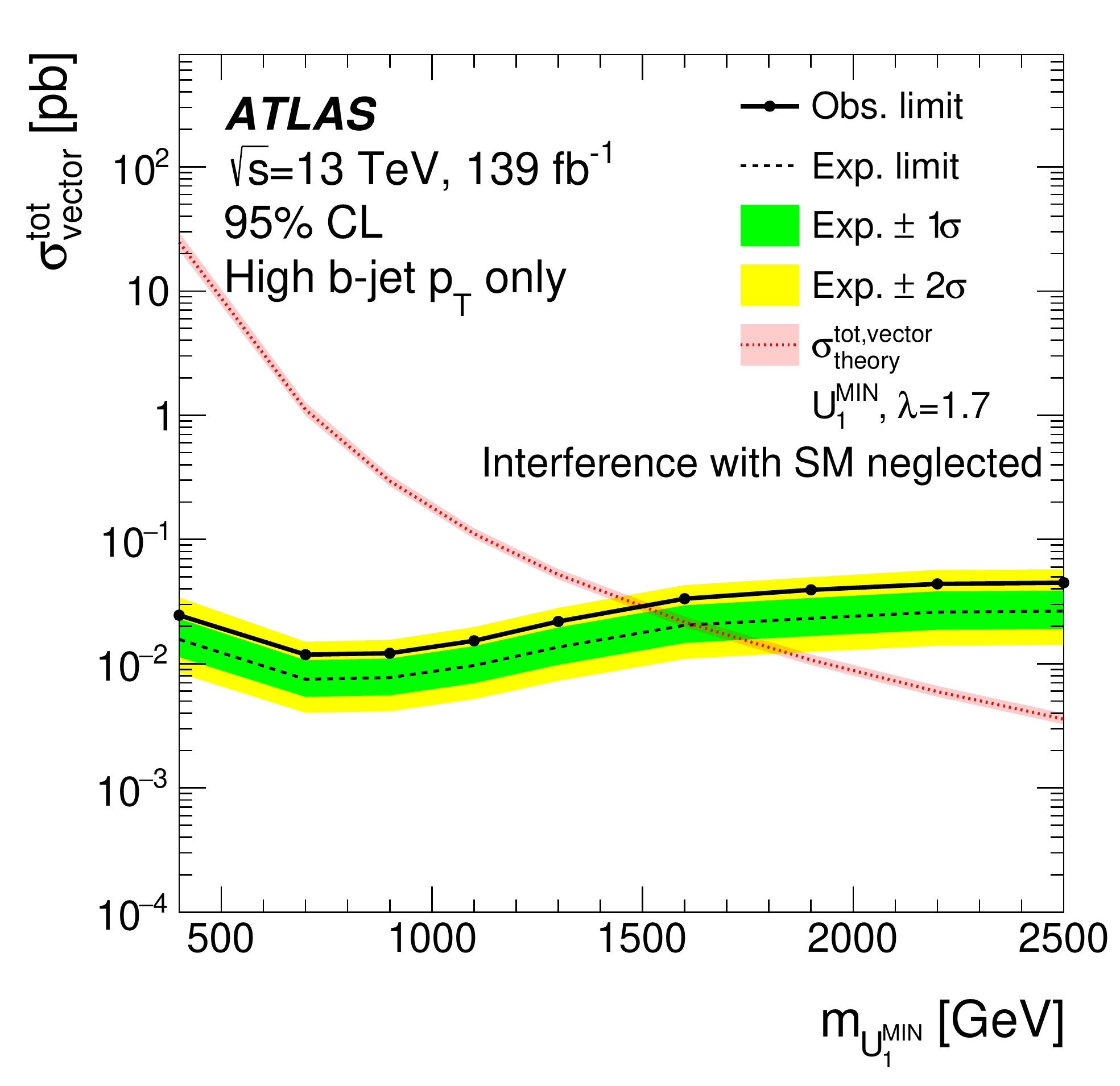}}
\subfloat[]{\includegraphics[width=0.33\textwidth]{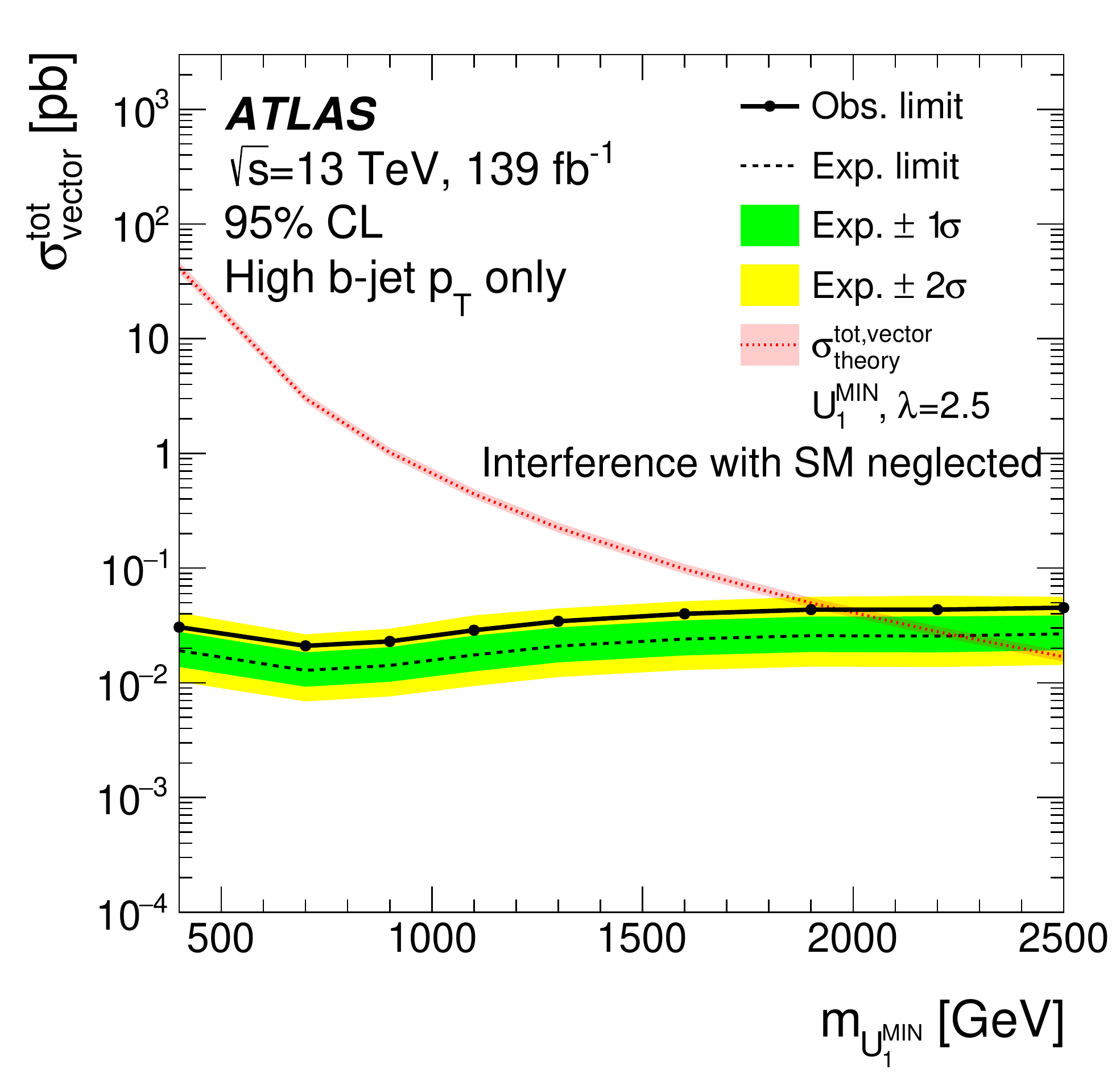}} \\
\caption{Observed (solid line) and expected (dashed line) 95$\%$ CL upper limits on the cross-section of single plus non-resonant plus pair vector LQ production from the combination of the high $b$-jet $\pt$ signal region for the $\lephad$ and $\hadhad$ channels. The dotted curve indicates the total theoretical predictions for singly, non-resonant and pair-produced vector LQ at LO\@. The top row shows the \YMLQ model ($\kappa=0$) with (a) $\lambda = 1.0$, (b) $\lambda = 1.7$ and (c) $\lambda = 2.5$. The bottom row shows the \minLQ model ($\kappa=1$) with (d) $\lambda = 1.0$, (e) $\lambda = 1.7$ and (f) $\lambda = 2.5$.
The interference of the non-resonant LQ production with SM processes is expected to be small in the high $b$-jet $\pt$ category and it is neglected.}
\label{fig:limit_singlepairvectorLQ}
\end{figure}

\begin{figure}[!ht]
\centering
\subfloat[]{\includegraphics[width=0.49\textwidth]{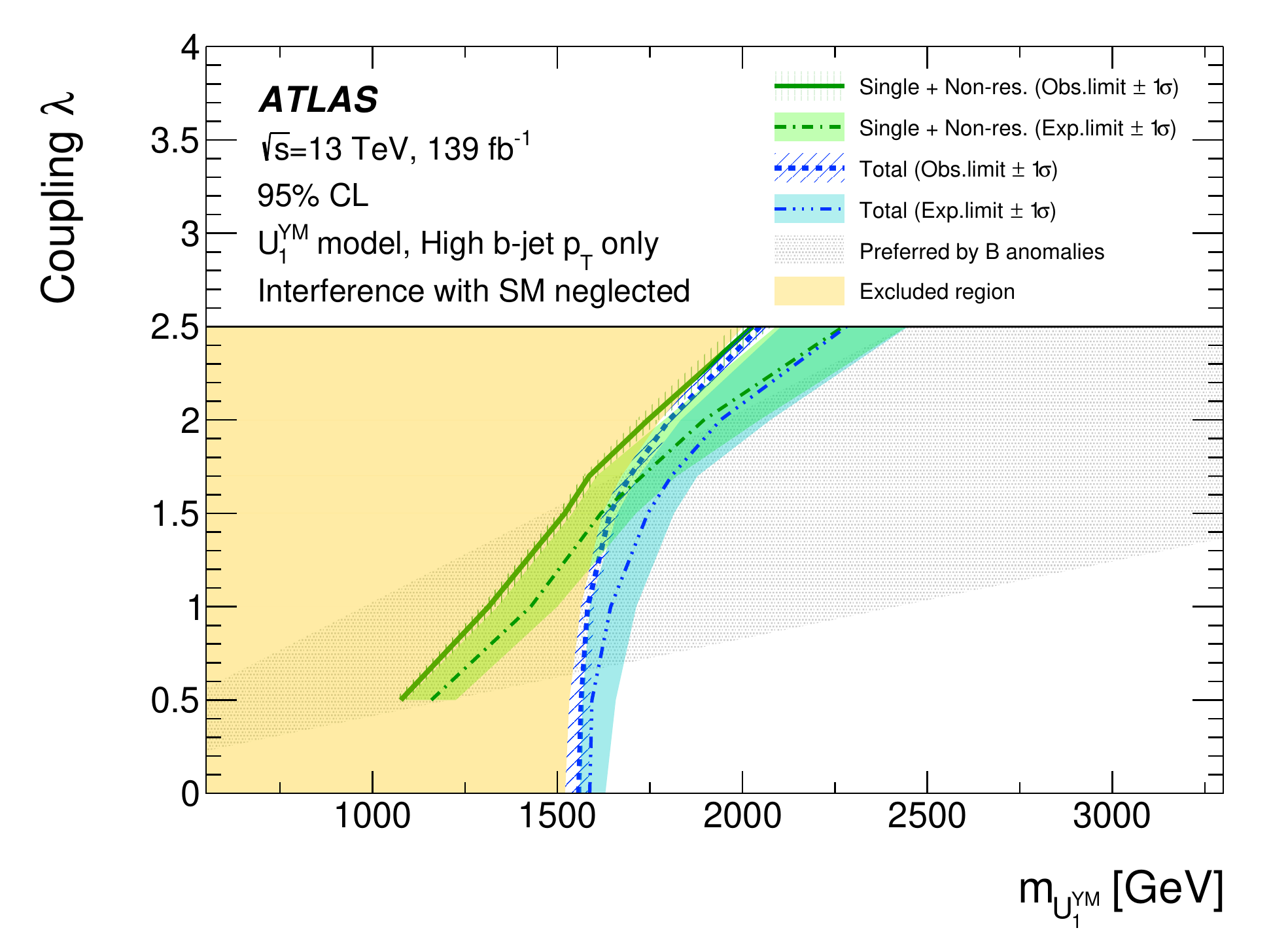}}
\subfloat[]{\includegraphics[width=0.49\textwidth]{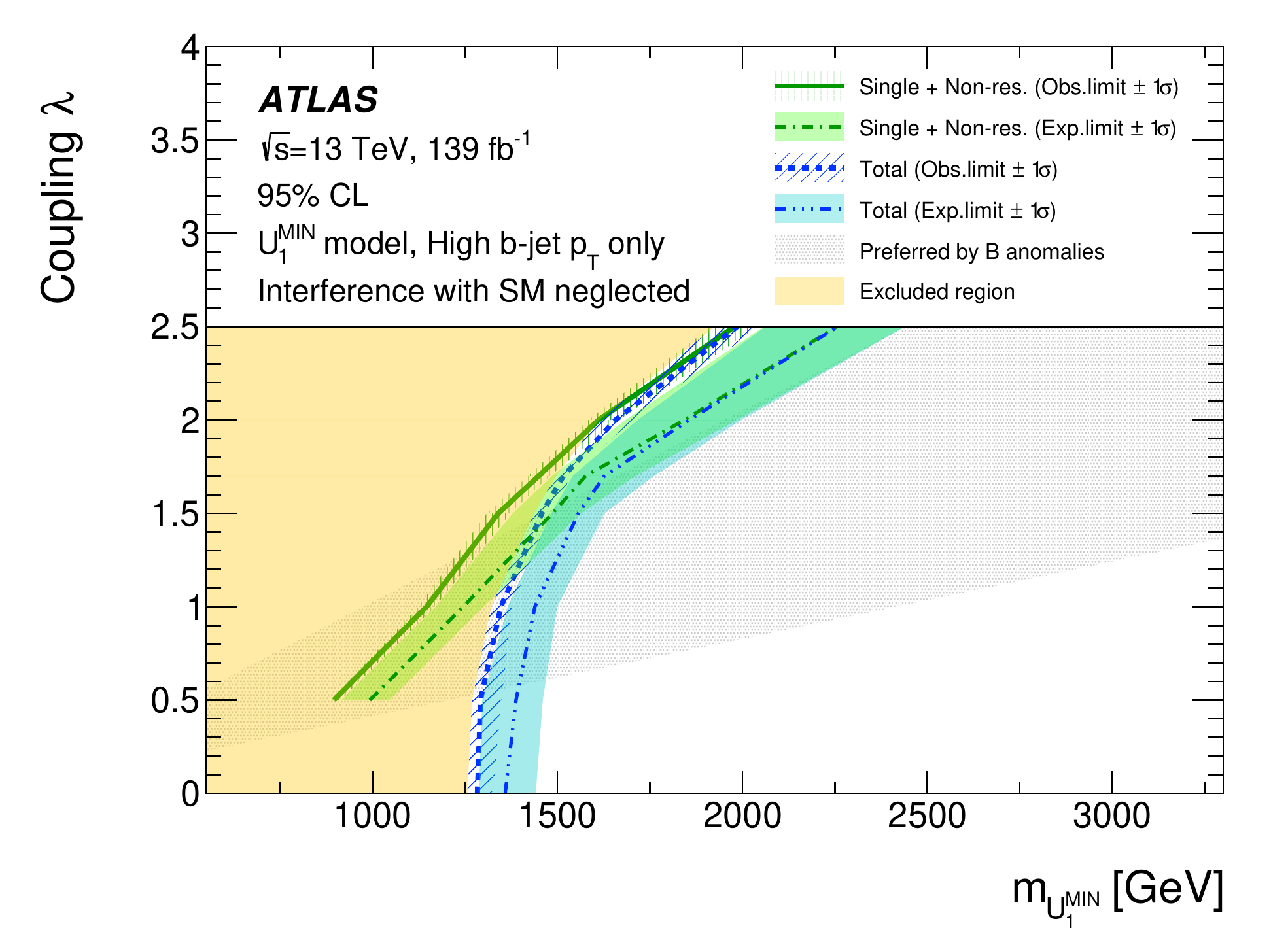}}
\caption{The two-dimensional 95$\%$ CL exclusion limits in the $\lambda-\mlq$ plane for singly plus non-resonant produced vector LQ (green lines) and for the sum, referred as \emph{Total}, of single plus non-resonant plus pair vector LQ production (blue lines), with (a) showing the case with $\kappa=0$ and (b) the case with $\kappa=1$. Regions to the left of the lines are excluded. The dotted area shows the preferred region where the chosen LQ model can explain observed $B$ anomalies~\cite{Aebischer:2022oqe}.
The interference of the non-resonant LQ production with SM processes is expected to be small in the high $b$-jet $\pt$ category and it is neglected.}
\label{fig:limit_vector_2D}
\end{figure}

The results are interpreted considering all LQ production modes in the $U_{1}$ model.
Several values of the coupling $\lambda$ are considered, each one with a value of the coupling parameter $\kappa$ of 0 or 1.
The exclusion limits for 
the single plus non-resonant plus pair vector LQ production are shown in Figure~\ref{fig:limit_singlepairvectorLQ}. The behaviour of the upper limits as a function of $\mlq$ reflects the signal acceptance times efficiency of the analysis.
Figure~\ref{fig:limit_vector_2D} shows the vector LQ limits in the $\lambda-\mlq$ plane for each of the $\kappa$ coupling values considered.
 
\begin{figure}[!h]
\centering
\subfloat[]{\includegraphics[width=0.33\textwidth]{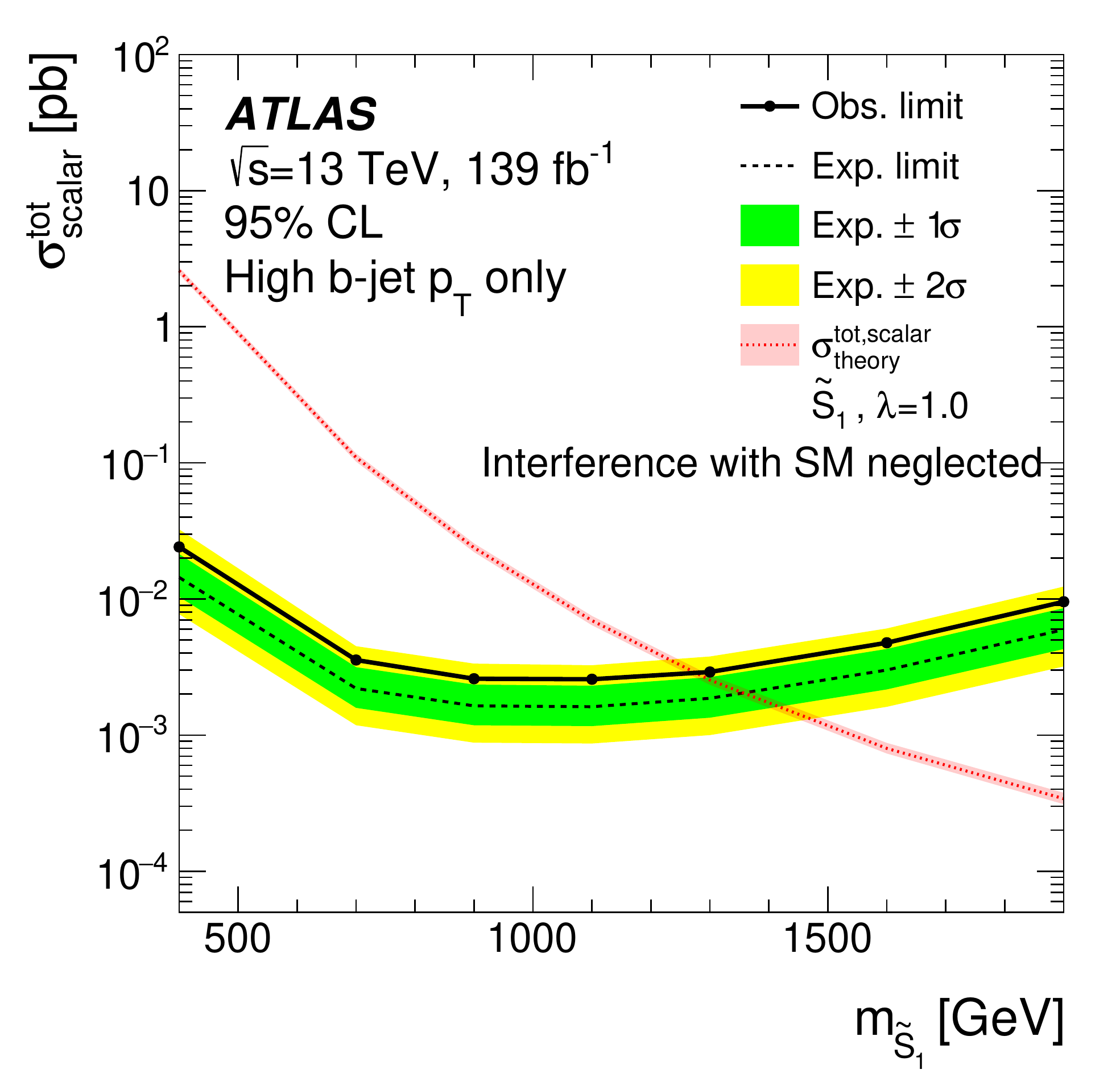}}
\subfloat[]{\includegraphics[width=0.33\textwidth]{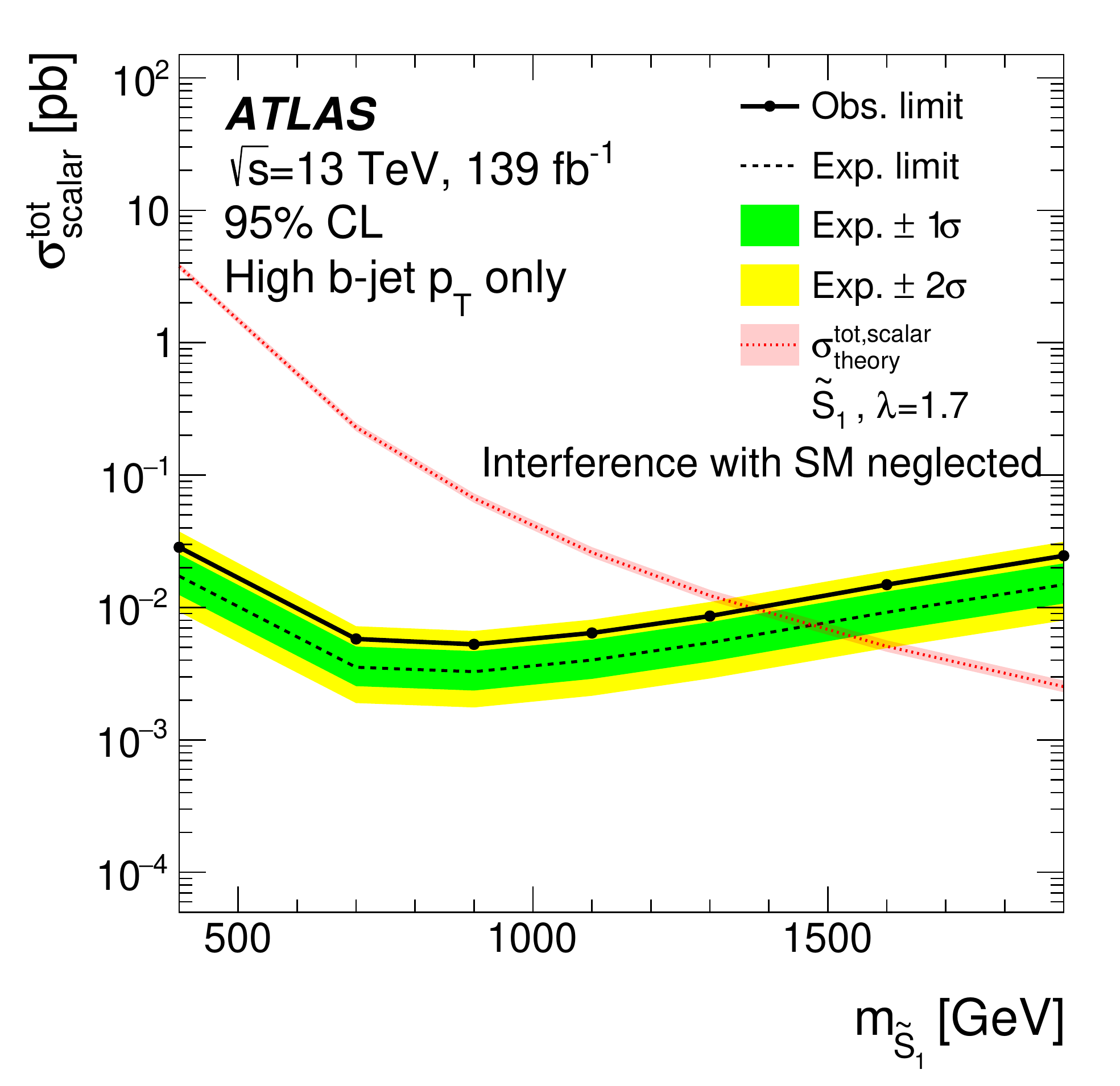}}
\subfloat[]{\includegraphics[width=0.33\textwidth]{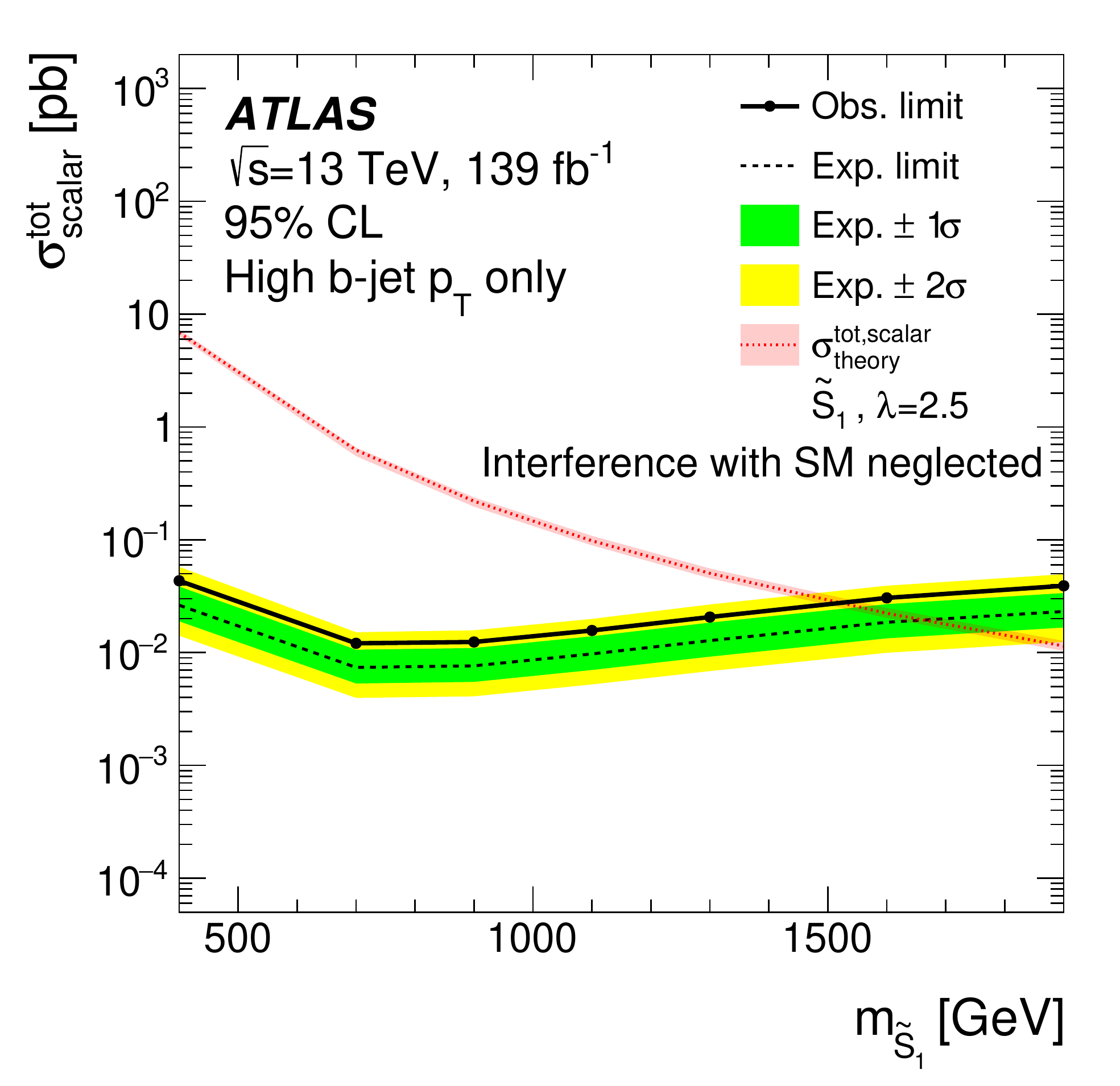}} \\
\caption{Observed (solid line) and expected (dashed line) 95$\%$ CL upper limits for (a)~$\lambda = 1.0$, (b)~$\lambda = 1.7$ and (c)~$\lambda = 2.5$ on the cross-section of single plus non-resonant plus pair $\widetilde{S_{1}}$ production hypotheses from the combination of the high $b$-jet $\pt$ signal region for the $\lephad$ and $\hadhad$ channels. The dotted curve indicates the total theoretical predictions for singly, non-resonant and pair-produced scalar  LQ. The prediction for singly plus non-resonant produced $\widetilde{S_{1}}$ (pair-produced $\widetilde{S_{1}}$) is calculated at LO (NNLO+NNLL).
The interference of the non-resonant LQ production with SM processes is expected to be small in the high $b$-jet $\pt$ category and it is neglected.}
\label{fig:limit_singlepairscalarLQ}
\end{figure}

\begin{figure}[!ht]
\centering
\includegraphics[width=0.6\textwidth]{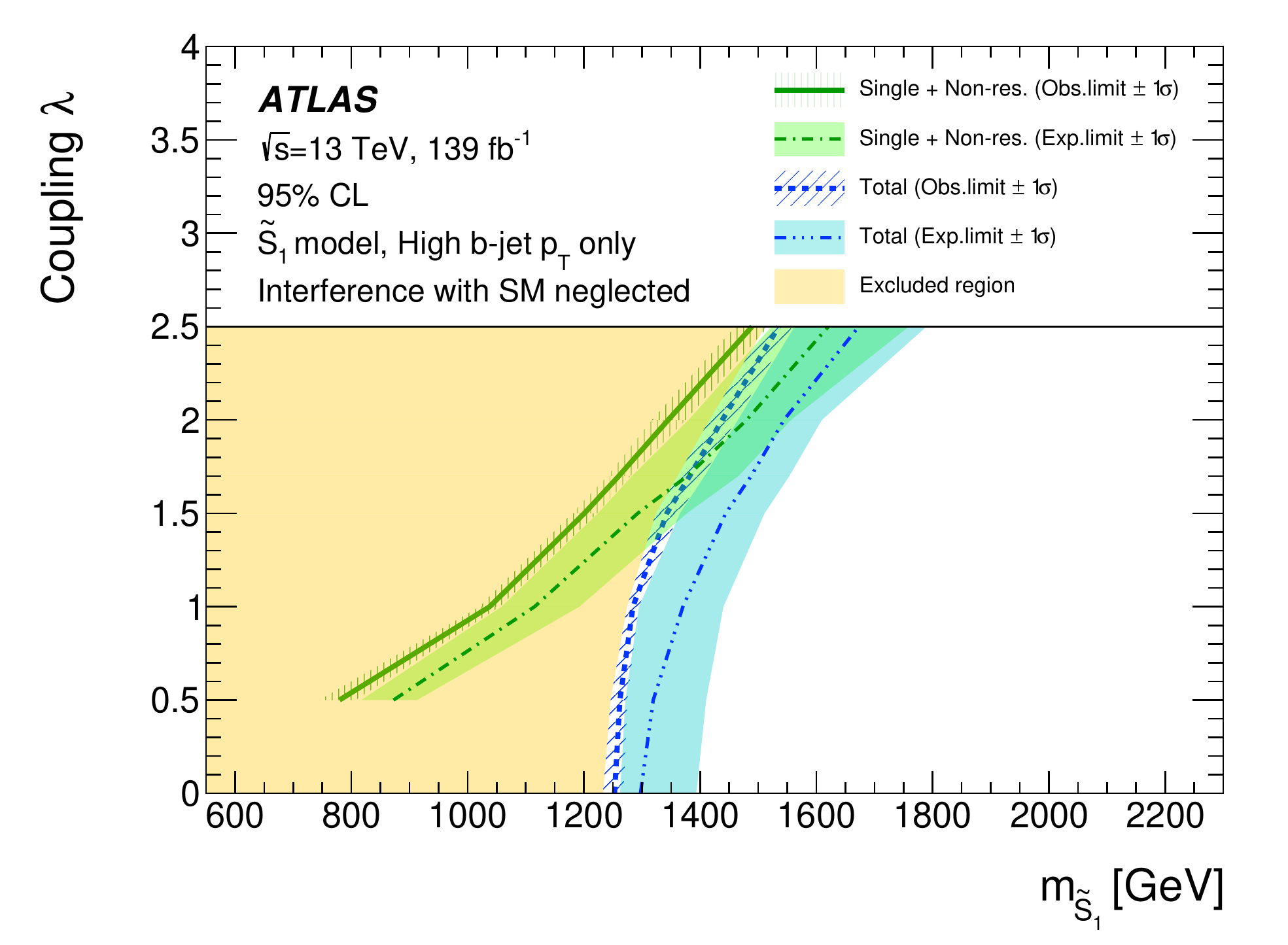}
\caption{The two-dimensional 95$\%$ CL exclusion limits in the $\lambda-\mlq$ plane for singly plus non-resonant produced $\widetilde{S_{1}}$ (green lines) and  for the sum, referred as \emph{Total}, of single plus non-resonant plus pair vector LQ production (blue lines). Regions to the left of the lines are excluded. The interference of the non-resonant LQ production with SM processes is expected to be small in the high $b$-jet $\pt$ category and it is neglected.}
\label{fig:limit_scalar_2D}
\end{figure}

The same procedure is used to interpret the results for single, non-resonant and pair production of scalar LQs from the $\widetilde{S_{1}}$ model. 
The single $\widetilde{S_{1}}$ production and the combined single plus non-resonant plus pair $\widetilde{S_{1}}$ production are considered.
The 95\% CL$_{\mathrm{s}}$ limits on the $\widetilde{S_{1}}$  production cross-section are derived as a function of LQ mass for various assumptions on the coupling $\lambda$ value.
The 
single plus non-resonant plus pair  $\widetilde{S_{1}}$ production  result is shown in Figure~\ref{fig:limit_singlepairscalarLQ}.
The exclusion limits in the $\lambda-\mlq$ plane are shown in Figure~\ref{fig:limit_scalar_2D}.
 
The observed and expected limits on the LQ mass for the various signal production modes considered are reported in Table~\ref{tab:limit_summary}.
This analysis is the first ATLAS result for the search of singly produced LQs in the $b\tau\tau$ final state.
Vector LQs in the $U_{1}$ Yang-Mills (Minimal coupling) model are excluded below masses of 1.56 (1.29)~\TeV{} for all $\lambda$ values; $\widetilde{S_{1}}$ masses below 1.26~\TeV{} are excluded for all $\lambda$ values.
The observed limits obtained are less stringent than the expected limits, which is mostly driven by the higher data yields relative to the predicted yields in the highest $\st$ bin in the $\hadhad$ channel.
Overall the $\hadhad$ channel is more sensitive than the $\lephad$ channel due to the smaller background and the larger signal to background ratio in the last $\st$ bin.
This difference arises from the larger \ttbar{} background for the  $b\tlep\thad$ final state in the most sensitive part of the $\st$ spectrum.

\begin{table}[!tp]
\small
\begin{center}
\caption{Observed (expected) 95$\%$ lower limits on the LQ mass in high $b$-jet $\pt$ signal region for the singly plus non-resonant and singly plus non-resonant plus pair produced LQs. All limits are reported in \TeV{}.}
\label{tab:limit_summary}
\begin{tabular}{l|lll}
\toprule
Model   & $\lambda=1.0$ & $\lambda=1.7$ & $\lambda=2.5$ \\
\midrule  \vspace{0.5mm}
Single$+$non-resonant \YMLQ production & 1.31 (1.43) & 1.59 (1.73) & 2.03 (2.27) \\  \vspace{0.5mm}
Single$+$non-resonant \minLQ production & 1.15 (1.24) & 1.45 (1.58) & 1.98 (2.26) \\  \vspace{0.5mm}
Single$+$non-resonant$+$pair \YMLQ production & 1.58 (1.64) &  1.70 (1.81) & 2.05 (2.28) \\  \vspace{0.5mm}
Single$+$non-resonant$+$pair \minLQ production &  1.35 (1.44) & 1.52 (1.63) & 1.99 (2.26) \\
\midrule   \vspace{0.5mm}
Single$+$non-resonant $\widetilde{S_{1}}$ production & 1.04 (1.11) &  1.26 (1.38) & 1.49 (1.62) \\
Single$+$non-resonant$+$pair $\widetilde{S_{1}}$ production & 1.28 (1.37) & 1.38 (1.49) & 1.53 (1.67) \\
\bottomrule
\end{tabular}
\end{center}
\end{table}
 
\FloatBarrier

\begin{figure}[htbp]
\centering
\subfloat[]{\includegraphics[width=0.8\textwidth]{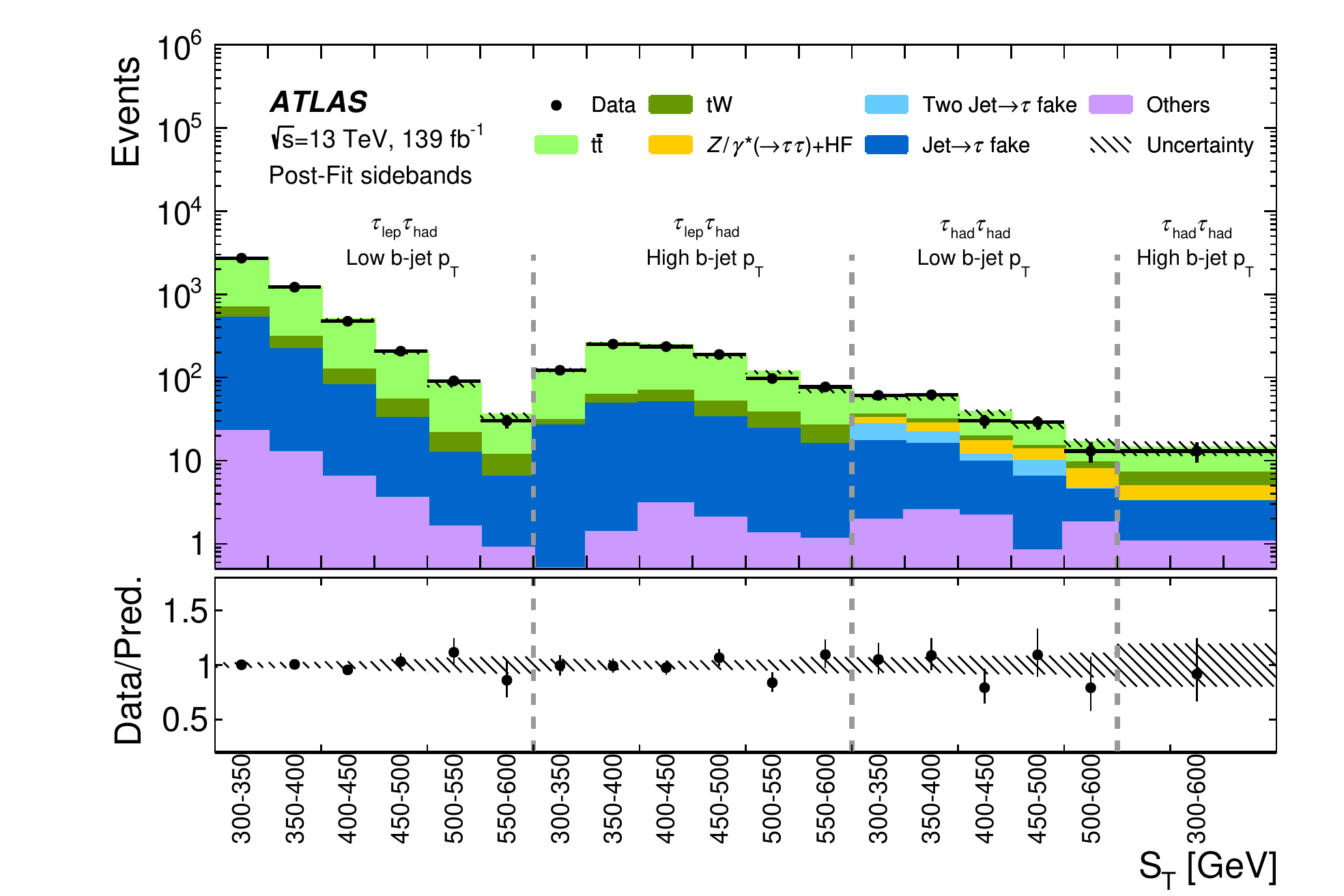}} \\
\subfloat[]{\includegraphics[width=0.8\textwidth]{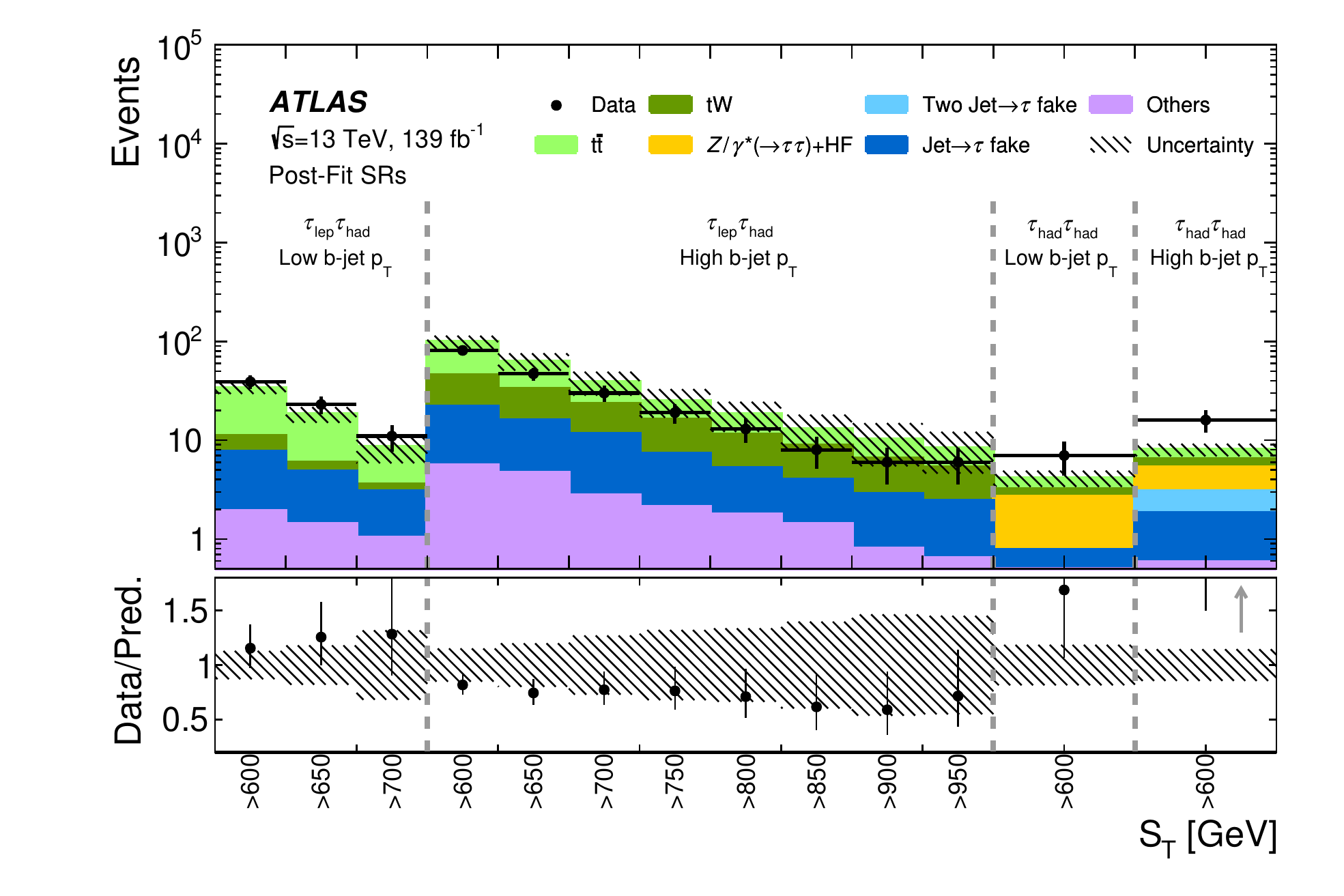}}
\caption{(a) Post-fit distributions of the discriminating variable $\st$ in the sideband regions for the high and low $b$-jet $\pt$ signal regions for the $\lephad$ and $\hadhad$ channels. The distributions are fit simultaneously considering the background-only hypothesis. (b) Observed and predicted yields of the background as a function of the $\st$ threshold used to define the signal regions for the high and low $b$-jet $\pt$ signal regions of the $\lephad$ and $\hadhad$ channels. The background predictions in the signal regions have been extracted by projecting the results from  the  fit of the sideband regions, assuming the background-only hypothesis in the context of the model-independent search.  The lower panel shows the ratio of the data to the background predictions.}
\label{fig:model_indep_postfit}
\end{figure}

An additional model-independent search considering both the high and low $b$-jet $\pt$ signal regions in the  $\lephad$ and $\hadhad$ channels is performed. For each of these regions, the events with $\st < 600$ \GeV{} form a sideband region, while a signal region is defined by counting the number of events with a $\st$ value  above a variable threshold. First the four sideband regions, one for each channel and for each $b$-jet $\pt$ signal region, are fit simultaneously considering the background-only hypothesis. Then, the fit results are used to scale the  predicted background contribution in the signal regions. In each signal region, the signal is obtained by counting the number of observed data events after subtracting the background prediction.
Figure~\ref{fig:model_indep_postfit} shows the post-fit distributions of the $\st$ variable in the sideband region and the background composition in each channel as a function of the $\st$ lower bound threshold used to define the signal regions.
 
\begin{figure}[htbp]
\centering
\includegraphics[width=0.8\textwidth]{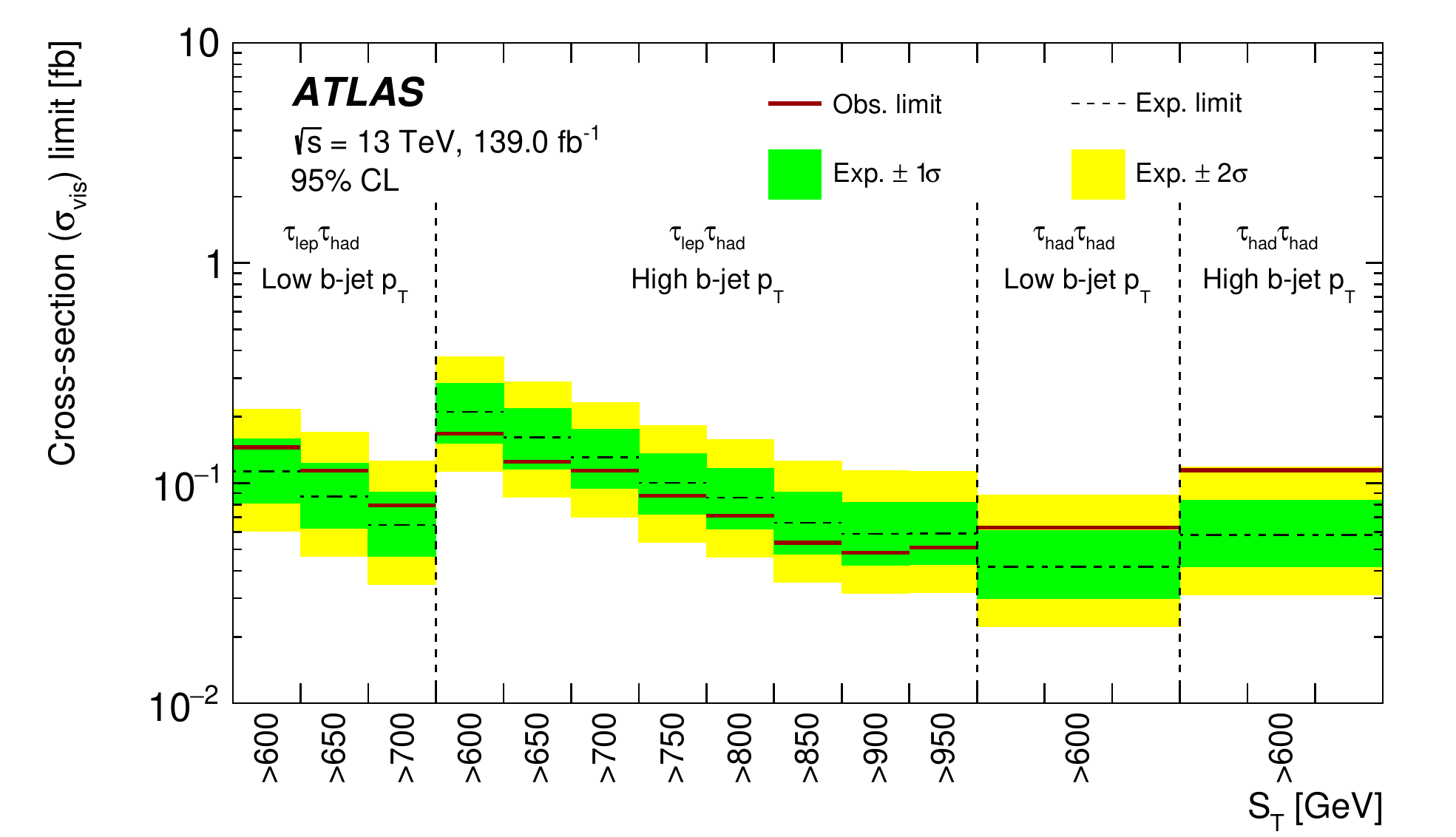}
\caption{Observed (solid line) and expected (dashed line) 95$\%$ upper limits on the visible cross-section, $\sigma_{\mathrm{vis}}$, obtained from the model-independent search by a signal-plus-background fit in the high and low $b$-jet $\pt$ signal regions for the $\lephad$ and $\hadhad$ channels.}
\label{fig:model_indep_limit}
\end{figure}

Since no significant excess is observed in any of the signal regions, a signal-plus-background fit is performed considering a generic signal in the signal region.
As for the LQs search, the parameter of interest of the statistical analysis is the signal strength $\mu$, and the results are translated into upper limits on the number of signal events and, dividing them by the integrated luminosity, they can be expressed in terms of  upper limits on the visible cross-section, $\sigma_{\mathrm{vis}}$.
Figure~\ref{fig:model_indep_limit} shows the limit values of the visible cross-section as a function of $\st$ lower bound threshold in each signal category.
The  visible cross-section limits can be reinterpreted as limits on specific physics models as long as
the selection efficiency and acceptance of the model (including any
uncertainties in these values) for a specific signal region definition used in
this analysis is known. By dividing the visible cross-section limits
given here by this efficiency and acceptance, upper limits on the
cross-section can be derived.
 
\FloatBarrier


\section{Conclusion}
\label{sec:conclusion}
 
A search for scalar and vector leptoquarks is performed in the $b\tau\tau$ final state using
$pp$ collision data at $\sqrt{s}=13$~\TeV{} recorded by the ATLAS detector at the LHC from 2015 to 2018 corresponding to an integrated luminosity of 139~fb$^{-1}$.
Final states including one leptonic and one hadronic $\tau$-lepton decay or two hadronic $\tau$-leptons decays are considered.
In each of these two final states, events are classified, based on the $\pt$ of the $b$-jet, in two signal regions of low and high $b$-jet $\pt$.
The benchmark model is $U_{1}$ for vector leptoquarks in the Yang--Mills or Minimal coupling scenarios with $\lambda$ between 0.5 and 2.5.
For scalar leptoquarks the benchmark model is $\widetilde{S_{1}}$, with values of the $\lambda$ parameter ranging between 0.5 and 2.5.

Upper limits at 95\% CL on the cross-section for leptoquarks produced via either single plus non-resonant production, or considering all production modes (including pair production), and decaying into $b\tau$ are set.
The results have been extracted considering only the high $b$-jet signal region and the combination of both final states.
For the Yang--Mills coupling, the observed (expected) lower limits on the leptoquark mass are 1.58 (1.64)~\TeV{} for $\lambda=1.0$, 1.70 (1.81)~\TeV{} for $\lambda=1.7$, and 2.05 (2.28)~\TeV{} for $\lambda=2.5$, considering all leptoquark production modes.
In the Minimal coupling  case,
considering all leptoquark production modes, the lower limits are 1.35 (1.44)~\TeV{} for $\lambda=1.0$, 1.52 (1.63)~\TeV{} for $\lambda=1.7$ and 1.99 (2.26)~\TeV{} for $\lambda=2.5$.
The observed (expected) lower limits,
considering all leptoquark production modes in the  $\widetilde{S_{1}}$ model, are
1.28 (1.37)~\TeV{} for $\lambda=1.0$, 1.38 (1.49)~\TeV{} for $\lambda=1.7$ and 1.53 (1.67)~\TeV{} for $\lambda=~2.5$.
 
An interpretation of the results in a model-independent scenario is also performed for each of the signal region categories. The 95\% confidence level limits on the visible cross-section vary between 0.17~fb and $4.8\cdot 10^{-2}$~fb as a function of the event variable $\st$ ranging from $\st>600$~\GeV{} to $\st>950$~\GeV{}.


\section*{Acknowledgements}


We thank CERN for the very successful operation of the LHC, as well as the
support staff from our institutions without whom ATLAS could not be
operated efficiently.
 
We acknowledge the support of
ANPCyT, Argentina;
YerPhI, Armenia;
ARC, Australia;
BMWFW and FWF, Austria;
ANAS, Azerbaijan;
CNPq and FAPESP, Brazil;
NSERC, NRC and CFI, Canada;
CERN;
ANID, Chile;
CAS, MOST and NSFC, China;
Minciencias, Colombia;
MEYS CR, Czech Republic;
DNRF and DNSRC, Denmark;
IN2P3-CNRS and CEA-DRF/IRFU, France;
SRNSFG, Georgia;
BMBF, HGF and MPG, Germany;
GSRI, Greece;
RGC and Hong Kong SAR, China;
ISF and Benoziyo Center, Israel;
INFN, Italy;
MEXT and JSPS, Japan;
CNRST, Morocco;
NWO, Netherlands;
RCN, Norway;
MEiN, Poland;
FCT, Portugal;
MNE/IFA, Romania;
MESTD, Serbia;
MSSR, Slovakia;
ARRS and MIZ\v{S}, Slovenia;
DSI/NRF, South Africa;
MICINN, Spain;
SRC and Wallenberg Foundation, Sweden;
SERI, SNSF and Cantons of Bern and Geneva, Switzerland;
MOST, Taiwan;
TENMAK, T\"urkiye;
STFC, United Kingdom;
DOE and NSF, United States of America.
In addition, individual groups and members have received support from
BCKDF, CANARIE, Compute Canada and CRC, Canada;
PRIMUS 21/SCI/017 and UNCE SCI/013, Czech Republic;
COST, ERC, ERDF, Horizon 2020 and Marie Sk{\l}odowska-Curie Actions, European Union;
Investissements d'Avenir Labex, Investissements d'Avenir Idex and ANR, France;
DFG and AvH Foundation, Germany;
Herakleitos, Thales and Aristeia programmes co-financed by EU-ESF and the Greek NSRF, Greece;
BSF-NSF and MINERVA, Israel;
Norwegian Financial Mechanism 2014-2021, Norway;
NCN and NAWA, Poland;
La Caixa Banking Foundation, CERCA Programme Generalitat de Catalunya and PROMETEO and GenT Programmes Generalitat Valenciana, Spain;
G\"{o}ran Gustafssons Stiftelse, Sweden;
The Royal Society and Leverhulme Trust, United Kingdom.
 
The crucial computing support from all WLCG partners is acknowledged gratefully, in particular from CERN, the ATLAS Tier-1 facilities at TRIUMF (Canada), NDGF (Denmark, Norway, Sweden), CC-IN2P3 (France), KIT/GridKA (Germany), INFN-CNAF (Italy), NL-T1 (Netherlands), PIC (Spain), ASGC (Taiwan), RAL (UK) and BNL (USA), the Tier-2 facilities worldwide and large non-WLCG resource providers. Major contributors of computing resources are listed in Ref.~\cite{ATL-SOFT-PUB-2021-003}.


\clearpage

\printbibliography

@Article{Evans:2008zzb,
      author         = "Evans, Lyndon and Bryant, Philip",
      title          = "{LHC Machine}",
      journal        = "JINST",
      volume         = "3",
      pages          = "S08001",
      doi            = "10.1088/1748-0221/3/08/S08001",
      year           = "2008",
      SLACcitation   = "%%CITATION = JINST,3,S08001;%%",
}

@Article{Cacciari:2008gp,
     author    = "Cacciari, Matteo and Salam, Gavin P. and Soyez, Gregory",
     title     = "{The anti-\(k_{t}\) jet clustering algorithm}",
     journal   = "JHEP",
     volume    = "04",
     year      = "2008",
     pages     = "063",
     eprint    = "0802.1189",
     archivePrefix = "arXiv",
     primaryClass  =  "hep-ph",
     doi       = "10.1088/1126-6708/2008/04/063",
     SLACcitation  = "%%CITATION = 0802.1189;%%"
}

@Article{Fastjet,
      author         = "Cacciari, Matteo and Salam, Gavin P. and Soyez, Gregory",
      title          = "{FastJet user manual}",
      journal        = "Eur. Phys. J. C",
      volume         = "72",
      year           = "2012",
      pages          = "1896",
      doi            = "10.1140/epjc/s10052-012-1896-2",
      eprint         = "1111.6097",
      archivePrefix  = "arXiv",
      primaryClass   = "hep-ph",
      reportNumber   = "CERN-PH-TH-2011-297",
      SLACcitation   = "%%CITATION = ARXIV:1111.6097;%%"
}

@Article{Butterworth:2015oua,
      author         = "Butterworth, Jon and others",
      title          = "{PDF4LHC recommendations for LHC Run II}",
      journal        = "J. Phys. G",
      volume         = "43",
      year           = "2016",
      pages          = "023001",
      doi            = "10.1088/0954-3899/43/2/023001",
      eprint         = "1510.03865",
      archivePrefix  = "arXiv",
      primaryClass   = "hep-ph",
      reportNumber   = "OUTP-15-17P, SMU-HEP-15-12, TIF-UNIMI-2015-14,
                        LCTS-2015-27, CERN-PH-TH-2015-249",
      SLACcitation   = "%%CITATION = ARXIV:1510.03865;%%"
}

@Article{Pumplin:2002vw,
     author    = "Pumplin, J. and others",
     title     = "{New Generation of Parton Distributions with Uncertainties from Global QCD Analysis}",
     journal   = "JHEP",
     volume    = "07",
     year      = "2002",
     pages     = "012",
     doi       = "10.1088/1126-6708/2002/07/012",
     eprint    = "hep-ph/0201195",
     archivePrefix = "arXiv",
     SLACcitation  = "%%CITATION = HEP-PH/0201195;%%"
}

@article{Lai:2010vv,
      author         = "Lai, H.-L. and others",
      title          = "{New parton distributions for collider physics}",
      journal        = "Phys. Rev. D",
      volume         = "82",
      pages          = "074024",
      doi            = "10.1103/PhysRevD.82.074024",
      year           = "2010",
      eprint         = "1007.2241",
      archivePrefix  = "arXiv",
      primaryClass   = "hep-ph",
      reportNumber   = "MSUHEP-100707, SMU-HEP-10-10",
      SLACcitation   = "%%CITATION = ARXIV:1007.2241;%%",
}

@article{Ball:2012cx,
      author         = "{NNPDF Collaboration} and Ball, Richard D. and others",
      title          = "{Parton distributions with LHC data}",
      journal        = "Nucl. Phys. B",
      volume         = "867",
      year           = "2013",
      pages          = "244",
      doi            = "10.1016/j.nuclphysb.2012.10.003",
      eprint         = "1207.1303",
      archivePrefix  = "arXiv",
      primaryClass   = "hep-ph",
      reportNumber   = "EDINBURGH-2012-08, IFUM-FT-997, FR-PHENO-2012-014,
                        RWTH-TTK-12-25, CERN-PH-TH-2012-037, SFB-CPP-12-47\,
                        --CERN-PH-TH-2012-037",
      SLACcitation   = "%%CITATION = ARXIV:1207.1303;%%"
}

@article{Ball:2014uwa,
      author         = "{The NNPDF Collaboration} and Ball, Richard D. and others",
      title          = "{Parton distributions for the LHC run II}",
      journal        = "JHEP",
      volume         = "04",
      year           = "2015",
      pages          = "040",
      doi            = "10.1007/JHEP04(2015)040",
      eprint         = "1410.8849",
      archivePrefix  = "arXiv",
      primaryClass   = "hep-ph",
      reportNumber   = "EDINBURGH-2014-15, IFUM-1034-FT, CERN-PH-TH-2013-253,
                        OUTP-14-11P, CAVENDISH-HEP-14-11",
      SLACcitation   = "%%CITATION = ARXIV:1410.8849;%%"
}

@article{Czakon:2012pz,
      author         = "Czakon, Michal and Mitov, Alexander",
      title          = "{NNLO corrections to top pair production at hadron
                        colliders: the quark-gluon reaction}",
      journal        = "JHEP",
      volume         = "01",
      pages          = "080",
      doi            = "10.1007/JHEP01(2013)080",
      year           = "2013",
      eprint         = "1210.6832",
      archivePrefix  = "arXiv",
      primaryClass   = "hep-ph",
      SLACcitation   = "%%CITATION = ARXIV:1210.6832;%%",
}

@Article{Sjostrand:2007gs,
	author    = "Sj{\"o}strand, T. and Mrenna, S. and Skands, P.",
	title     = "{A brief introduction to PYTHIA 8.1}",
	journal   = "Comput. Phys. Commun.",
	volume    = "178",
	year      = "2008",
	pages     = "852-867",
	eprint    = "0710.3820",
	archivePrefix = "arXiv",
	primaryClass  =  "hep-ph",
	doi       = "10.1016/j.cpc.2008.01.036",
	SLACcitation  = "%%CITATION = 0710.3820;%%"
}

@article{Sjostrand:2014zea,
      author         = "Sj{\"o}strand, Torbj{\"o}rn and Ask, Stefan and Christiansen,
                        Jesper R. and Corke, Richard and Desai, Nishita and Ilten,
                        Philip and Mrenna, Stephen and Prestel, Stefan and
                        Rasmussen, Christine O. and Skands, Peter Z.",
      title          = "{An introduction to PYTHIA 8.2}",
      journal        = "Comput. Phys. Commun.",
      volume         = "191",
      year           = "2015",
      pages          = "159",
      doi            = "10.1016/j.cpc.2015.01.024",
      eprint         = "1410.3012",
      archivePrefix  = "arXiv",
      primaryClass   = "hep-ph",
      reportNumber   = "LU-TP-14-36, MCNET-14-22, CERN-PH-TH-2014-190,
                        FERMILAB-PUB-14-316-CD, DESY-14-178, SLAC-PUB-16122,
                        --FERMILAB-PUB-14-316-CD",
      SLACcitation   = "%%CITATION = ARXIV:1410.3012;%%"
}

@Article{Lange:2001uf,
      author         = "Lange, D. J.",
      title          = "{The EvtGen particle decay simulation package}",
      booktitle      = "{Proceedings, 7th International Conference on B physics
                        at hadron machines (BEAUTY 2000)}",
      journal        = "Nucl. Instrum. Meth. A",
      volume         = "462",
      year           = "2001",
      pages          = "152",
      doi            = "10.1016/S0168-9002(01)00089-4",
      SLACcitation   = "%%CITATION = NUIMA,A462,152;%%"
}

@article{Golonka:2005pn,
      author         = "Golonka, Piotr and Was, Zbigniew",
      title          = "{PHOTOS Monte Carlo: a precision tool for QED corrections
                        in \(Z\) and \(W\) decays}",
      journal        = "Eur. Phys. J. C",
      volume         = "45",
      year           = "2006",
      pages          = "97-107",
      doi            = "10.1140/epjc/s2005-02396-4",
      eprint         = "hep-ph/0506026",
      archivePrefix  = "arXiv",
      XprimaryClass   = "hep-ph",
      reportNumber   = "IFJPAN-V-05-01, CERN-PH-TH-2005-091",
      SLACcitation   = "%%CITATION = HEP-PH/0506026;%%"
}

@Article{Frixione:2008yi,
     author    = "Frixione, Stefano and Laenen, Eric and Motylinski, Patrick and White, Chris and Webber, Bryan R.",
     title     = "{Single-top hadroproduction in association with a \(W\) boson}",
     journal   = "JHEP",
     volume    = "07",
     year      = "2008",
     pages     = "029",
     eprint    = "0805.3067",
     archivePrefix = "arXiv",
     primaryClass  =  "hep-ph",
     doi       = "10.1088/1126-6708/2008/07/029"
}

@Article{Alwall:2014hca,
      author         = "Alwall, J. and Frederix, R. and Frixione, S. and Hirschi,
                        V. and Maltoni, F. and Mattelaer, O. and Shao, H. -S. and
                        Stelzer, T. and Torrielli, P. and Zaro, M.",
      title          = "{The automated computation of tree-level and
                        next-to-leading order differential cross sections, and
                        their matching to parton shower simulations}",
      journal        = "JHEP",
      volume         = "07",
      year           = "2014",
      pages          = "079",
      doi            = "10.1007/JHEP07(2014)079",
      eprint         = "1405.0301",
      archivePrefix  = "arXiv",
      primaryClass   = "hep-ph",
      reportNumber   = "CERN-PH-TH-2014-064, CP3-14-18, LPN14-066, MCNET-14-09,
                        ZU-TH-14-14",
      SLACcitation   = "%%CITATION = ARXIV:1405.0301;%%"
}

@Article{Nason:2004rx,
      author         = "Nason, Paolo",
      title          = "{A new method for combining NLO QCD with shower Monte Carlo algorithms}",
      journal        = "JHEP",
      volume         = "11",
      pages          = "040",
      doi            = "10.1088/1126-6708/2004/11/040",
      year           = "2004",
      eprint         = "hep-ph/0409146",
      archivePrefix  = "arXiv",
}

@Article{Frixione:2007vw,
      author         = "Frixione, Stefano and Nason, Paolo and Oleari, Carlo",
      title          = "{Matching NLO QCD computations with parton shower
                        simulations: the POWHEG method}",
      journal        = "JHEP",
      volume         = "11",
      pages          = "070",
      doi            = "10.1088/1126-6708/2007/11/070",
      year           = "2007",
      eprint         = "0709.2092",
      archivePrefix  = "arXiv",
      primaryClass   = "hep-ph",
}

@Article{Alioli:2010xd,
      author         = "Alioli, Simone and Nason, Paolo and Oleari, Carlo and Re,
                        Emanuele",
      title          = "{A general framework for implementing NLO calculations in
                        shower Monte Carlo programs: the POWHEG BOX}",
      journal        = "JHEP",
      volume         = "06",
      pages          = "043",
      doi            = "10.1007/JHEP06(2010)043",
      year           = "2010",
      eprint         = "1002.2581",
      archivePrefix  = "arXiv",
      primaryClass   = "hep-ph",
}

@article{Alioli:2008gx,
      author         = "Alioli, Simone and Nason, Paolo and Oleari, Carlo and Re,
                        Emanuele",
      title          = "{NLO vector-boson production matched with shower in
                        POWHEG}",
      journal        = "JHEP",
      volume         = "07",
      year           = "2008",
      pages          = "060",
      doi            = "10.1088/1126-6708/2008/07/060",
      eprint         = "0805.4802",
      archivePrefix  = "arXiv",
      primaryClass   = "hep-ph",
      reportNumber   = "NSF-KITP-08-74",
      SLACcitation   = "%%CITATION = ARXIV:0805.4802;%%"
}

@article{Aliev:2010zk,
      author         = "Aliev, M. and Lacker, H. and Langenfeld, U. and Moch, S.
                        and Uwer, P. and Wiedermann, M.",
      title          = "{HATHOR -- HAdronic Top and Heavy quarks crOss section
                        calculatoR}",
      journal        = "Comput. Phys. Commun.",
      volume         = "182",
      year           = "2011",
      pages          = "1034-1046",
      doi            = "10.1016/j.cpc.2010.12.040",
      eprint         = "1007.1327",
      archivePrefix  = "arXiv",
      primaryClass   = "hep-ph",
      reportNumber   = "DESY-10-091, HU-EP-10-33, SFB-CPP-10-60",
      SLACcitation   = "%%CITATION = ARXIV:1007.1327;%%"
}

@article{Beneke:2011mq,
      author         = "Beneke, M. and Falgari, P. and Klein, S. and Schwinn, C.",
      title          = "{Hadronic top-quark pair production with NNLL threshold
                        resummation}",
      journal        = "Nucl. Phys. B",
      volume         = "855",
      year           = "2012",
      pages          = "695-741",
      doi            = "10.1016/j.nuclphysb.2011.10.021",
      eprint         = "1109.1536",
      archivePrefix  = "arXiv",
      primaryClass   = "hep-ph",
      reportNumber   = "TTK-11-38, ITP-UU-11-26, SPIN-11-19, FR-PHENO-2011-015,
                        SFB-CPP-11-49",
      SLACcitation   = "%%CITATION = ARXIV:1109.1536;%%"
}

@article{Cacciari:2011hy,
      author         = "Cacciari, Matteo and Czakon, Michal and Mangano,
                        Michelangelo and Mitov, Alexander and Nason, Paolo",
      title          = "{Top-pair production at hadron colliders with
                        next-to-next-to-leading logarithmic soft-gluon
                        resummation}",
      journal        = "Phys. Lett. B",
      volume         = "710",
      pages          = "612-622",
      doi            = "10.1016/j.physletb.2012.03.013",
      year           = "2012",
      eprint         = "1111.5869",
      primaryClass   = "hep-ph",
      archivePrefix  = "arXiv",
      reportNumber   = "CERN-PH-TH-2011-277, TTK-11-54",
      SLACcitation   = "%%CITATION = ARXIV:1111.5869;%%",
}

@article{Czakon:2012zr,
      author         = "Czakon, Michal and Mitov, Alexander",
      title          = "{NNLO corrections to top-pair production at hadron
                        colliders: the all-fermionic scattering channels}",
      journal        = "JHEP",
      volume         = "12",
      pages          = "054",
      doi            = "10.1007/JHEP12(2012)054",
      year           = "2012",
      eprint         = "1207.0236",
      archivePrefix  = "arXiv",
      primaryClass   = "hep-ph",
      SLACcitation   = "%%CITATION = ARXIV:1207.0236;%%",
}

@article{Czakon:2013goa,
      author         = "Czakon, Michal and Fiedler, Paul and Mitov, Alexander",
      title          = "{Total Top-Quark Pair-Production Cross Section at
                        Hadron Colliders Through \(O(\alpha_S^4)\)}",
      journal        = "Phys. Rev. Lett.",
      volume         = "110",
      pages          = "252004",
      doi            = "10.1103/PhysRevLett.110.252004",
      year           = "2013",
      eprint         = "1303.6254",
      archivePrefix  = "arXiv",
      primaryClass   = "hep-ph",
      reportNumber   = "CERN-PH-TH-2013-056, TTK-13-08",
      SLACcitation   = "%%CITATION = ARXIV:1303.6254;%%",
}

@article{Baernreuther:2012ws,
      author         = "B{\"a}rnreuther, Peter and Czakon, Michal and Mitov,
                        Alexander",
      title          = "{Percent-Level-Precision Physics at the
                  Tevatron: Next-to-Next-to-Leading Order QCD
                  Corrections to \(q \bar{q} \to t \bar{t} + X\)}",
      journal        = "Phys. Rev. Lett.",
      volume         = "109",
      pages          = "132001",
      doi            = "10.1103/PhysRevLett.109.132001",
      year           = "2012",
      eprint         = "1204.5201",
      archivePrefix  = "arXiv",
      primaryClass   = "hep-ph",
      SLACcitation   = "%%CITATION = ARXIV:1204.5201;%%",
}

@article{Frixione:2007nw,
      author         = "Frixione, Stefano and Ridolfi, Giovanni and Nason, Paolo",
      title          = "{A positive-weight next-to-leading-order Monte Carlo for
                        heavy flavour hadroproduction}",
      journal        = "JHEP",
      volume         = "09",
      pages          = "126",
      doi            = "10.1088/1126-6708/2007/09/126",
      year           = "2007",
      eprint         = "0707.3088",
      archivePrefix  = "arXiv",
      primaryClass   = "hep-ph",
      SLACcitation   = "%%CITATION = ARXIV:0707.3088;%%",
}

@Article{Hoeche:2009rj,
      author         = "H{\"o}che, Stefan and Krauss, Frank and Schumann, Steffen and Siegert, Frank",
      title          = "{QCD matrix elements and truncated showers}",
      journal        = "JHEP",
      volume         = "05",
      pages          = "053",
      doi            = "10.1088/1126-6708/2009/05/053",
      year           = "2009",
      eprint         = "0903.1219",
      archivePrefix  = "arXiv",
      primaryClass   = "hep-ph",
}

@Article{Gleisberg:2008fv,
      author         = "Gleisberg, Tanju and H{\"o}che, Stefan",
      title          = "{Comix, a new matrix element generator}",
      journal        = "JHEP",
      volume         = "12",
      pages          = "039",
      doi            = "10.1088/1126-6708/2008/12/039",
      year           = "2008",
      eprint         = "0808.3674",
      archivePrefix  = "arXiv",
      primaryClass   = "hep-ph",
}

@Article{Schumann:2007mg,
      author         = "Schumann, Steffen and Krauss, Frank",
      title          = "{A parton shower algorithm based on Catani--Seymour dipole factorisation}",
      journal        = "JHEP",
      volume         = "03",
      pages          = "038",
      doi            = "10.1088/1126-6708/2008/03/038",
      year           = "2008",
      eprint         = "0709.1027",
      archivePrefix  = "arXiv",
      primaryClass   = "hep-ph",
}

@article{Hoeche:2012yf,
      author         = "H{\"o}che, Stefan and Krauss, Frank and Sch{\"o}nherr, Marek and
                        Siegert, Frank",
      title          = "{QCD matrix elements + parton showers. The NLO case}",
      journal        = "JHEP",
      volume         = "04",
      year           = "2013",
      pages          = "027",
      doi            = "10.1007/JHEP04(2013)027",
      eprint         = "1207.5030",
      archivePrefix  = "arXiv",
      primaryClass   = "hep-ph",
      reportNumber   = "SLAC-PUB-15191, IPPP-12-52, DCPT-12-104, LPN12-081,
                        FR-PHENO-2012-017, MCNET-12-09, --FR-PHENO-2012-017",
      SLACcitation   = "%%CITATION = ARXIV:1207.5030;%%"
}

@article{Bothmann:2019yzt,
      author = "Bothmann, Enrico and others",
      title = "{Event generation with Sherpa 2.2}",
      journal = "SciPost Phys.",
      volume = "7",
      year = "2019",
      number = "3",
      pages = "034",
      doi = "10.21468/SciPostPhys.7.3.034",
      reportNumber = "FERMILAB-PUB-19-218-T, SLAC-PUB-17433, IPPP/19/42, MCNET-19-11",
      eprint = "1905.09127",
      archivePrefix = "arXiv",
      primaryClass = "hep-ph",
}

@article{Hoeche:2011fd,
      author         = "H{\"o}che, Stefan and Krauss, Frank and Sch{\"o}nherr, Marek and
                        Siegert, Frank",
      title          = "{A critical appraisal of NLO+PS matching methods}",
      journal        = "JHEP",
      volume         = "09",
      year           = "2012",
      pages          = "049",
      doi            = "10.1007/JHEP09(2012)049",
      eprint         = "1111.1220",
      archivePrefix  = "arXiv",
      primaryClass   = "hep-ph",
      reportNumber   = "SLAC-PUB-14661, IPPP-11-67, DCPT-11-134, LPN11-58,
                        FR-PHENO-2011-019, MCNET-11-24",
      SLACcitation   = "%%CITATION = ARXIV:1111.1220;%%"
}

@article{Catani:2001cc,
      author         = "Catani, S. and Krauss, F. and Webber, B. R. and Kuhn, R.",
      title          = "{QCD Matrix Elements + Parton Showers}",
      journal        = "JHEP",
      volume         = "11",
      year           = "2001",
      pages          = "063",
      doi            = "10.1088/1126-6708/2001/11/063",
      eprint         = "hep-ph/0109231",
      archivePrefix  = "arXiv",
      reportNumber   = "CERN-TH-2000-367, CAVENDISH-HEP-00-03",
      SLACcitation   = "%%CITATION = HEP-PH/0109231;%%"
}

@article{Davidson:2010ew,
      author         = "Davidson, N. and Przedzinski, T. and Was, Z.",
      title          = "{PHOTOS Interface in C++: Technical and physics documentation}",
      journal        = "Comput. Phys. Commun.",
      volume         = "199",
      year           = "2016",
      pages          = "86-101",
      doi            = "10.1016/j.cpc.2015.09.013",
      eprint         = "1011.0937",
      archivePrefix  = "arXiv",
      primaryClass   = "hep-ph",
      reportNumber   = "CERN-PH-TH-2010-261, IFJPAN-IV-2010-6",
      SLACcitation   = "%%CITATION = ARXIV:1011.0937;%%",
}

@article{Cascioli:2011va,
      author         = "Cascioli, Fabio and Maierh{\"o}fer, Philipp and Pozzorini, Stefano",
      title          = "{Scattering Amplitudes with Open Loops}",
      journal        = "Phys. Rev. Lett.",
      volume         = "108",
      year           = "2012",
      pages          = "111601",
      doi            = "10.1103/PhysRevLett.108.111601",
      eprint         = "1111.5206",
      archivePrefix  = "arXiv",
      primaryClass   = "hep-ph",
      reportNumber   = "ZU-TH-23-11, LPN11-66",
      SLACcitation   = "%%CITATION = ARXIV:1111.5206;%%"
}

@article{Buccioni:2019sur,
    author = "Buccioni, Federico and Lang, Jean-Nicolas and Lindert, Jonas M. and Maierh{\"o}fer, Philipp and Pozzorini, Stefano and Zhang, Hantian and Zoller, Max F.",
    title = "{OpenLoops 2}",
    eprint = "1907.13071",
    archivePrefix = "arXiv",
    primaryClass = "hep-ph",
    reportNumber = "IPPP/19/62, FR-PHENO-2019-12, PSI-PR-19-15, ZU-TH 37/19",
    doi = "10.1140/epjc/s10052-019-7306-2",
    journal = "Eur. Phys. J. C",
    volume = "79",
    number = "10",
    pages = "866",
    year = "2019"
}

@article{Denner:2016kdg,
      author         = "Denner, Ansgar and Dittmaier, Stefan and Hofer, Lars",
      title          = "{\textsc{Collier}: A fortran-based complex one-loop library in
                        extended regularizations}",
      journal        = "Comput. Phys. Commun.",
      volume         = "212",
      year           = "2017",
      pages          = "220-238",
      doi            = "10.1016/j.cpc.2016.10.013",
      eprint         = "1604.06792",
      archivePrefix  = "arXiv",
      primaryClass   = "hep-ph",
      reportNumber   = "FR-PHENO-2016-003, ICCUB-16-016",
      SLACcitation   = "%%CITATION = ARXIV:1604.06792;%%"
}

@article{Czakon:2011xx,
  author         = "Czakon, Michal and Mitov, Alexander",
  title          = "{Top++: A program for the calculation of the top-pair cross-section at hadron colliders}",
  journal        = "Comput. Phys. Commun.",
  volume         = "185",
  year           = "2014",
  pages          = "2930",
  doi            = "10.1016/j.cpc.2014.06.021",
  eprint         = "1112.5675",
  archivePrefix  = "arXiv",
  primaryClass   = "hep-ph",
  reportNumber   = "CERN-PH-TH-2011-303, TTK-11-58",
}

@article{Agostinelli:2002hh,
      author         = "Agostinelli, S. and others",
      title          = "{\textsc{Geant4} -- a simulation toolkit}",
      journal        = "Nucl. Instrum. Meth. A",
      volume         = "506",
      year           = "2003",
      pages          = "250",
      doi            = "10.1016/S0168-9002(03)01368-8",
      reportNumber   = "SLAC-PUB-9350, FERMILAB-PUB-03-339",
      SLACcitation   = "%%CITATION = NUIMA,A506,250;%%"
}

@Article{Read:2002hq,
    Author = {Read, Alexander L.},
    Journal = {J. Phys. G},
    Pages = {2693},
    doi = "10.1088/0954-3899/28/10/313",
    Title = "{Presentation of search results: the \(CL_S\) technique}",
    Volume = {28},
    Year = {2002}
}

@Article{Cowan:2010js,
   Author = {Cowan, Glen and Cranmer, Kyle and Gross, Eilam and Vitells, Ofer},
   Title = {{Asymptotic formulae for likelihood-based tests of new physics}},
   Journal = {Eur. Phys. J. C},
   Volume = {71},
   Year = {2011},
   Pages = {1554},
   doi = "10.1140/epjc/s10052-011-1554-0",
   Eprint = "1007.1727",
   Archiveprefix = {arXiv},
   Primaryclass = {physics.data-an},
   related = "Cowan:2010js-err",
   relatedstring  = "Erratum:",
}

@Article{PERF-2007-01,
    author         = "{ATLAS Collaboration}",
    title          = "{The ATLAS Experiment at the CERN Large Hadron Collider}",
    journal        = "JINST",
    volume         = "3",
    year           = "2008",
    pages          = "S08003",
    doi            = "10.1088/1748-0221/3/08/S08003",
    primaryClass   = "hep-ex",
}

@Article{SOFT-2010-01,
    author         = "{ATLAS Collaboration}",
    title          = "{The ATLAS Simulation Infrastructure}",
    journal        = "Eur. Phys. J. C",
    volume         = "70",
    year           = "2010",
    pages          = "823",
    doi            = "10.1140/epjc/s10052-010-1429-9",
    eprint         = "1005.4568",
    archivePrefix  = "arXiv",
    primaryClass   = "physics.ins-det",
}

@Article{STDM-2012-23,
    author         = "{ATLAS Collaboration}",
    title          = "{Measurement of the \(Z/\gamma^*\) boson transverse momentum distribution in \(pp\) collisions at \(\sqrt{s} = 7\,\text{TeV}\) with the ATLAS detector}",
    journal        = "JHEP",
    volume         = "09",
    year           = "2014",
    pages          = "145",
    doi            = "10.1007/JHEP09(2014)145",
    reportNumber   = "CERN-PH-EP-2014-075",
    eprint         = "1406.3660",
    archivePrefix  = "arXiv",
    primaryClass   = "hep-ex",
}

@Article{PERF-2014-03,
    author         = "{ATLAS Collaboration}",
    title          = "{Performance of pile-up mitigation techniques for jets in \(pp\) collisions at \(\sqrt{s} = 8\,\text{TeV}\) using the ATLAS detector}",
    journal        = "Eur. Phys. J. C",
    volume         = "76",
    year           = "2016",
    pages          = "581",
    doi            = "10.1140/epjc/s10052-016-4395-z",
    reportNumber   = "CERN-PH-EP-2015-206",
    eprint         = "1510.03823",
    archivePrefix  = "arXiv",
    primaryClass   = "hep-ex",
}

@Article{PERF-2014-07,
    author         = "{ATLAS Collaboration}",
    title          = "{Topological cell clustering in the ATLAS calorimeters and its performance in LHC Run~1}",
    journal        = "Eur. Phys. J. C",
    volume         = "77",
    year           = "2017",
    pages          = "490",
    doi            = "10.1140/epjc/s10052-017-5004-5",
    reportNumber   = "CERN-PH-EP-2015-304",
    eprint         = "1603.02934",
    archivePrefix  = "arXiv",
    primaryClass   = "hep-ex",
}

@Article{PERF-2016-07,
    author         = "{ATLAS Collaboration}",
    title          = "{Performance of missing transverse momentum reconstruction with the ATLAS detector using proton--proton collisions at \(\sqrt{s} = 13\,\text{TeV}\)}",
    journal        = "Eur. Phys. J. C",
    volume         = "78",
    year           = "2018",
    pages          = "903",
    doi            = "10.1140/epjc/s10052-018-6288-9",
    reportNumber   = "CERN-EP-2017-274",
    eprint         = "1802.08168",
    archivePrefix  = "arXiv",
    primaryClass   = "hep-ex",
}

@Article{TRIG-2016-01,
    author         = "{ATLAS Collaboration}",
    title          = "{Performance of the ATLAS trigger system in 2015}",
    journal        = "Eur. Phys. J. C",
    volume         = "77",
    year           = "2017",
    pages          = "317",
    doi            = "10.1140/epjc/s10052-017-4852-3",
    reportNumber   = "CERN-EP-2016-241",
    eprint         = "1611.09661",
    archivePrefix  = "arXiv",
    primaryClass   = "hep-ex",
}

@Article{EXOT-2017-30,
    author         = "{ATLAS Collaboration}",
    title          = "{Searches for third-generation scalar leptoquarks in \(\sqrt{s} = 13\,\text{TeV}\) \(pp\) collisions with the ATLAS detector}",
    journal        = "JHEP",
    volume         = "06",
    year           = "2019",
    pages          = "144",
    doi            = "10.1007/JHEP06(2019)144",
    reportNumber   = "CERN-EP-2019-026",
    eprint         = "1902.08103",
    archivePrefix  = "arXiv",
    primaryClass   = "hep-ex",
}

@Article{DAPR-2018-01,
    author         = "{ATLAS Collaboration}",
    title          = "{ATLAS data quality operations and performance for 2015--2018 data-taking}",
    journal        = "JINST",
    volume         = "15",
    year           = "2020",
    pages          = "P04003",
    doi            = "10.1088/1748-0221/15/04/P04003",
    reportNumber   = "CERN-EP-2019-207",
    eprint         = "1911.04632",
    archivePrefix  = "arXiv",
    primaryClass   = "physics.ins-det",
}

@Article{EGAM-2018-01,
    author         = "{ATLAS Collaboration}",
    title          = "{Electron and photon performance measurements with the ATLAS detector using the 2015--2017 LHC proton--proton collision data}",
    journal        = "JINST",
    volume         = "14",
    year           = "2019",
    pages          = "P12006",
    doi            = "10.1088/1748-0221/14/12/P12006",
    reportNumber   = "CERN-EP-2019-145",
    eprint         = "1908.00005",
    archivePrefix  = "arXiv",
    primaryClass   = "hep-ex",
}

@Article{MUON-2018-03,
    author         = "{ATLAS Collaboration}",
    title          = "{Muon reconstruction and identification efficiency in ATLAS using the full Run~2 \(pp\) collision data set at \(\sqrt{s} = 13\,\text{TeV}\)}",
    journal        = "Eur. Phys. J. C",
    volume         = "81",
    year           = "2021",
    pages          = "578",
    doi            = "10.1140/epjc/s10052-021-09233-2",
    reportNumber   = "CERN-EP-2020-199",
    eprint         = "2012.00578",
    archivePrefix  = "arXiv",
    primaryClass   = "hep-ex",
}

@Article{PIX-2018-001,
    author         = "Abbott, B. and others",
    title          = "{Production and integration of the ATLAS Insertable B-Layer}",
    journal        = "JINST",
    volume         = "13",
    year           = "2018",
    pages          = "T05008",
    doi            = "10.1088/1748-0221/13/05/T05008",
    eprint         = "1803.00844",
    archivePrefix  = "arXiv",
    primaryClass   = "physics.ins-det",
}

@Article{SUSY-2018-12,
    author         = "{ATLAS Collaboration}",
    title          = "{Search for a scalar partner of the top quark in the all-hadronic \(t\bar{t}\) plus missing transverse momentum final state at \(\sqrt{s} = 13\,\text{TeV}\) with the ATLAS detector}",
    journal        = "Eur. Phys. J. C",
    volume         = "80",
    year           = "2020",
    pages          = "737",
    doi            = "10.1140/epjc/s10052-020-8102-8",
    reportNumber   = "CERN-EP-2020-044",
    eprint         = "2004.14060",
    archivePrefix  = "arXiv",
    primaryClass   = "hep-ex",
}

@Article{SUSY-2018-34,
    author         = "{ATLAS Collaboration}",
    title          = "{Search for new phenomena in final states with \(b\)-jets and missing transverse momentum in \(\sqrt{s} = 13\,\text{TeV}\) \(pp\) collisions with the ATLAS detector}",
    journal        = "JHEP",
    volume         = "05",
    year           = "2021",
    pages          = "093",
    doi            = "10.1007/JHEP05(2021)093",
    reportNumber   = "CERN-EP-2021-001",
    eprint         = "2101.12527",
    archivePrefix  = "arXiv",
    primaryClass   = "hep-ex",
}

@Article{TOPQ-2018-15,
    author         = "{ATLAS Collaboration}",
    title          = "{Measurements of top-quark pair differential and double-differential cross-sections in the \(\ell\)+jets channel with \(pp\) collisions at \(\sqrt{s} = 13\,\text{TeV}\) using the ATLAS detector}",
    journal        = "Eur. Phys. J. C",
    volume         = "79",
    year           = "2019",
    pages          = "1028",
    doi            = "10.1140/epjc/s10052-019-7525-6",
    reportNumber   = "CERN-EP-2019-149",
    eprint         = "1908.07305",
    archivePrefix  = "arXiv",
    primaryClass   = "hep-ex",
    related        = "TOPQ-2018-15-err",
    relatedstring  = "Erratum:",
}

@Article{TOPQ-2018-17,
    author         = "{ATLAS Collaboration}",
    title          = "{Measurement of the \(t\bar{t}\) production cross-section and lepton differential distributions in \(e\mu\) dilepton events from \(pp\) collisions at \(\sqrt{s} = 13\,\text{TeV}\) with the ATLAS detector}",
    journal        = "Eur. Phys. J. C",
    volume         = "80",
    year           = "2020",
    pages          = "528",
    doi            = "10.1140/epjc/s10052-020-7907-9",
    reportNumber   = "CERN-EP-2019-203",
    eprint         = "1910.08819",
    archivePrefix  = "arXiv",
    primaryClass   = "hep-ex",
}

@Article{TRIG-2018-01,
    author         = "{ATLAS Collaboration}",
    title          = "{Performance of the ATLAS muon triggers in Run~2}",
    journal        = "JINST",
    volume         = "15",
    year           = "2020",
    pages          = "P09015",
    doi            = "10.1088/1748-0221/15/09/p09015",
    reportNumber   = "CERN-EP-2020-031",
    eprint         = "2004.13447",
    archivePrefix  = "arXiv",
    primaryClass   = "hep-ex",
}

@Article{TRIG-2018-05,
    author         = "{ATLAS Collaboration}",
    title          = "{Performance of electron and photon triggers in ATLAS during LHC Run~2}",
    journal        = "Eur. Phys. J. C",
    volume         = "80",
    year           = "2020",
    pages          = "47",
    doi            = "10.1140/epjc/s10052-019-7500-2",
    reportNumber   = "CERN-EP-2019-169",
    eprint         = "1909.00761",
    archivePrefix  = "arXiv",
    primaryClass   = "hep-ex",
}

@Article{EXOT-2019-13,
    author         = "{ATLAS Collaboration}",
    title          = "{Search for pairs of scalar leptoquarks decaying into quarks and electrons or muons in \(\sqrt{s} = 13\,\text{TeV}\) \(pp\) collisions with the ATLAS detector}",
    journal        = "JHEP",
    volume         = "10",
    year           = "2020",
    pages          = "112",
    doi            = "10.1007/JHEP10(2020)112",
    reportNumber   = "CERN-EP-2020-084",
    eprint         = "2006.05872",
    archivePrefix  = "arXiv",
    primaryClass   = "hep-ex",
}

@Article{EXOT-2019-15,
    author         = "{ATLAS Collaboration}",
    title          = "{Search for pair production of third-generation scalar leptoquarks decaying into a top quark and a \(\tau\)-lepton in \(pp\) collisions at \(\sqrt{s} = 13\,\text{TeV}\) with the ATLAS detector}",
    journal        = "JHEP",
    volume         = "06",
    year           = "2021",
    pages          = "179",
    doi            = "10.1007/JHEP06(2021)179",
    reportNumber   = "CERN-EP-2020-241",
    eprint         = "2101.11582",
    archivePrefix  = "arXiv",
    primaryClass   = "hep-ex",
}

@Article{EXOT-2019-19,
    author         = "{ATLAS Collaboration}",
    title          = "{Search for pair production of scalar leptoquarks decaying into first- or second-generation leptons and top quarks in proton--proton collisions at \(\sqrt{s} = 13\,\text{TeV}\) with the ATLAS detector}",
    journal        = "Eur. Phys. J. C",
    volume         = "81",
    year           = "2021",
    pages          = "313",
    doi            = "10.1140/epjc/s10052-021-09009-8",
    reportNumber   = "CERN-EP-2020-160",
    eprint         = "2010.02098",
    archivePrefix  = "arXiv",
    primaryClass   = "hep-ex",
}

@Article{FTAG-2019-07,
    author         = "{ATLAS Collaboration}",
    title          = "{ATLAS flavour-tagging algorithms for the LHC Run~2 \(pp\) collision dataset}",
    year           = "2022",
    reportNumber   = "CERN-EP-2022-226",
    eprint         = "2211.16345",
    archivePrefix  = "arXiv",
    primaryClass   = "physics.data-an",
}

@Article{SUSY-2019-18,
    author         = "{ATLAS Collaboration}",
    title          = "{Search for new phenomena in \(pp\) collisions in final states with tau leptons, \(b\)-jets, and missing transverse momentum with the ATLAS detector}",
    journal        = "Phys. Rev. D",
    volume         = "104",
    year           = "2021",
    pages          = "112005",
    doi            = "10.1103/PhysRevD.104.112005",
    reportNumber   = "CERN-EP-2021-118",
    eprint         = "2108.07665",
    archivePrefix  = "arXiv",
    primaryClass   = "hep-ex",
}

@Article{EXOT-2022-39,
    author         = "{ATLAS Collaboration}",
    title          = "{Search for leptoquarks decaying into the b\(\tau\) final state in \(pp\) collisions at \(\sqrt{s} = 13\,\text{TeV}\) with the ATLAS detector}",
    year           = "2023",
    reportNumber   = "CERN-EP-2023-033",
    eprint         = "2305.15962",
    archivePrefix  = "arXiv",
    primaryClass   = "hep-ex",
}

@Booklet{ATL-SOFT-PUB-2021-001,
    author         = "{ATLAS Collaboration}",
    title          = "{The ATLAS Collaboration Software and Firmware}",
    howpublished   = "{ATL-SOFT-PUB-2021-001}",
    url            = "https://cds.cern.ch/record/2767187",
    year           = "2021",
}

@Booklet{ATL-SOFT-PUB-2021-003,
    author         = "{ATLAS Collaboration}",
    title          = "{ATLAS Computing Acknowledgements}",
    howpublished   = "{ATL-SOFT-PUB-2021-003}",
    url            = "https://cds.cern.ch/record/2776662",
    year           = "2021",
}

@Report{ATLAS-TDR-19,
    author         = "{ATLAS Collaboration}",
    title          = "{ATLAS Insertable B-Layer: Technical Design Report}",
    type           = "ATLAS-TDR-19; CERN-LHCC-2010-013",
    year           = "2010",
    url            = "https://cds.cern.ch/record/1291633",
    related        = "ATLAS-TDR-19-addm",
    relatedstring  = "Addendum:",
}

@Article{CMS-B2G-16-027,
    author         = "{CMS Collaboration}",
    title          = "{Search for Leptoquarks Coupled to Third-Generation Quarks in Proton--Proton Collisions at \(\sqrt{s} = 13\,\text{TeV}\)}",
    journal        = "Phys. Rev. Lett.",
    volume         = "121",
    year           = "2018",
    pages          = "241802",
    doi            = "10.1103/PhysRevLett.121.241802",
    reportNumber   = "CERN-EP-2018-233",
    eprint         = "1809.05558",
    archivePrefix  = "arXiv",
    primaryClass   = "hep-ex",
}

@Article{CMS-B2G-16-028,
    author         = "{CMS Collaboration}",
    title          = "{Search for third-generation scalar leptoquarks decaying to a top quark and a \(\tau\) lepton at \(\sqrt{s} = 13\,\text{TeV}\)}",
    journal        = "Eur. Phys. J. C",
    volume         = "78",
    year           = "2018",
    pages          = "707",
    doi            = "10.1140/epjc/s10052-018-6143-z",
    reportNumber   = "CERN-EP-2018-019",
    eprint         = "1803.02864",
    archivePrefix  = "arXiv",
    primaryClass   = "hep-ex",
}

@Article{CMS-EXO-17-015,
    author         = "{CMS Collaboration}",
    title          = "{Search for dark matter in events with a leptoquark and missing transverse momentum in proton--proton collisions at \(13\,\text{TeV}\)}",
    journal        = "Phys. Lett. B",
    volume         = "795",
    year           = "2019",
    pages          = "76",
    doi            = "10.1016/j.physletb.2019.05.046",
    reportNumber   = "CERN-EP-2018-278",
    eprint         = "1811.10151",
    archivePrefix  = "arXiv",
    primaryClass   = "hep-ex",
}

@Article{CMS-EXO-17-016,
    author         = "{CMS Collaboration}",
    title          = "{Search for heavy neutrinos and third-generation leptoquarks in hadronic states of two \(\tau\) leptons and two jets in proton--proton collisions at \(\sqrt{s} = 13\,\text{TeV}\)}",
    journal        = "JHEP",
    volume         = "03",
    year           = "2019",
    pages          = "170",
    doi            = "10.1007/JHEP03(2019)170",
    reportNumber   = "CERN-EP-2018-272",
    eprint         = "1811.00806",
    archivePrefix  = "arXiv",
    primaryClass   = "hep-ex",
}

@Booklet{ATLAS-CONF-2017-029,
    author         = "{ATLAS Collaboration}",
    title          = "{Measurement of the tau lepton reconstruction and identification performance in the ATLAS experiment using \(pp\) collisions at \(\sqrt{s} = 13~\text{TeV}\)}",
    howpublished   = "{ATLAS-CONF-2017-029}",
    url            = "https://cds.cern.ch/record/2261772",
    year           = "2017",
}

@Booklet{ATLAS-CONF-2017-061,
    author         = "{ATLAS Collaboration}",
    title          = "{The ATLAS Tau Trigger in Run~2}",
    howpublished   = "{ATLAS-CONF-2017-061}",
    url            = "https://cds.cern.ch/record/2274201",
    year           = "2017",
}

@Booklet{ATLAS-CONF-2019-021,
    author         = "{ATLAS Collaboration}",
    title          = "{Luminosity determination in \(pp\) collisions at \(\sqrt{s} = 13\,\text{TeV}\) using the ATLAS detector at the LHC}",
    howpublished   = "{ATLAS-CONF-2019-021}",
    url            = "https://cds.cern.ch/record/2677054",
    year           = "2019",
}

@Booklet{ATL-PHYS-PUB-2014-021,
    author         = "{ATLAS Collaboration}",
    title          = "{ATLAS Pythia~8 tunes to \(7~\text{TeV}\) data}",
    howpublished   = "{ATL-PHYS-PUB-2014-021}",
    url            = "https://cds.cern.ch/record/1966419",
    year           = "2014",
}

@Booklet{ATL-PHYS-PUB-2015-026,
    author         = "{ATLAS Collaboration}",
    title          = "{Vertex Reconstruction Performance of the ATLAS Detector at \(\sqrt{s} = 13~\text{TeV}\)}",
    howpublished   = "{ATL-PHYS-PUB-2015-026}",
    url            = "https://cds.cern.ch/record/2037717",
    year           = "2015",
}

@Booklet{ATL-PHYS-PUB-2016-017,
    author         = "{ATLAS Collaboration}",
    title          = "{The Pythia~8 A3 tune description of ATLAS minimum bias and inelastic measurements incorporating the Donnachie--Landshoff diffractive model}",
    howpublished   = "{ATL-PHYS-PUB-2016-017}",
    url            = "https://cds.cern.ch/record/2206965",
    year           = "2016",
}

@Booklet{ATL-PHYS-PUB-2016-020,
    author         = "{ATLAS Collaboration}",
    title          = "{Studies on top-quark Monte Carlo modelling for Top2016}",
    howpublished   = "{ATL-PHYS-PUB-2016-020}",
    url            = "https://cds.cern.ch/record/2216168",
    year           = "2016",
}

@Booklet{ATL-PHYS-PUB-2017-006,
    author         = "{ATLAS Collaboration}",
    title          = "{ATLAS simulation of boson plus jets processes in Run~2}",
    howpublished   = "{ATL-PHYS-PUB-2017-006}",
    url            = "https://cds.cern.ch/record/2261937",
    year           = "2017",
}

@Booklet{ATL-PHYS-PUB-2017-013,
    author         = "{ATLAS Collaboration}",
    title          = "{Optimisation and performance studies of the ATLAS \(b\)-tagging algorithms for the 2017-18 LHC run}",
    howpublished   = "{ATL-PHYS-PUB-2017-013}",
    url            = "https://cds.cern.ch/record/2273281",
    year           = "2017",
}

@Booklet{ATL-PHYS-PUB-2019-033,
    author         = "{ATLAS Collaboration}",
    title          = "{Identification of hadronic tau lepton decays using neural networks in the ATLAS experiment}",
    howpublished   = "{ATL-PHYS-PUB-2019-033}",
    url            = "https://cds.cern.ch/record/2688062",
    year           = "2019",
}

@Article{Dimopoulos1979,
  author = "S. Dimopoulos and L. Susskind",
  title = "{Mass without scalars}",
  journal = "Nucl. Phys. B",
  volume = "155",
  number = {1},
  pages = {237-252},
  year = {1979},
  doi = "10.1016/0550-3213(79)90364-X"
}

@article{Dimopoulos:1979sp,
    author = "Dimopoulos, Savas",
    editor = "Zichichi, A.",
    title = "{Technicolored signatures}",
    reportNumber = "PRINT-79-0962 (STANFORD)",
    doi = "10.1016/0550-3213(80)90277-1",
    journal = "Nucl. Phys. B",
    volume = "168",
    pages = "69--92",
    year = "1980"
}

@article{Farhi:1980xs,
    author = "Farhi, Edward and Susskind, Leonard",
    title = "{Technicolour}",
    reportNumber = "CERN-TH-2975",
    doi = "10.1016/0370-1573(81)90173-3",
    journal = "Phys. Rept.",
    volume = "74",
    pages = "277",
    year = "1981"
}

@article{Schrempp:1984nj,
    author = "Schrempp, Barbara and Schrempp, Fridger",
    title = "{Light leptoquarks}",
    reportNumber = "DESY-84-117",
    doi = "10.1016/0370-2693(85)91450-9",
    journal = "Phys. Lett. B",
    volume = "153",
    pages = "101--107",
    year = "1985"
}

@article{Pati:1973uk,
    author = "Pati, Jogesh C. and Salam, Abdus",
    title = "{Unified Lepton-Hadron Symmetry and a Gauge Theory of the Basic Interactions}",
    reportNumber = "IC-73-41-INT-REP",
    doi = "10.1103/PhysRevD.8.1240",
    journal = "Phys. Rev. D",
    volume = "8",
    pages = "1240--1251",
    year = "1973"
}

@Article{PatiSalam1974,
  author = "J. C. Pati and A Salam",
  title = "{Lepton number as the fourth "color"}",
  journal = "Phys. Rev. D",
  volume = "10",
  year = "1974",
  pages = "275",
  doi = "10.1103/PhysRevD.10.275"
}

@Article{GeorgiGlashow1974,
  author = "H. Georgi and S. L. Glashow",
  title = "{Unity of All Elementary-Particle Forces}",
  journal = "Phys. Rev. Lett.",
  volume = "32",
  year = "1974",
  pages = "438",
  doi = "10.1103/PhysRevLett.32.438"
}

@article{BaBar:1,
    author = "{BaBar Collaboration}",
    collaboration = "BaBar",
    title = "{Evidence for an Excess of $\bar{B} \to D^{(*)} \tau^-\bar{\nu}_\tau$ Decays}",
    eprint = "1205.5442",
    archivePrefix = "arXiv",
    primaryClass = "hep-ex",
    reportNumber = "BABAR-PUB-12-012, SLAC-PUB-15028",
    doi = "10.1103/PhysRevLett.109.101802",
    journal = "Phys. Rev. Lett.",
    volume = "109",
    pages = "101802",
    year = "2012"
}

@article{BaBar:2,
    author = "{BaBar Collaboration}",
    collaboration = "BaBar",
    title = "{Measurement of an excess of $\bar{B} \to D^{(*)}\tau^- \bar{\nu}_\tau$ decays and implications for charged Higgs bosons}",
    eprint = "1303.0571",
    archivePrefix = "arXiv",
    primaryClass = "hep-ex",
    reportNumber = "BABAR-PUB-13-001, SLAC-PUB-15381",
    doi = "10.1103/PhysRevD.88.072012",
    journal = "Phys. Rev. D",
    volume = "88",
    number = "7",
    pages = "072012",
    year = "2013"
}

@article{Belle:1,
    author = "{Belle Collaboration}",
    collaboration = "Belle",
    title = "{Measurement of the branching ratio of $\bar{B} \to D^{(\ast)} \tau^- \bar{\nu}_\tau$ relative to $\bar{B} \to D^{(\ast)} \ell^- \bar{\nu}_\ell$ decays with hadronic tagging at Belle}",
    eprint = "1507.03233",
    archivePrefix = "arXiv",
    primaryClass = "hep-ex",
    reportNumber = "KEK-REPORT-2015-18",
    doi = "10.1103/PhysRevD.92.072014",
    journal = "Phys. Rev. D",
    volume = "92",
    number = "7",
    pages = "072014",
    year = "2015"
}

@article{LHCb:2,
    author = "{LHCb Collaboration}",
    collaboration = "LHCb",
    title = "{Measurement of the Ratio of Branching Fractions $\mathcal{B}(\bar{B}^0 \to D^{*+}\tau^{-}\bar{\nu}_{\tau})/\mathcal{B}(\bar{B}^0 \to D^{*+}\mu^{-}\bar{\nu}_{\mu})$}",
    eprint = "1506.08614",
    archivePrefix = "arXiv",
    primaryClass = "hep-ex",
    reportNumber = "CERN-PH-EP-2015-150, LHCB-PAPER-2015-025",
    doi = "10.1103/PhysRevLett.115.111803",
    journal = "Phys. Rev. Lett.",
    volume = "115",
    number = "11",
    pages = "111803",
    year = "2015",
    related = "LHCb:2-err",
    relatedstring  = "Erratum:",
}

@Article{Muong22021,
  author = "{Muon g-2 Collaboration}",
  title = "{Measurement of the Positive Muon Anomalous Magnetic Moment to 0.46~ppm}",
  journal = "Phys. Rev. Lett.",
  volume = "126",
  number = "141801",
  year = "2021",
  doi = "10.1103/PhysRevLett.126.141801"
}

@article{Greljo:2021xmg,
    author = "Greljo, Admir and Stangl, Peter and Thomsen, Anders Eller",
    title = "{A model of muon anomalies}",
    eprint = "2103.13991",
    archivePrefix = "arXiv",
    primaryClass = "hep-ph",
    doi = "10.1016/j.physletb.2021.136554",
    journal = "Phys. Lett. B",
    volume = "820",
    pages = "136554",
    year = "2021"
}

@Article{DorsnerLQ2016,
  author = "I. Dor\v{s}ner and S. Fajfer and A. Greljo and J. F. Kamenik and N. Ko\v{s}nik",
  title = "Physics of leptoquarks in precision experiments and at particle colliders",
  journal = "Phys. Rep.",
  volume = "641",
  pages = "1-68",
  year = "2016",
  doi = "10.1016/j.physrep.2016.06.001"
}

@article{U1Model,
    author = {Baker, Michael J. and Fuentes-Martin, Javier and Isidori, Gino and Konig, Matthias},
    title = "{High-$p_{T}$ signatures in vector--leptoquark models}",
    eprint = "1901.10480",
    archivePrefix = "arXiv",
    primaryClass = "hep-ph",
    reportNumber = "ZU-TH-19/05",
    doi = "10.1140/epjc/s10052-019-6853-x",
    journal = "Eur. Phys. J. C",
    volume = "79",
    number = "4",
    pages = "334",
    year = "2019"
}

@article{Avoni:2018iuv,
  author         = "{G. Avoni et al.}",
      title          = "{The new LUCID-2 detector for luminosity measurement and
                        monitoring in ATLAS}",
      journal        = "JINST",
      volume         = "13",
      year           = "2018",
      number         = "07",
      pages          = "P07017",
      doi            = "10.1088/1748-0221/13/07/P07017",
      SLACcitation   = "%%CITATION = JINST,13,P07017;%%"
}

@article{LQToolbox,
    author = "I. Dor\v{s}ner and A. Greljo",
    title = "{Leptoquark toolbox for precision collider studies}",
    eprint = "1801.07641",
    archivePrefix = "arXiv",
    primaryClass = "hep-ph",
    doi = "10.1007/JHEP05(2018)126",
    journal = "{JHEP}",
    volume = "05",
    pages = "126",
    year = "2018"
}

@article{ATLAS:2017ghe,
    author = "{ATLAS Collaboration}",
    collaboration = "ATLAS",
    title = "{Jet reconstruction and performance using particle flow with the ATLAS Detector}",
    eprint = "1703.10485",
    archivePrefix = "arXiv",
    primaryClass = "hep-ex",
    reportNumber = "CERN-EP-2017-024",
    doi = "10.1140/epjc/s10052-017-5031-2",
    journal = "Eur. Phys. J. C",
    volume = "77",
    number = "7",
    pages = "466",
    year = "2017"
}

@article{Mandal_2016,
    author = "Mandal, Tanumoy and Mitra, Subhadip and Seth, Satyajit",
    title = "{Pair production of scalar leptoquarks at the LHC to NLO parton shower accuracy}",
    eprint = "1506.07369",
    archivePrefix = "arXiv",
    primaryClass = "hep-ph",
    reportNumber = "HRI-RECAPP-2015-013, MITP-15-043",
    doi = "10.1103/PhysRevD.93.035018",
    journal = "Phys. Rev. D",
    volume = "93",
    number = "3",
    pages = "035018",
    year = "2016"
}

@article{Kr_mer_2005,
    author = {Kr\"amer, M. and Plehn, T. and Spira, M. and Zerwas, P. M.},
    title = "{Pair production of scalar leptoquarks at the CERN LHC}",
    eprint = "hep-ph/0411038",
    archivePrefix = "arXiv",
    reportNumber = "CERN-PH-TH-2004-207, DESY-04-200, EDINBURGH-2004-28, FERMILAB-PUB-04-298-T, PSI-PR-04-12",
    doi = "10.1103/PhysRevD.71.057503",
    journal = "Phys. Rev. D",
    volume = "71",
    pages = "057503",
    year = "2005"
}

@article{Kr_mer_1997,
    author = {Kr\"amer, M. and Plehn, T. and Spira, M. and Zerwas, P. M.},
    title = "{Pair Production of Scalar Leptoquarks at the Fermilab Tevatron}",
    eprint = "hep-ph/9704322",
    archivePrefix = "arXiv",
    reportNumber = "RAL-97-017, RAL-97-17, DESY-97-063, CERN-TH-97-067, CERN-TH-97-67",
    doi = "10.1103/PhysRevLett.79.341",
    journal = "Phys. Rev. Lett.",
    volume = "79",
    pages = "341--344",
    year = "1997"
}

@article{Beenakker_2016,
    author = {Beenakker, Wim and Borschensky, Christoph and Kr\"amer, Michael and Kulesza, Anna and Laenen, Eric},
    title = "{NNLL-fast: predictions for coloured supersymmetric particle production at the LHC with threshold and Coulomb resummation}",
    eprint = "1607.07741",
    archivePrefix = "arXiv",
    primaryClass = "hep-ph",
    reportNumber = "MS-TP-16-17, NIKHEF-2016-037, TTK-16-28",
    doi = "10.1007/JHEP12(2016)133",
    journal = "JHEP",
    volume = "12",
    pages = "133",
    year = "2016"
}

@article{Beenakker_1998,
    author = {Beenakker, W. and Kr\"amer, M. and Plehn, T. and Spira, M. and Zerwas, P. M.},
    title = "{Stop production at hadron colliders}",
    eprint = "hep-ph/9710451",
    archivePrefix = "arXiv",
    reportNumber = "DESY-97-214, CERN-TH-97-177, RAL-TR-97-056",
    doi = "10.1016/S0550-3213(98)00014-5",
    journal = "Nucl. Phys. B",
    volume = "515",
    pages = "3--14",
    year = "1998"
}

@article{Beenakker_2010,
    author = {Beenakker, Wim and Brensing, Silja and Kr\"amer, Michael and Kulesza, Anna and Laenen, Eric and Niessen, Irene},
    title = "{Supersymmetric top and bottom squark production at hadron colliders}",
    eprint = "1006.4771",
    archivePrefix = "arXiv",
    primaryClass = "hep-ph",
    reportNumber = "CERN-PH-TH-2010-142, ITFA-2010-015, ITP-UU-10-17, NIKHEF-2010-016, TTK-10-33",
    doi = "10.1007/JHEP08(2010)098",
    journal = "JHEP",
    volume = "08",
    pages = "098",
    year = "2010"
}

@article{Beenakker_2016_v1,
    author = {Beenakker, Wim and Borschensky, Christoph and Heger, Raphael and Kr\"amer, Michael and Kulesza, Anna and Laenen, Eric},
    title = "{NNLL resummation for stop pair-production at the LHC}",
    eprint = "1601.02954",
    archivePrefix = "arXiv",
    primaryClass = "hep-ph",
    reportNumber = "MS-TP-16-01, TTK-16-01, NIKHEF-2016-002",
    doi = "10.1007/JHEP05(2016)153",
    journal = "JHEP",
    volume = "05",
    pages = "153",
    year = "2016"
}

@article{Borschensky_2020,
    author = {Borschensky, Christoph and Fuks, Benjamin and Kulesza, Anna and Schwartl\"ander, Daniel},
    title = "{Scalar leptoquark pair production at hadron colliders}",
    eprint = "2002.08971",
    archivePrefix = "arXiv",
    primaryClass = "hep-ph",
    reportNumber = "MS-TP-20-09",
    doi = "10.1103/PhysRevD.101.115017",
    journal = "Phys. Rev. D",
    volume = "101",
    number = "11",
    pages = "115017",
    year = "2020"
}

@article{White_2009,
     doi = {10.1088/1126-6708/2009/11/074},
     year = {2009},
     month = {11},
     publisher = {Springer Science and Business Media {LLC}
},
     volume = {11},
     pages = {074--074},
     number = {11},
     author = {Chris D White and Stefano Frixione and Eric Laenen and Fabio Maltoni},
     title = {Isolating Wt production at the LHC},
     journal = {JHEP},
     archivePrefix = {arXiv},
     eprint = {0908.0631},
}

@article{BUCHMULLER1987442,
    author = "Buchmuller, W. and Ruckl, R. and Wyler, D.",
    title = "{Leptoquarks in lepton-quark collisions}",
    reportNumber = "DESY-86-150, ITP-UH-14-86",
    doi = "10.1016/0370-2693(87)90637-X",
    journal = "Phys. Lett. B",
    volume = "191",
    pages = "442--448",
    year = "1987",
    related = "BUCHMULLER1987442-err",
    relatedstring  = "Erratum:",
}

@article{PhysRevD.99.055028,
  title = {${R}_{{D}^{(*)}}$ motivated ${\mathcal{S}}_{1}$ leptoquark scenarios: Impact of interference on the exclusion limits from LHC data},
  author = {Mandal, Tanumoy and Mitra, Subhadip and Raz, Swapnil},
  journal = {Phys. Rev. D},
  volume = {99},
  issue = {5},
  pages = {055028},
  numpages = {16},
  year = {2019},
  month = {3},
  publisher = {American Physical Society},
  doi = {10.1103/PhysRevD.99.055028},
  archivePrefix = {arXiv},
  eprint = {1811.03561}
}

@article{Bessaa:2014jya,
    author = "Bessaa, Assia and Davidson, Sacha",
    title = "{Constraints on $t$ -channel leptoquark exchange from LHC contact interaction searches}",
    eprint = "1409.2372",
    archivePrefix = "arXiv",
    primaryClass = "hep-ph",
    doi = "10.1140/epjc/s10052-015-3313-0",
    journal = "Eur. Phys. J. C",
    volume = "75",
    number = "2",
    pages = "97",
    year = "2015"
}

@article{Aebischer:2022oqe,
    author = "Aebischer, Jason and Isidori, Gino and Pesut, Marko and Stefanek, Ben A. and Wilsch, Felix",
    title = "{Confronting the vector leptoquark hypothesis with new low- and high-energy data}",
    eprint = "2210.13422",
    archivePrefix = "arXiv",
    DOI= "10.1140/epjc/s10052-023-11304-5",
    journal = {Eur. Phys. J. C},
    year = 2023,
    volume = 83,
    number = 2,
    pages = "153",

}

@article{Buttazzo_2017,
	doi     = {10.1007/jhep11(2017)044}, 
	journal = "JHEP",
        year    = "2017",
	volume  = "11",
	pages   = "044",
	author  = {Dario Buttazzo and Admir Greljo and Gino Isidori and David Marzocca},
	title   = {B-physics anomalies: a guide to combined explanations},

}

@article{Borsanyi_2021,
    author = "Borsanyi, Sz. and others",
    title = "{Leading hadronic contribution to the muon magnetic moment from lattice QCD}",
    eprint = "2002.12347",
    archivePrefix = "arXiv",
    primaryClass = "hep-lat",
    doi = "10.1038/s41586-021-03418-1",
    journal = "Nature",
    volume = "593",
    number = "7857",
    pages = "51--55",
    year = "2021"
}

@article{Winter:2003tt_fixed,
      author         = "Winter, Jan-Christopher and Krauss, Frank and Soff,
                        Gerhard",
      title          = "{A modified cluster-hadronisation model}",
      journal        = "Eur. Phys. J. C",
      volume         = "36",
      year           = "2004",
      pages          = "381-395",
      doi            = "10.1140/epjc/s2004-01960-8",
      eprint         = "hep-ph/0311085",
      archivePrefix  = "arXiv",
      XprimaryClass   = "hep-ph",
      reportNumber   = "CERN-TH-2003-272",
      SLACcitation   = "%%CITATION = HEP-PH/0311085;%%"
}

@article{ATLAS:2023uox,
    author = "{ATLAS Collaboration}",
    title = "{Search for pair production of third-generation leptoquarks decaying into a bottom quark and a $\tau$-lepton with the ATLAS detector}",
    eprint = "2303.01294",
    archivePrefix = "arXiv",
    primaryClass = "hep-ex",
    reportNumber = "CERN-EP-2022-267",
    month = "3",
    year = "2023"
}

@article{Kant:2014ohav2,
      author         = "Kant, P. and Kind, O. M. and Kintscher, T. and Lohse, T.
                        and Martini, T. and Mölbitz, S. and Rieck, P. and Uwer, P.",
      title          = "{\textsc{HatHor} for single top-quark production: Updated
                        predictions and uncertainty estimates for single top-quark
                        production in hadronic collisions}",
      journal        = "Comput. Phys. Commun.",
      volume         = "191",
      year           = "2015",
      pages          = "74-89",
      doi            = "10.1016/j.cpc.2015.02.001",
      eprint         = "1406.4403",
      archivePrefix  = "arXiv",
      primaryClass   = "hep-ph",
      reportNumber   = "HU-EP-14-22",
      SLACcitation   = "%%CITATION = ARXIV:1406.4403;%%"
}

@Article{JETM-2018-05-corr,
    author         = "{ATLAS Collaboration}",
    title          = "{Jet energy scale and resolution measured in proton--proton collisions at \(\sqrt{s} = 13\,\text{TeV}\) with the ATLAS detector}",
    journal        = "Eur. Phys. J. C",
    volume         = "81",
    year           = "2021",
    pages          = "689",
    doi            = "10.1140/epjc/s10052-021-09402-3",
    reportNumber   = "CERN-EP-2020-083",
    eprint         = "2007.02645",
    archivePrefix  = "arXiv",
    primaryClass   = "hep-ex",
}

@Article{CMS-EXO-19-015-corr,
    author         = "{CMS Collaboration}",
    title          = "{Search for singly and pair-produced leptoquarks coupling to third-generation fermions in proton--proton collisions at \(\sqrt{s} = 13\,\text{TeV}\)}",
    journal        = "Phys. Lett. B",
    volume         = "819",
    year           = "2021",
    pages          = "136446",
    doi            = "10.1016/j.physletb.2021.136446",
    reportNumber   = "CERN-EP-2020-216",
    eprint         = "2012.04178",
    archivePrefix  = "arXiv",
    primaryClass   = "hep-ex",
}

\clearpage
 
 
\begin{flushleft}
\hypersetup{urlcolor=black}
{\Large The ATLAS Collaboration}

\bigskip

\AtlasOrcid[0000-0002-6665-4934]{G.~Aad}$^\textrm{\scriptsize 90b}$,
\AtlasOrcid[0000-0002-5888-2734]{B.~Abbott}$^\textrm{\scriptsize 120}$,
\AtlasOrcid[0000-0002-1002-1652]{K.~Abeling}$^\textrm{\scriptsize 55}$,
\AtlasOrcid[0000-0001-5763-2760]{N.J.~Abicht}$^\textrm{\scriptsize 49}$,
\AtlasOrcid[0000-0002-8496-9294]{S.H.~Abidi}$^\textrm{\scriptsize 29}$,
\AtlasOrcid[0000-0002-9987-2292]{A.~Aboulhorma}$^\textrm{\scriptsize 35e}$,
\AtlasOrcid[0000-0001-5329-6640]{H.~Abramowicz}$^\textrm{\scriptsize 151}$,
\AtlasOrcid[0000-0002-1599-2896]{H.~Abreu}$^\textrm{\scriptsize 150}$,
\AtlasOrcid[0000-0003-0403-3697]{Y.~Abulaiti}$^\textrm{\scriptsize 117}$,
\AtlasOrcid[0000-0003-0762-7204]{A.C.~Abusleme~Hoffman}$^\textrm{\scriptsize 137a}$
% ...
\bigskip
\\

% ...
$^{29}$Physics Department, Brookhaven National Laboratory, Upton NY; United States of America.\\
% ...
$^{35}$$^{(a)}$Facult\'e des Sciences Ain Chock, R\'eseau Universitaire de Physique des Hautes Energies - Universit\'e Hassan II, Casablanca;$^{(b)}$Facult\'{e} des Sciences, Universit\'{e} Ibn-Tofail, K\'{e}nitra;$^{(c)}$Facult\'e des Sciences Semlalia, Universit\'e Cadi Ayyad, LPHEA-Marrakech;$^{(d)}$LPMR, Facult\'e des Sciences, Universit\'e Mohamed Premier, Oujda;$^{(e)}$Facult\'e des sciences, Universit\'e Mohammed V, Rabat;$^{(f)}$Institute of Applied Physics, Mohammed VI Polytechnic University, Ben Guerir; Morocco.\\
% ...
$^{49}$Fakult\"{a}t Physik , Technische Universit{\"a}t Dortmund, Dortmund; Germany.\\
% ...
$^{55}$II. Physikalisches Institut, Georg-August-Universit\"{a}t G\"ottingen, G\"ottingen; Germany.\\
% ...
$^{90}$$^{(a)}$L2IT, Universit\'e de Toulouse, CNRS/IN2P3, UPS, Toulouse;$^{(b)}$CPPM, Aix-Marseille Universit\'e, CNRS/IN2P3, Marseille; France.\\
% ...
$^{117}$Department of Physics, New York University, New York NY; United States of America.\\
% ...
$^{120}$Homer L. Dodge Department of Physics and Astronomy, University of Oklahoma, Norman OK; United States of America.\\
% ...
$^{137}$$^{(a)}$Departamento de F\'isica, Pontificia Universidad Cat\'olica de Chile, Santiago;$^{(b)}$Millennium Institute for Subatomic physics at high energy frontier (SAPHIR), Santiago;$^{(c)}$Instituto de Investigaci\'on Multidisciplinario en Ciencia y Tecnolog\'ia, y Departamento de F\'isica, Universidad de La Serena;$^{(d)}$Universidad Andres Bello, Department of Physics, Santiago;$^{(e)}$Instituto de Alta Investigaci\'on, Universidad de Tarapac\'a, Arica;$^{(f)}$Departamento de F\'isica, Universidad T\'ecnica Federico Santa Mar\'ia, Valpara\'iso; Chile.\\
% ...
$^{150}$Department of Physics, Technion, Israel Institute of Technology, Haifa; Israel.\\
$^{151}$Raymond and Beverly Sackler School of Physics and Astronomy, Tel Aviv University, Tel Aviv; Israel.\\
% ...
$^{a}$ Also Affiliated with an institute covered by a cooperation agreement with CERN.\\
$^{b}$ Also at An-Najah National University, Nablus; Palestine.\\
$^{c}$ Also at APC, Universit\'e Paris Cit\'e, CNRS/IN2P3, Paris; France.\\
$^{d}$ Also at Borough of Manhattan Community College, City University of New York, New York NY; United States of America.\\
$^{e}$ Also at Center for High Energy Physics, Peking University; China.\\
$^{f}$ Also at Center for Interdisciplinary Research and Innovation (CIRI-AUTH), Thessaloniki; Greece.\\
% ...

\end{flushleft}


\end{document}